\documentclass[journal,twoside]{IEEEtran}
\usepackage{amsmath,epsfig,graphicx,amsbsy,amssymb,citesort,url}
\usepackage{subeqn,dsfont,algorithm,algorithmic}

\def\real    { \mathbb{R} }
\def \A {\mathcal{M}}
\def \aalg {\mathbb{M}}
\def \aaprox {\mathfrak{M}}
\def \taprox {\mathfrak{T}}
\def \saprox {\mathfrak{S}}
\def \T {\mathcal{T}}
\def \R {\mathcal{R}}

\def \S {\mathcal{S}}
\def \n {n} 
\def \trans {^T} 
\def \pinv {^\dag} 

\def \bpsi {\psi}
\def \bnu {\nu}
\def \X {X}
\def \x {x}
\def \s {s}
\def \y {y}
\def \u {u}
\def \w {w}
\def \r {r}
\def \b {b}
\def \n {n}

\def \e {e}
\def \xhat {\widehat{\x}}

\def \xbar {\bar{\x}}
\def \Xbar {\widetilde{\X}}
\def \xtilde {\widetilde{\x}}
\def \Xtilde {\widetilde{\X}}

\def \balpha {\alpha}

\newcommand{\abs}[1]{\left\vert#1\right\vert}

\newtheorem{THEO}{Theorem}
\newtheorem{LEMM}{Lemma}

\newtheorem{DEFI}{Definition}

\newtheorem{PROP}{Proposition}
\newcommand{\bigo}[1]{\mathcal{O}\left(#1\right)}

\newcommand{\qed}{{\unskip\nobreak\hfil\penalty50\hskip2em\vadjust{}
           \nobreak\hfil$\Box$\parfillskip=0pt\finalhyphendemerits=0\par}}

\begin{document}

\title{Model-Based Compressive Sensing}

\author{Richard G. Baraniuk,~\IEEEmembership{Fellow,~IEEE,} 
Volkan Cevher,~\IEEEmembership{Member,~IEEE,} \\
Marco F. Duarte,~\IEEEmembership{Member,~IEEE,} and 
Chinmay Hegde,~\IEEEmembership{Student Member,~IEEE}
\thanks{Manuscript received August 26, 2008; revised July 2, 2009.
The authors are listed alphabetically.  
This work was supported by grants NSF CCF-0431150, 
CCF-0728867, CNS-0435425, and CNS-0520280, 
DARPA/ONR N66001-08-1-2065, ONR N00014-07-1-0936, 
N00014-08-1-1067, N00014-08-1-1112, and
N00014-08-1-1066, AFOSR FA9550-07-1-0301, ARO MURI 
W911NF-07-1-0185, and the Texas Instruments Leadership 
University Program, and was completed while M. F. Duarte 
was a Ph.D. student at Rice University. The material in this 
paper was presented in part at the SIAM Conference on 
Imaging Science, San Diego, CA, June 2008, and at the 
Conference on Information Sciences
and Systems (CISS), Baltimore, MD, March 2009.}
\thanks{R. G. Baraniuk, V. Cevher, and C. Hegde are with Rice University, Houston, TX 77005 USA (e-mail: richb@rice.edu; vcevher@rice.edu; chinmay@rice.edu).}%
\thanks{V. Cevher is also with Ecole Polytechnique F\'ed\'erale de Laussane, Laussane, Switzerland (e-mail: volkan.cevher@epfl.ch).}
\thanks{M. F. Duarte is with Princeton University, Princeton, NJ 08544 USA (e-mail: mduarte@princeton.edu).}}

\markboth{Baraniuk \MakeLowercase{\textit{et al.}}: Model-based Compressive Sensing}%
{Baraniuk \MakeLowercase{\textit{et al.}}: Model-based Compressive Sensing}

\maketitle

\begin{abstract}
Compressive sensing (CS) is an alternative to Shannon/Nyquist
sampling for the acquisition of sparse or compressible signals that can
be well approximated by just $K\ll N$ elements from an
$N$-dimensional basis. Instead of taking periodic samples, CS
measures inner products with $M<N$ random vectors and then recovers
the signal via a sparsity-seeking optimization or greedy algorithm.
Standard CS dictates that robust signal recovery is
possible from $M=\bigo{K\log(N/K)}$ measurements. 
It is possible to substantially
decrease $M$ without sacrificing robustness by leveraging more
realistic signal models that go beyond simple sparsity and
compressibility by including structural dependencies between the 
values and locations of the signal coefficients. This paper introduces a model-based
CS theory that parallels the conventional theory and provides concrete
guidelines on how to create model-based recovery algorithms with
provable performance guarantees. A highlight is the introduction of a
new class of structured compressible signals along with a new sufficient
condition for robust structured compressible signal recovery that we dub
the restricted amplification property, which is the natural
counterpart to the restricted isometry property of conventional CS.
Two examples integrate two relevant signal models --- wavelet trees 
and block sparsity --- into two state-of-the-art CS recovery algorithms and
prove that they offer robust recovery from just $M=\bigo{K}$
measurements. Extensive numerical simulations demonstrate the
validity and applicability of our new theory and algorithms.
\end{abstract}

\begin{IEEEkeywords}\noindent
Compressive sensing, sparsity, signal model, union of subspaces,
wavelet tree, block sparsity
\end{IEEEkeywords}
%

\section{Introduction}

\IEEEPARstart{W}{e are} in the midst of a digital revolution that is enabling the
development and deployment of new sensors and sensing systems with
ever increasing fidelity and resolution.  The theoretical foundation
is the Shannon/Nyquist sampling theorem, which states that a signal's 
information is preserved if it is uniformly sampled at a rate at least two 
times faster than its Fourier bandwidth.
Unfortunately, in many important and emerging applications, the
resulting Nyquist rate can be so high that we end up with too many
samples and must compress in order to store or transmit them. In
other applications the cost of signal acquisition
is prohibitive, either because of a high cost per sample, or because
state-of-the-art samplers cannot achieve the high sampling rates
required by Shannon/Nyquist. Examples include radar imaging and
exotic imaging modalities outside visible wavelengths.

Transform compression systems reduce the effective dimensionality of
an $N$-dimensional signal $x$ by re-representing it in terms of a
sparse or compressible set of coefficients $\alpha$ in a basis expansion
$x=\Psi\alpha$, with $\Psi$ an $N\times N$ basis matrix. By sparse
we mean that only $K\ll N$ of the coefficients $\alpha$ are nonzero
and need to be stored or transmitted.  By compressible we mean that
the coefficients $\alpha$, when sorted, decay rapidly enough to zero
that $\alpha$ can be well-approximated as $K$-sparse. The sparsity
and compressibility properties are pervasive in many signal classes of interest. 
For example, smooth signals and images are compressible in the Fourier 
basis, while piecewise smooth signals and images are compressible in a
wavelet basis \cite{Mallatbook}; the JPEG and JPEG2000 standards are
examples of practical transform compression systems based on these
bases.

{\em Compressive sensing} (CS) provides an alternative to
Shannon/Nyquist sampling when the signal under acquisition is known
to be sparse or compressible~\cite{DonohoCS,CandesCS,richbCS}. In
CS, we measure not periodic signal samples but rather inner products
with $M \ll N$ measurement vectors.  In matrix notation, the
measurements $y=\Phi x=\Phi\Psi\alpha$, where the rows of the
$M\times N$ matrix $\Phi$ contain the measurement vectors. While the
matrix $\Phi\Psi$ is rank deficient, and hence loses information in
general, it can be shown to preserve the information in sparse and
compressible signals if it satisfies the so-called {\em restricted
isometry property} (RIP) \cite{CandesCS}. Intriguingly, a large
class of random matrices have the RIP with high probability. To
recover the signal from the compressive measurements $\y$, we search
for the sparsest coefficient vector $\alpha$ that agrees with the
measurements. To date, research in CS has focused primarily on
reducing both the number of measurements $M$ (as a function of $N$
and $K$) and on increasing the robustness and reducing the
computational complexity of the recovery algorithm. Today's
state-of-the-art CS systems can robustly recover $K$-sparse and compressible
signals from just $M=\bigo{K \log(N/K)}$ noisy measurements using
polynomial-time optimization solvers or greedy algorithms.

While this represents significant progress from Nyquist-rate
sampling, our contention in this paper is that it is possible to do
even better by more fully leveraging concepts from state-of-the-art
signal compression and processing algorithms. In many such
algorithms, the key ingredient is a more realistic {\em structured sparsity
model} that goes beyond simple sparsity by codifying the
inter-dependency {\em structure} among the signal coefficients
$\alpha$.\footnote{Obviously, sparsity and compressibility
correspond to simple signal models where each coefficient is treated
independently; for example in a sparse model, the fact that the
coefficient $\alpha_i$ is large has no bearing on the size of any
$\alpha_j$, $j\neq i$. We will reserve the use of the term ``model''
for situations where we are enforcing structured dependencies between the values
and the locations of the coefficients $\alpha_i$.} 
For instance, modern wavelet image coders
exploit not only the fact that most of the wavelet coefficients of a
natural image are small but also the fact that the values and
locations of the large coefficients have a particular structure.
Coding the coefficients according to a structured sparsity model
enables these algorithms to compress images close to the maximum
amount possible -- significantly better than a na\"{i}ve coder that
just processes each large coefficient independently.
We have previously developed a new CS recovery algorithm that 
promotes structure in the sparse representation by tailoring the
recovered signal according to a sparsity-promoting probabilistic 
model, such as an Ising graphical model~\cite{csmrf}. Such 
probabilistic models favor certain configurations for the magnitudes 
and indices of the significant coefficients of the signal.

In this paper, we expand on this concept by introducing a 
model-based CS theory that parallels the conventional theory and 
provides concrete guidelines on how to create structured signal recovery 
algorithms with provable performance guarantees. By reducing the number
of degrees of freedom of a sparse/compressible signal by permitting 
only certain configurations of the large and zero/small coefficients, 
structured sparsity models provide two immediate benefits to CS. 
First, they enable us to reduce, in some cases significantly, the 
number of measurements $M$ required to stably recover a signal. 
Second, during signal recovery, they enable us to better differentiate 
true signal information from recovery artifacts, which leads to a more 
robust recovery.

To precisely quantify the benefits of model-based CS, we introduce
and study several new theoretical concepts that could be of more
general interest. We begin with structured sparsity models for 
$K$-sparse signals and make precise how the structure reduces
the number of potential sparse signal supports in $\alpha$. Then
using the {\em model-based restricted isometry property} from
\cite{samplingunion,dosamplingunion}, we prove that such {\em
structured sparse signals} can be robustly recovered from noisy
compressive measurements. Moreover, we quantify the required number
of measurements $M$ and show that for some structured sparsity 
models $M$ is independent
of $N$. These results unify and generalize the limited related work
to date on structured sparsity models for strictly sparse signals
\cite{samplingunion,dosamplingunion,Hassibi,EldarUSS,DCS}. We then
introduce the notion of a {\em structured compressible signal}, whose
coefficients $\alpha$ are no longer strictly sparse but have a
structured power-law decay. To establish that structured compressible
signals can be robustly recovered from compressive measurements, we
generalize the standard RIP to a new {\em restricted amplification
property} (RAmP).  Using the RAmP, we show that the required
number of measurements $M$ for recovery of structured compressible 
signals is independent of $N$.

To take practical advantage of this new theory, we demonstrate how
to integrate structured sparsity models into two state-of-the-art CS recovery
algorithms, CoSaMP \cite{CoSaMP} and iterative hard thresholding
(IHT)~\cite{NowakEM,DaubechiesThresholding,CandesPSR,IHT2,IHT}. 
The key modification is surprisingly simple:  we
merely replace the nonlinear sparse approximation step in these greedy
algorithms with a structured sparse approximation. Thanks to our new
theory, both new model-based recovery algorithms have provable
robustness guarantees for both structured sparse and structured compressible
signals.

\begin{figure*}[!t]
\centering
\begin{tabular}{cccc}
{\includegraphics[width=0.2\hsize]{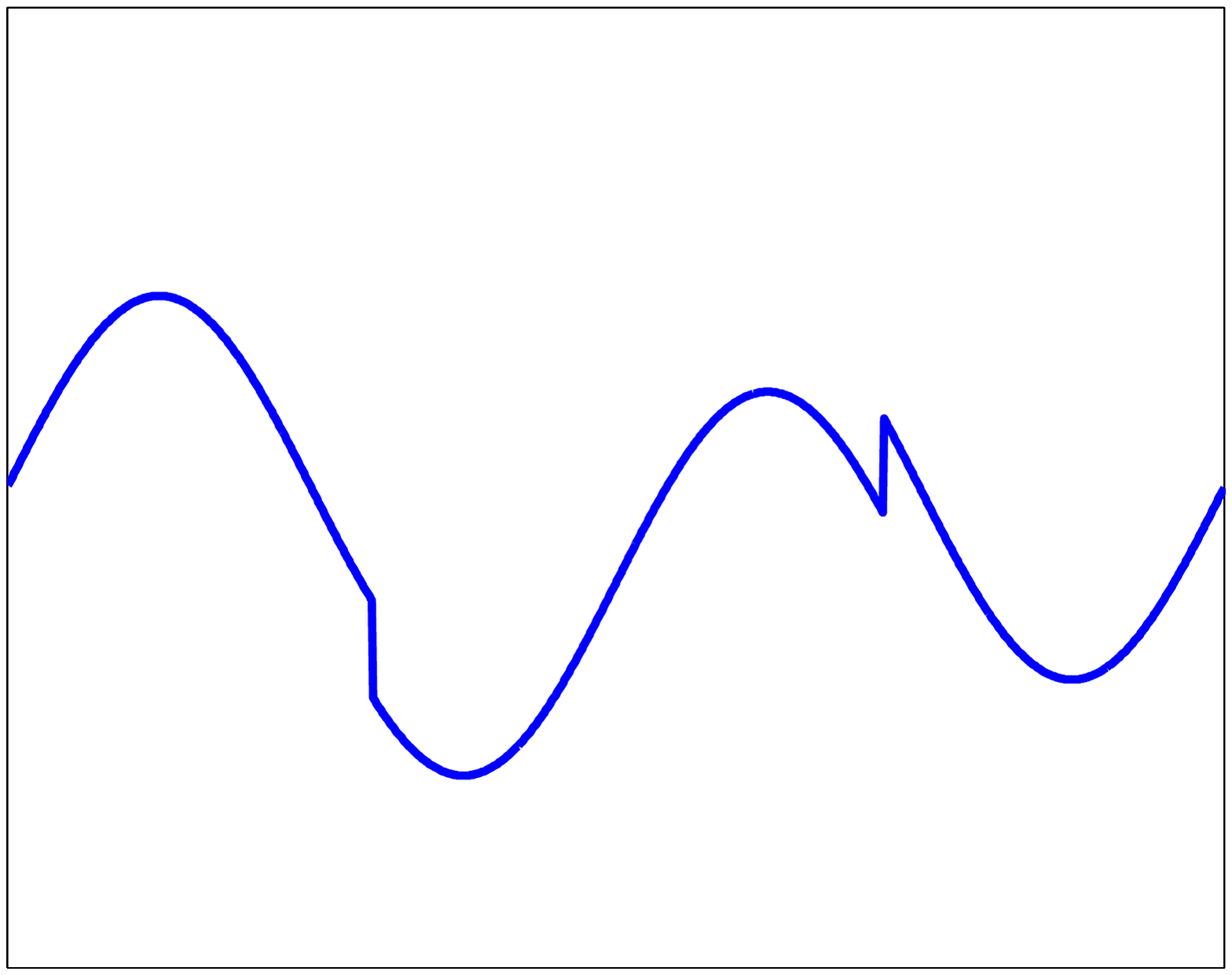}}&
{\includegraphics[width=0.2\hsize]{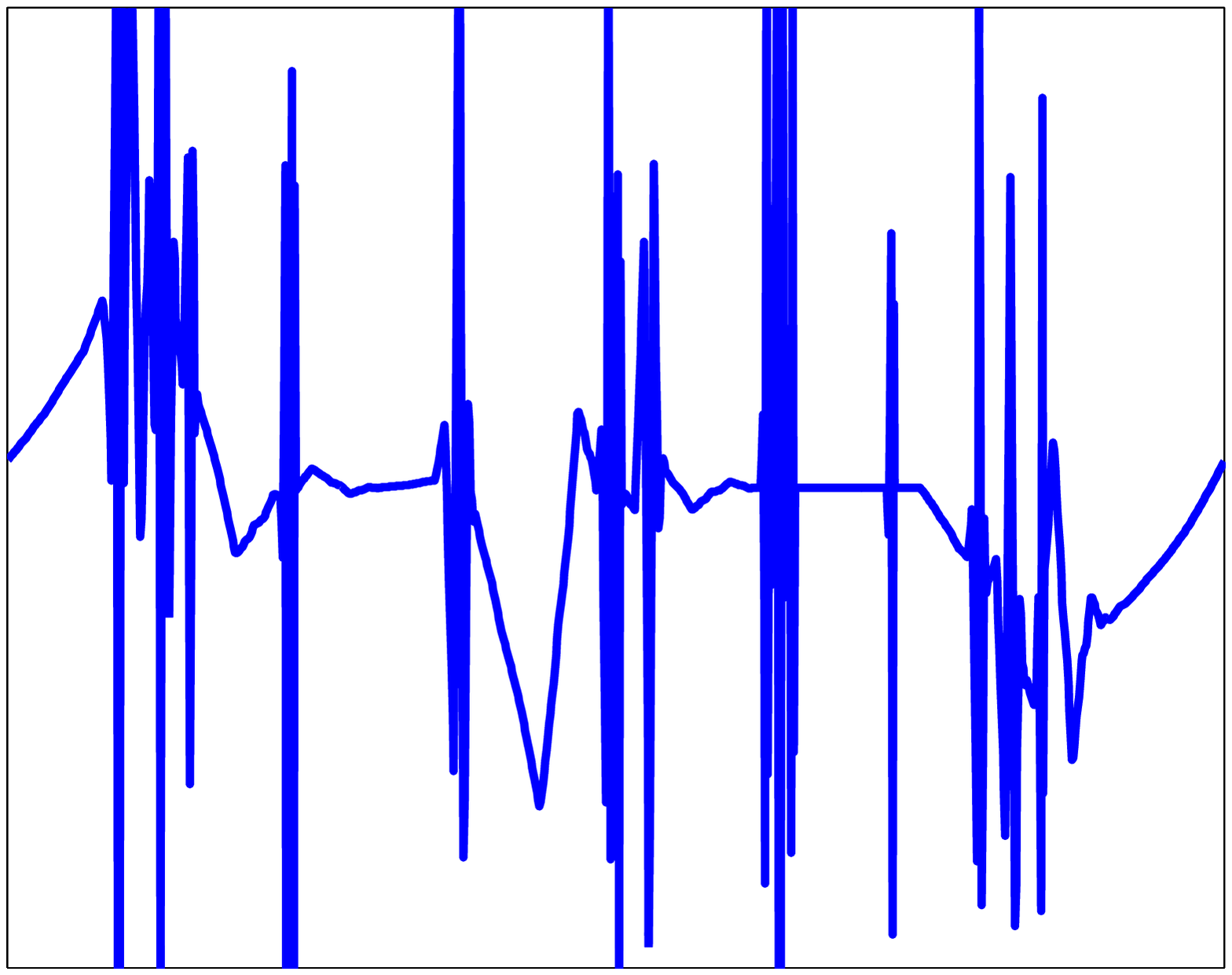}}&
{\includegraphics[width=0.2\hsize]{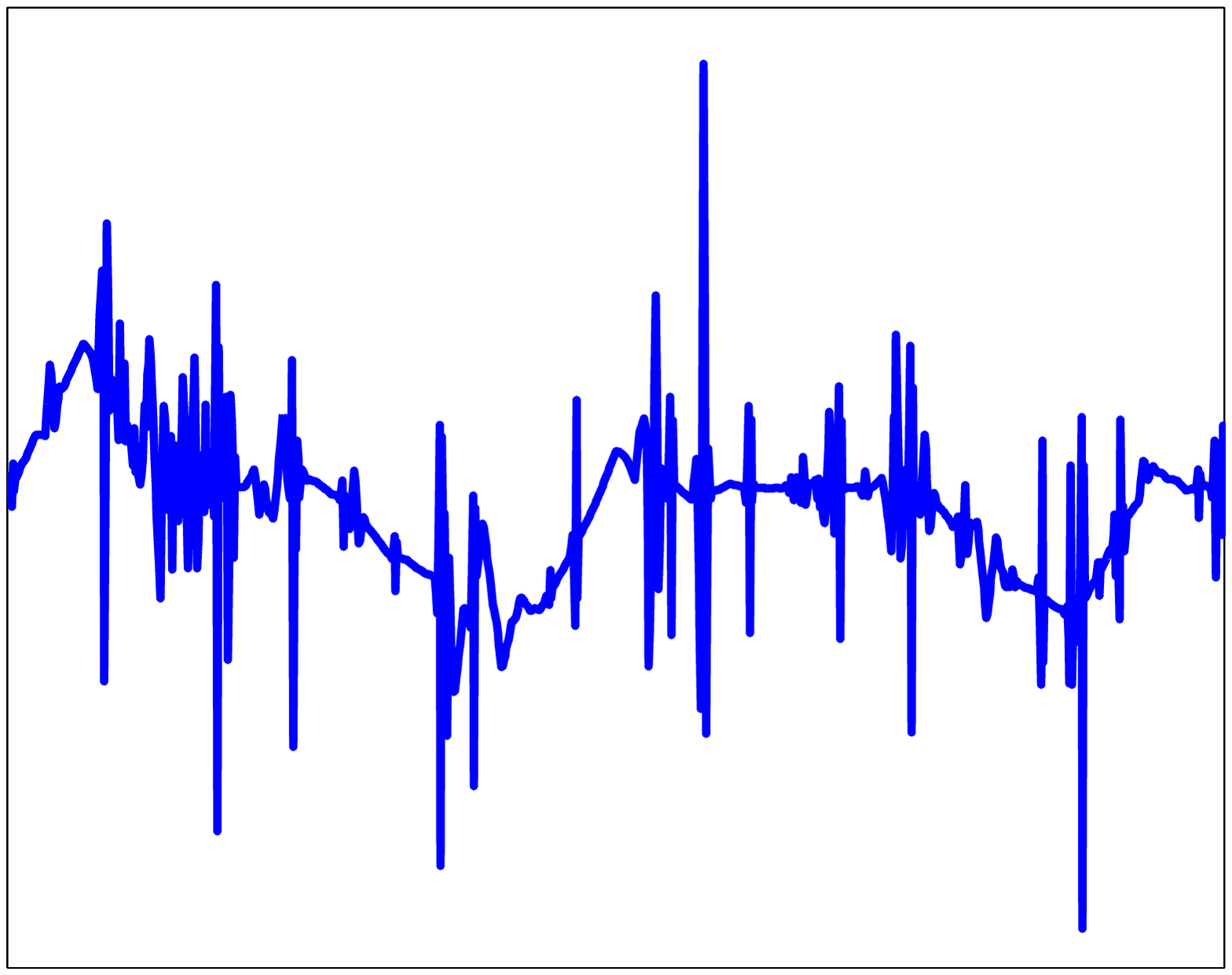}} &
{\includegraphics[width=0.2\hsize]{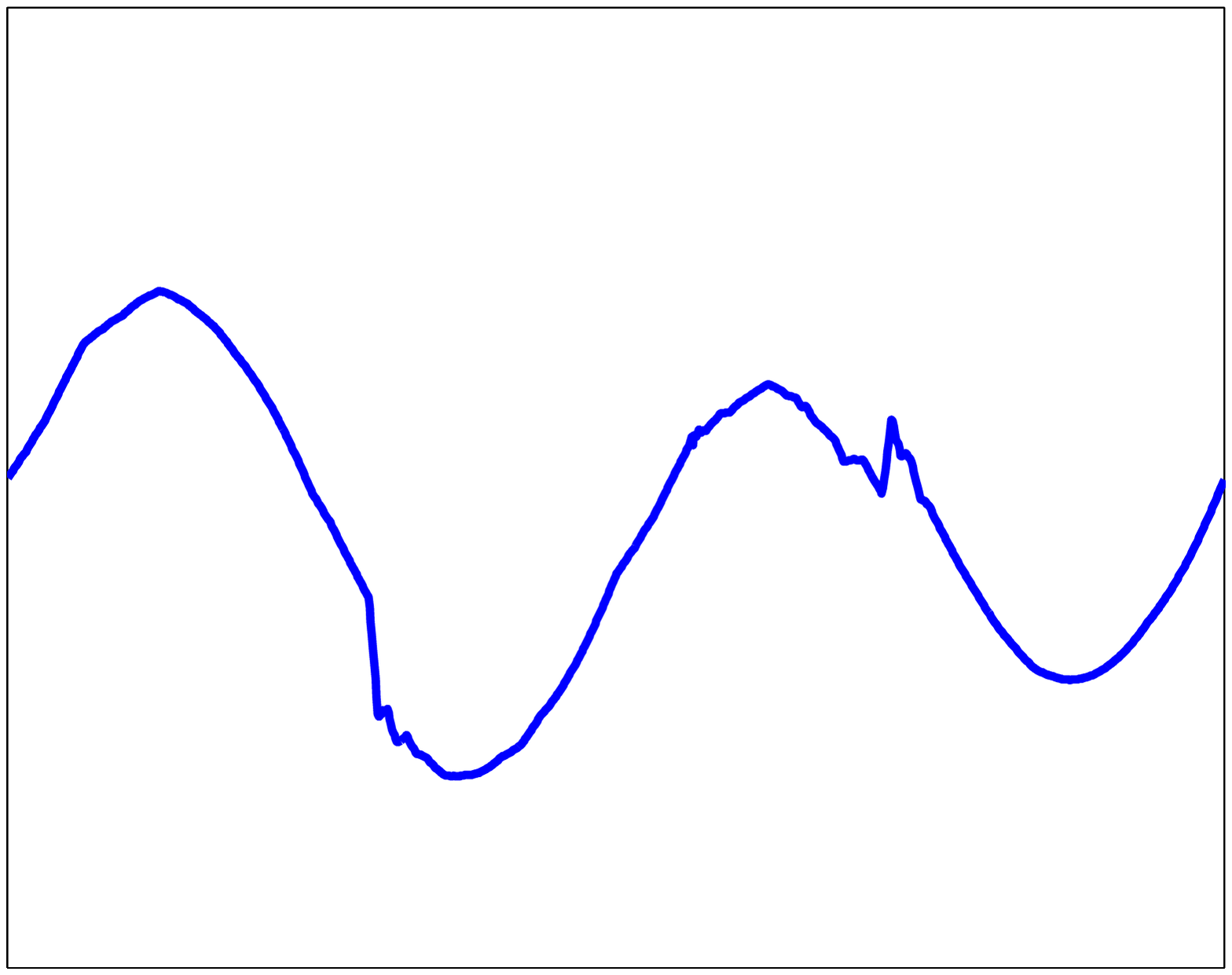}}\\
(a) test signal & (b) CoSaMP &
(c) $\ell_1$-norm min. &  (d) model-based recovery\\
& (RMSE $=1.123$) & (RMSE $=0.751$) & (RMSE $=0.037$)
\end{tabular}
\caption{\sl Example performance of structured signal
recovery. (a) Piecewise smooth {\em HeaviSine} test signal of length
$N=1024$.  This signal is compressible under a connected wavelet
tree model. Signal recovered from $M=80$ random Gaussian
measurements using (b) the iterative recovery algorithm CoSaMP, (c)
standard $\ell_1$-norm minimization via linear programming, and 
(d) the wavelet tree-based CoSaMP algorithm from Section \ref{sec:trees}.  
In all figures, root mean-squared error (RMSE) values are normalized 
with respect to the $\ell_2$ norm of the signal. \label{fig:wvcomparison}}
\end{figure*}

To validate our theory and algorithms and demonstrate their
general applicability and utility, we present two specific instances
of model-based CS and conduct a range of simulation experiments. The
first structured sparsity model accounts for the fact that the large wavelet
coefficients of piecewise smooth signals and images tend to live on
a rooted, connected {\em tree structure} \cite{HMT}. Using the fact
that the number of such trees is much smaller than $N \choose K$,
the number of $K$-sparse signal supports in $N$ dimensions, we prove
that a tree-based CoSaMP algorithm needs only $M=\bigo K$
measurements to robustly recover tree-sparse and tree-compressible
signals. This provides a significant reduction against the standard CS 
requirement $M = \bigo{K\log (N/K)}$ as the signal length $N$ increases. 
Figure \ref{fig:wvcomparison} indicates the potential
performance gains on a tree-compressible, piecewise smooth signal.

The second structured sparsity model accounts for the fact that 
the large coefficients of many sparse signals cluster together 
\cite{Hassibi,EldarUSS}.
Such a so-called {\em block sparse} model is equivalent to a {\em
joint sparsity} model for an ensemble of $J$, length-$N$ signals
\cite{DCS}, where the supports of the signals' large coefficients
are shared across the ensemble. Using the fact that the number of
clustered supports is much smaller than $JN \choose JK$, we prove
that a block-based CoSaMP algorithm needs only $M=\bigo{JK+
K \log (\frac{N}{K})}$ measurements to robustly recover
block-sparse and block-compressible signals.  In contrast, standard 
CS requires $M = \bigo{JK \log(N/K)}$; block sparsity reduces the 
dependence of $M$ on the signal length $N$, particularly for large 
block sizes $J$.

Our new theory and methods relate to a small body of previous work
aimed at integrating structured sparsity into CS. Several 
groups have developed structured sparse signal recovery algorithms
\cite{samplingunion,dosamplingunion,Hassibi,RichBUMD05,MarcoSPARS05,LaDoICIP,DuarteICASSP08,DoLaCAMSAP,WaveletBCS,BreslerSIAM};
however, their approaches have either been ad hoc or focused on a
single structured sparsity model. Most previous work on unions of subspaces
\cite{samplingunion,dosamplingunion,BreslerSIAM} has focused exclusively on
strictly sparse signals and has considered neither compressibility nor feasible 
recovery algorithms. A related CS modeling framework for structured sparse
and compressible signals~\cite{EldarUSS} collects the $N$ samples 
of a signal into $D$ groups, $D \le N$, and allows signals where $K$ out of 
$D$ groups have nonzero coefficients.
This framework is immediately applicable to block-sparse signals and
signal ensembles with common sparse supports.
While \cite{EldarUSS} provides recovery algorithms, measurement
bounds, and recovery guarantees similar to those provided in
Section~\ref{sec:ensembles}, our proposed framework has the ability
to focus on arbitrary subsets of the $D \choose K$ groups that yield
more elaborate structures, such as connected subtrees for wavelet
coefficients.
To the best of our knowledge, our general algorithmic framework for
model-based recovery, the concept of a model-compressible 
signal, and the associated RAmP are new to the literature.

This paper is organized as follows. A review of the CS theory in
Section~\ref{sec:back} lays out the foundational concepts that we
extend to the model-based case in subsequent sections.  
Section~\ref{sec:models} develops the concept of structured sparse signals 
and introduces the concept of structured compressible signals. We also
quantify how structured sparsity models improve the measurement and 
recovery process by exploiting the model-based RIP for structured sparse 
signals and by introducing the RAmP for structured compressible signals. 
Section~\ref{sec:recovery} indicates how to tune CoSaMP to incorporate 
structured sparsity models and establishes its robustness properties for
structured sparse and structured compressible signals; the 
modifications to the IHT algorithm are very similar, so we defer them to 
an appendix to reduce redundancy. Sections~\ref{sec:trees} 
and \ref{sec:ensembles} then specialize our theory to the special cases of 
wavelet tree and block sparse signal models, respectively, and report on 
a series of numerical experiments that validate our theoretical claims.  We 
conclude with a discussion in Section~\ref{sec:concl}. To make the paper 
more readable, all proofs are relegated to a series of appendices.

\section{Background on Compressive Sensing}
\label{sec:back}

\subsection{Sparse and compressible signals}
\label{sec:normalsparse}

Given a basis $\{\psi_i\}_{i=1}^N$, we can represent every signal
$\x\in\real^N$  in terms of $N$ coefficients $\{\alpha_i\}_{i=1}^N$ as
$\x = \sum_{i=1}^N \alpha_i\psi_i$;
stacking the $\psi_i$ as columns into the $N\times N$ matrix $\Psi$,
we can write succinctly that $\x= \Psi \alpha$. In the sequel, we will
assume without loss of generality that the
signal $\x$ is sparse or compressible in the canonical domain so
that the sparsity basis $\Psi$ is the identity and $\alpha=\x$.

A signal $\x$ is $K$-{\em sparse} if only $K\ll N$ entries of $\x$ are 
nonzero.  We call the set of indices corresponding to the nonzero 
entries the {\em support} of $x$ and denote it by $\textrm{supp}(x)$. 
The set of all $K$-sparse signals is the union of the $N \choose K$, 
$K$-dimensional subspaces aligned with the coordinate axes
in $\real^N$. We denote this union of subspaces by $\Sigma_K$.

Many natural and manmade signals are not strictly sparse, but can be
approximated as such; we call such signals {\em compressible}.
Consider a signal $x$ whose coefficients, when sorted in order of
decreasing magnitude, decay according to the power law
\begin{equation}
\abs{\x_{{\mathcal I}(i)}} \le G \, i^{-1/r}, ~~~ i=1,\ldots,N,
\label{eq:comp}
\end{equation}
where ${\mathcal I}$ indexes the sorted coefficients. Thanks to the
rapid decay of their coefficients, such signals are
well-approximated by $K$-sparse signals. Let $\x_K \in \Sigma_K$
represent the best $K$-term approximation of $\x$, which is obtained
by keeping just the first $K$ terms in $\x_{{\mathcal I}(i)}$ from
(\ref{eq:comp}). Denote the error of this approximation in the
$\ell_p$ norm as
\begin{equation}
\sigma_K(\x)_p := \min_{\bar{\x} \in \Sigma_K} \|\x-\bar{\x}
\|_p = \|\x-\x_K \|_p,
\end{equation}
where the $\ell_p$ norm of the vector $\x$ is defined as $\|\x\|_p =
\left(\sum_{i=1}^N |\x_i|^p \right)^{1/p}$ for $0<p<\infty$. Then,
for $r < p$, we have that
\begin{equation}
\sigma_K(\x)_p \le \left(rs\right)^{-1/p}G K^{-s}, \label{eq:kta}
\end{equation}
with $s = \frac{1}{r}-\frac{1}{p}$. That is, when measured in the
$\ell_p$ norm, the signal's best approximation error has a power-law
decay with exponent $s$ as $K$ increases. In the sequel we let $p = 2$, 
yielding $s = 1/r-1/2$, and we dub a signal that obeys (\ref{eq:kta}) an 
{\em $s$-compressible} signal.

The approximation of compressible signals by sparse signals is the
basis of {\em transform coding} as is used in algorithms like JPEG
and JPEG2000 \cite{Mallatbook}.  In this framework, we acquire the
full $N$-sample signal $\x$; compute the complete set of transform
coefficients $\alpha$ via $\alpha=\Psi^{-1}\x$; locate the $K$
largest coefficients and discard the $(N-K)$ smallest coefficients;
and encode the $K$ values and locations of the largest coefficients.
While a widely accepted standard, this sample-then-compress
framework suffers from three inherent inefficiencies. First, we must
start with a potentially large number of samples $N$ even if the
ultimate desired $K$ is small. Second, the encoder must compute all
of the $N$ transform coefficients $\alpha$, even though it will
discard all but $K$ of them. Third, the encoder faces the overhead
of encoding the locations of the large coefficients.

\subsection{Compressive measurements and the restricted isometry
property}

Compressive sensing (CS) integrates the signal acquisition and
compression steps into a single process
\cite{DonohoCS,CandesCS,richbCS}.  In CS we do not acquire $\x$
directly but rather acquire $M<N$ linear measurements $\y=\Phi\x$
using an $M\times N$ measurement matrix $\Phi$. We then recover $\x$
by exploiting its sparsity or compressibility.
Our goal is to push $M$ as close as possible to $K$ in order to
perform as much signal ``compression'' during acquisition as
possible.

In order to recover a good estimate of $\x$ (the $K$ largest
$\x_i$'s, for example) from the $M$ compressive measurements, the
measurement matrix $\Phi$ should satisfy the {\em restricted
isometry property} (RIP) \cite{CandesCS}.
\begin{DEFI}
An $M\times N$ matrix $\Phi$ has the {\em $K$-restricted isometry
property} ($K$-RIP) with constant $\delta_K$ if, for all
$\x \in \Sigma_K$,
\begin{equation}
(1-\delta_K) \|\x\|_2^2 \le \|\Phi\x\|_2^2 \le (1+\delta_K)
\|\x\|_2^2.
\end{equation}
\end{DEFI}

In words, the $K$-RIP ensures that all submatrices of $\Phi$ of size
$M \times K$ are close to an isometry, and therefore distance (and
information) preserving.  Practical recovery algorithms typically
require that $\Phi$ have a slightly stronger $2K$-RIP, $3K$-RIP, or
higher-order RIP in order to preserve distances between $K$-sparse
vectors (which are $2K$-sparse in general), three-way sums of
$K$-sparse vectors (which are $3K$-sparse in general), and other
higher-order structures.

While checking whether a measurement matrix $\Phi$ satisfies the
$K$-RIP is an NP-Complete problem in general~\cite{natarajan}, random
matrices whose entries are independent and identically distributed (i.i.d.) 
Gaussian, Rademacher ($\pm 1$), or more generally
subgaussian\footnote{A random variable $X$ is called subgaussian if
there exists $c>0$ such that $\mathbb{E}\left(e^{Xt}\right) \le
e^{c^2t^2/2}$ for all $t \in \real$. Examples include the Gaussian, Bernoulli,
and Rademacher random variables, as well as any bounded random
variable.~\cite{subgaussian}} work with high probability
provided $M=\bigo{K \log(N/K)}$. These random matrices also have a
so-called {\em universality} property in that, for any choice of orthonormal basis
matrix $\Psi$, $\Phi\Psi$ has the $K$-RIP with high probability.  This is useful 
when the signal is sparse not in the canonical domain but in basis $\Psi$. 
A random $\Phi$ corresponds to an intriguing data acquisition protocol in 
which each measurement $y_j$ is a randomly weighted linear combination 
of the entries of $x$.

\subsection{Recovery algorithms}
\label{sec:normalrecovery}

Since there are infinitely many signal coefficient vectors
$\x^\prime$ that produce the same set of compressive measurements
$\y=\Phi\x$, to recover the ``right'' signal we exploit our a priori
knowledge of its sparsity or compressibility.  For example, we could
seek the sparsest $\x$ that agrees with the measurements $\y$:
\begin{equation}
\widehat{\x} = \arg\min_{\x^\prime} \|\x^\prime\|_0 \textrm{~s.t.~}\y=\Phi\x^\prime,
\label{eq:L0}
\end{equation}
where the $\ell_0$ ``norm'' of a vector counts its number of nonzero
entries. While this optimization can recover a $K$-sparse signal
from just $M=2K$ compressive measurements, it is unfortunately a
combinatorial, NP-hard problem~\cite{natarajan}; furthermore, the recovery is not
stable in the presence of noise~\cite{richbCS}.

\sloppy Practical, stable recovery algorithms rely on the RIP (and
therefore require at least $M=\bigo{K \log(N/K)}$ measurements);
they can be grouped into two camps.  The first approach convexifies
the $\ell_0$ ``norm'' minimization (\ref{eq:L0}) to the $\ell_1$-norm minimization
\begin{equation}
\widehat{\x} = \arg\min_{\x^\prime} \|\x^\prime\|_1 \textrm{~s.t.~}\y=\Phi\x^\prime.
\label{eq:L1}
\end{equation}
This corresponds to a linear program that can be solved in polynomial time 
\cite{CandesCS,DonohoCS}. Adaptations to deal with additive noise in
$\y$ or $\x$ include basis pursuit with denoising
(BPDN)~\cite{BPDN}, complexity-based regularization~\cite{Nowak},
and the Dantzig Selector~\cite{CandesDS}.

The second approach finds the sparsest $\x$ agreeing with the
measurements $\y$ through an iterative, greedy search. Algorithms
such as matching pursuit, orthogonal matching pursuit \cite{OMP},
StOMP~\cite{STOMP}, iterative hard thresholding 
(IHT)~\cite{NowakEM,DaubechiesThresholding,CandesPSR,IHT2,IHT},
CoSaMP~\cite{CoSaMP}, and Subspace Pursuit (SP)~\cite{SP} all
revolve around a best $L$-term approximation for the estimated
signal, with $L$ varying for each algorithm; typically $L$ is $\bigo K$.

\subsection{Performance bounds on signal recovery}

Given $M=\bigo{K \log(N/K)}$ compressive measurements, a number of
different CS signal recovery algorithms, including all of the
$\ell_1$-norm minimization techniques mentioned above and the 
CoSaMP, SP, and IHT iterative techniques, offer provably stable signal 
recovery with performance close to optimal $K$-term approximation (recall
(\ref{eq:kta}))~\cite{CandesCS,DonohoCS,IHT,CoSaMP}. 
For a random $\Phi$, all results hold with high probability.

For a noise-free, $K$-sparse signal, these algorithms offer perfect
recovery, meaning that the signal $\xhat$ recovered from the
compressive measurements $\y=\Phi x$ is exactly $\xhat=x$.

For a $K$-sparse signal $\x$ whose measurements are corrupted by
noise $n$ of bounded norm (that is, we measure $y=\Phi x + n$)
the mean-squared error of the signal $\xhat$ is
\begin{equation}
\| x - \xhat \|_2 \leq C \| n \|_2,
\end{equation}
with $C$ a small constant.

For an $s$-compressible signal $\x$ whose measurements are corrupted
by noise $n$ of bounded norm, the mean-squared error of the
recovered signal $\xhat$ is
\begin{equation}
\| x - \xhat \|_2 \leq C_1 \|x-x_K\|_2 + C_2\frac{1}{\sqrt{K}}
\|x-x_K\|_1 + C_3 \|n\|_2.
\label{eq:cosampguarantee}
\end{equation}
Using (\ref{eq:kta}) we can simplify this expression to
\begin{equation}
\| x - \xhat \|_2 \leq \frac{C_1GK^{-s}}{\sqrt{2s}} +
\frac{C_2GK^{-s}}{s-1/2} + C_3 \|n\|_2.
\end{equation}
For the recovery algorithm (\ref{eq:L1}), we obtain a bound very similar to (\ref{eq:cosampguarantee}), albeit with the $\ell_2$-norm error component removed~\cite{CandesRIP}.

\section{Structured Sparsity and Compressibility}
\label{sec:models}

While many natural and manmade signals and images can be described
to first-order as sparse or compressible, the support of their large
coefficients often has an underlying inter-dependency structure.
This phenomenon has received only limited attention by the CS
community to date
\cite{samplingunion,dosamplingunion,Hassibi,EldarUSS,MarcoSPARS05,LaDoICIP,DuarteICASSP08,DoLaCAMSAP,WaveletBCS}.
In this section, we introduce a model-based theory of CS that
captures such structure.  A model reduces the degrees of freedom of
a sparse/compressible signal by permitting only certain
configurations of supports for the large coefficient. As we will
show, this allows us to reduce, in some cases significantly, the
number of compressive measurements $M$ required to stably recover a
signal.

\subsection{Structured sparse signals} \label{sec:msparse}

Recall from Section \ref{sec:normalsparse} that a $K$-sparse signal
vector $\x$ lives in $\Sigma_K \subset \real^N$, which is a union of
$N \choose K$ subspaces of dimension $K$. Other than its
$K$-sparsity, there are no further constraints on the support or
values of its coefficients.  A {\em structured sparsity model} endows the
$K$-sparse signal $\x$ with additional structure that allows certain
$K$-dimensional subspaces in $\Sigma_K$ and disallows others
\cite{samplingunion,dosamplingunion}.

To state a formal definition of a structured sparsity model, let $\x\vert_\Omega$
represent the entries of $\x$ corresponding to the set of indices
$\Omega \subseteq \{1,\ldots,N\}$, and let $\Omega^C$ denote the
complement of the set $\Omega$.
\begin{DEFI}
\label{def:model} A {\em structured sparsity model} $\A_K$ is defined as the
union of $m_K$ canonical $K$-dimensional subspaces
\begin{equation}
\A_K = \bigcup_{m=1}^{m_K}  {\mathcal X}_m,~\text{s.t.}~
{\mathcal X}_m:= \{\x : \x\vert_{\Omega_m} \in \real^{K},
\x\vert_{\Omega_m^C} = 0\}, \nonumber
\end{equation}
where $\{\Omega_1,\ldots,\Omega_{m_K}\}$ is the set containing 
all allowed supports, with $|\Omega_m| = K$ for each 
$m=1,\ldots,m_K$, and each subspace $\mathcal{X}_m$ contains 
all signals $\x$ with $\mathrm{supp}(\x) \subseteq \Omega_m$.
\end{DEFI}
Signals from $\A_K$ are called {\em $K$-structured sparse}. Clearly,
$\A_K \subseteq \Sigma_K$ and contains $m_K\leq {N \choose K}$
subspaces.

In Sections~\ref{sec:trees} and~\ref{sec:ensembles} below we
consider two concrete structured sparsity models. The first model accounts for
the fact that the large wavelet coefficients of piecewise smooth
signals and images tend to live on a rooted, connected {\em tree
structure} \cite{HMT}.  The second model accounts for the fact that
the large coefficients of sparse signals often {\em cluster 
together into blocks} \cite{Hassibi,EldarUSS,DCS}.

\subsection{Model-based RIP} \label{sec:mrip}

If we know that the signal $\x$ being acquired is $K$-structured sparse,
then we can relax the RIP constraint on the CS measurement matrix
$\Phi$ and still achieve stable recovery from the compressive
measurements $\y=\Phi\x$ \cite{samplingunion,dosamplingunion}.

\begin{DEFI}~\cite{samplingunion,dosamplingunion}
An $M\times N$ matrix $\Phi$ has the {\em $\A_K$-restricted isometry
property} ($\A_K$-RIP) with constant $\delta_{\A_K}$ if, for all $\x
\in \A_K$, we have
\begin{equation}
(1-\delta_{\A_K}) \|\x\|_2^2 \le \|\Phi\x\|_2^2 \le
(1+\delta_{\A_K}) \|\x\|_2^2. \label{eq:modelrip}
\end{equation}
\end{DEFI}

Blumensath and Davies~\cite{samplingunion} have quantified the
number of measurements $M$ necessary for a random CS matrix to have
the $\A_K$-RIP with a given probability.

\begin{THEO}~\cite{samplingunion}
Let $\A_K$ be the union of $m_K$ subspaces of $K$-dimensions in
$\real^N$. Then, for any $t>0$ and any
\begin{equation}
M \ge \frac{2}{c \delta_{\A_K}^2}\left(\ln(2 m_K) + K \ln
\frac{12}{\delta_{\A_K}}+t \right), \label{eq:blum} \nonumber
\end{equation}
where $c$ is a positive constant, an $M\times N$ i.i.d.\ subgaussian random matrix has the $\A_K$-RIP
with constant $\delta_{\A_K}$ with probability at least $1-e^{-t}$.
\label{theo:munion}
\end{THEO}

This bound can be used to recover the conventional CS result by
substituting $m_K = {N \choose K} \approx (Ne/K)^K$. Similarly, 
as the number of subspaces $m_K$ that arise from the structure 
imposed can be significantly smaller than the standard 
${N \choose K}$, the number of rows needed for a random matrix 
to have the $\A_K$-RIP can be significantly lower than the number 
of rows needed for the standard RIP. The $\A_K$-RIP
property is sufficient for robust recovery of structured sparse signals,
as we show below in Section~\ref{sec:msr}.

\subsection{Structured compressible signals} \label{sec:mcomp}

Just as compressible signals are ``nearly $K$-sparse'' and thus live
close to the union of subspaces $\Sigma_K$ in $\real^N$,
structured compressible signals are ``nearly $K$-structured sparse'' and live
close to the restricted union of subspaces $\A_K$. In this section,
we make this new concept rigorous.  Recall from (\ref{eq:kta}) that
we defined compressible signals in terms of the decay of their
$K$-term approximation error.

The $\ell_2$ error incurred by approximating $x\in\real^N$ by the
best structured sparse approximation in $\A_K$ is given by
\begin{equation}\label{eq:modelapprox} 
\sigma_{\A_K}(\x) := \inf_{\xbar \in \A_K} \|\x - \xbar\|_2. 
\nonumber
\end{equation}
We define $\aalg_B(\x,K)$ as the algorithm that obtains the best 
{\em $K$-term structured sparse approximation} of $\x$ in the 
union of subspaces $\A_K$:
\begin{equation}
\aalg(\x,K) = \arg \min_{\xbar \in \A_K} \|\x-\xbar\|_2. \nonumber
\end{equation}
This implies that  $\|\x-\aalg(\x,K)\|_2 = \sigma_{\A_K}(\x)$.
The decay of this approximation error defines the structured compressibility of
a signal.

\begin{DEFI}
The set of {\em $s$-structured compressible signals} is defined as
\begin{equation}\nonumber
\aaprox_s = \left\{\x \in \real^N: \sigma_{\A_K}(\x) \le G K^{-s},
1\le K \le N, G < \infty\right\}.
\end{equation}
Define $|\x|_{\aaprox_s}$ as the smallest value of $G$ for which
this condition holds for $\x$ and $s$.
\end{DEFI}

We say that $\x \in \aaprox_s$ is an {\em $s$-structured compressible
signal} under the structured sparsity model $\A_K$. These approximation classes
have been characterized for certain structured sparsity models; see
Section~\ref{sec:trees} for an example. We will select the value of $s$ 
for which the distance between the approximation errors 
$\sigma_{\A_K}(\x)$ and the corresponding bounds $GK^{-1/s}$ is minimal.

\subsection{Nested model approximations and residual subspaces} \label{sec:rsubs}

In conventional CS, the same requirement (RIP) is a sufficient
condition for the stable recovery of both sparse and compressible
signals. In model-based recovery, however, the class of structured 
compressible signals is much larger than that of structured sparse 
signals, since the union of subspaces defined by structured sparse 
signals does not contain all canonical $K$-dimensional subspaces. 

To address this difference, we introduce some additional tools to 
develop a {\em sufficient} condition for the stable recovery of structured 
compressible signals. We will pay particular attention to structured 
sparsity models $\A_K$ that generate {\em nested approximations}, 
since they are more amenable to analysis and computation.

\begin{DEFI}
A structured sparsity model $\A = \{\A_1, \A_2, \ldots\}$ has the {\em nested
approximation property} (NAP)
if $\mathrm{supp}(\aalg(\x,K)) \subset \mathrm{supp}(\aalg(\x,K'))$
for all $K < K'$ and for all $\x \in \real^N$.
\end{DEFI}

In words, a structured sparsity model generates nested approximations 
if the support of the best $K'$-term structured sparse approximation 
contains the support of the best $K$-term structured sparse approximation 
for all $K < K'$. An important example of a NAP-generating 
structured sparse model is the standard compressible signal model of
\eqref{eq:kta}.

When a structured sparsity model obeys the NAP, the support of 
the difference between the best $jK$-term structured sparse 
approximation and the best $(j+1)K$-term structured sparse 
approximation of a signal can be shown to lie in a small union of 
subspaces, thanks to the structure enforced
by the model. This structure is captured by the set of subspaces
that are included in each subsequent approximation, as defined
below.

\begin{DEFI}
The $j^{th}$ set of {\em residual subspaces} of size $K$ is defined as
$\R_{j,K}(\A) = \{\u \in \real^N: \u = \aalg(\x,jK)-\aalg(\x,(j-1)K)~\textrm{for some}~\x \in \real^N\}$,
for $j = 1,\ldots,\lceil N/K\rceil$.
\end{DEFI}

Under the NAP, each structured compressible signal $\x$ can be 
partitioned into its best $K$-term structured sparse approximation 
$\x_{T_1}$, the additional components present in the best $2K$-term 
structured sparse approximation $\x_{T_2}$,
and so on, with $\x = \sum_{j=1}^{\lceil N/K \rceil} \x_{T_j}$ and
$\x_{T_j} \in \R_{j,K}(\A)$ for each $j$.  Each signal partition
$\x_{T_j}$ is a $K$-sparse signal, and thus $\R_{j,K}(\A)$ is a
union of subspaces of dimension $K$. We will denote by $R_j$ the
number of subspaces that compose $\R_{j,K}(\A)$ and omit the
dependence on $\A$ in the sequel for brevity.

Intuitively, the norms of the partitions $\|\x_{T_j}\|_2$ decay as
$j$ increases for signals that are structured compressible. As
the next subsection shows, this observation is instrumental in
relaxing the isometry restrictions on the measurement matrix $\Phi$
and bounding the recovery
error for $s$-structured compressible signals when the model obeys the
NAP.

\subsection{The restricted amplification property (RAmP)}
\label{sec:compcsmat}

For exactly $K$-structured sparse signals, we discussed in
Section~\ref{sec:mrip} that the number of compressive measurements
$M$ required for a random matrix to have the $\A_K$-RIP is
determined by the number of canonical subspaces $m_K$
via~\eqref{eq:blum}.  Unfortunately, such structured sparse concepts and
results do not immediately extend to structured compressible signals.
Thus, we develop a generalization of the $\A_K$-RIP that we will use
to quantify the stability of recovery for structured compressible
signals.

One way to analyze the robustness of compressible signal recovery in
conventional CS is to consider the tail of the signal outside its
$K$-term approximation as contributing additional ``noise'' to the
measurements of size $\|\Phi (\x-\x_K)\|_2$ \cite{CandesRIP,CoSaMP,IHT}.
Consequently, the conventional $K$-sparse recovery performance
result can be applied with the augmented noise $\n + \Phi(\x -
\x_K)$.

This technique can also be used to quantify the robustness of
structured compressible signal recovery.  The key quantity we must
control is the amplification of the structured sparse approximation
residual through $\Phi$.  The following property is a new
generalization of the RIP and model-based RIP.

\begin{DEFI}
A matrix $\Phi$ has the {\em $(\epsilon_K,r)$-restricted
amplification property} (RAmP) for the residual subspaces $\R_{j,K}$
of model $\A$ if
\begin{equation}
\|\Phi \u\|_2^2 \le (1+\epsilon_K)j^{2r}\|\u\|_2^2
\label{eq:ramp}
\end{equation}
for any $\u \in \R_{j,K}$ for each $1\le j \le \lceil N/K \rceil$.
\end{DEFI}

The regularity parameter $r>0$ caps the growth rate of the
amplification of $u \in \R_{j,K}$ as a function of $j$. Its value
can be chosen so that the growth in amplification with $j$ balances
the decay of the norm in each residual subspace $\R_{j,K}$ with $j$.

We can quantify the number of compressive measurements $M$ required
for a random measurement matrix $\Phi$ to have the RAmP with high
probability; we prove the following in
Appendix~\ref{app:gaussiancomp}.

\begin{THEO}
Let $\Phi$ be an $M\times N$ matrix with i.i.d.\ subgaussian entries
and let the set of residual subspaces $\R_{j,K}$ of the structured 
sparsity model $\A$ contain $R_j$ subspaces of dimension $K$ 
for each $ 1\le j \le \lceil N/K \rceil$. If
\begin{equation}
M \ge \max_{1\le j \le \lceil N/K \rceil}
\frac{2K+4\ln
\frac{R_jN}{K}+2t}{\left(j^r\sqrt{1+\epsilon_K}-1\right)^2}, \label{eq:gaussiancomp}
\end{equation}
then the matrix $\Phi$ has the $(\epsilon_K,r)$-RAmP with
probability $1-e^{-t}$. \label{theo:gaussiancomp}
\end{THEO}

The order of the bound of Theorem~\ref{theo:gaussiancomp} is lower than $\bigo{K
\log(N/K)}$ as long as the number of subspaces $R_j$ grows slower
than $N^K$.

Armed with the RaMP, we can state the following result, which will
provide robustness for the recovery of structured compressible signals;
see Appendix~\ref{app:phiresidual} for the proof.
\begin{THEO}
Let $\x \in \aaprox_s$ be an $s$-structured compressible signal under a
structured sparsity model $\A$ that obeys the NAP. If $\Phi$ has the
$(\epsilon_K,r)$-RAmP and $r=s-1$, then we have
\begin{equation}
\|\Phi(\x -\aalg(\x,K))\|_2 \le C_s\sqrt{1+\epsilon_K}K^{-s}\ln
\left\lceil \frac{N}{K} \right\rceil|\x|_{\aaprox_s},
\label{eq:phiresidual} \nonumber
\end{equation}
where $C_s$ is a constant that depends only on $s$.
\label{theo:phiresidual}
\end{THEO}

\section{Model-Based Signal Recovery Algorithms}
\label{sec:recovery}

To take practical advantage of our new theory for model-based CS, we
demonstrate how to integrate structured sparsity models into two 
state-of-the-art CS recovery algorithms, CoSaMP \cite{CoSaMP} 
(in this section) and iterative hard thresholding 
(IHT)~\cite{NowakEM,DaubechiesThresholding,CandesPSR,IHT2,IHT} 
(in Appendix~\ref{app:addalgs} to avoid repetition). The key 
modification is simple: we merely
replace the best $K$-term sparse approximation step in these greedy
algorithms with a best $K$-term structured sparse approximation. Since at
each iteration we need only search over the $m_K$ subspaces of
$\A_K$ rather than ${N \choose K}$ subspaces of $\Sigma_K$, fewer
measurements will be required for the same degree of robust signal
recovery.  Or, alternatively, using the same number of measurements,
more accurate recovery can be achieved.  

After presenting the modified CoSaMP algorithm, we prove robustness 
guarantees for both structured sparse and structured compressible signals.
To this end, we must define an enlarged union of subspaces that includes 
sums of elements in the structured sparsity model.

\begin{DEFI} The {\em $B$-order sum} for the set $\A_K$, with
$B>1$ an integer, is defined as
\begin{equation}
\A^B_K = \left\{\x = \sum_{r=1}^B x^{(r)},~{\rm with}~\x^{(r)} \in
\A_K \right\}. \nonumber
\end{equation}
\end{DEFI}

Define $\aalg_B(\x,K)$ as the algorithm that obtains the best
approximation of $\x$ in the enlarged union of subspaces $\A^B_K$:
\begin{equation}
\aalg_B(\x,K) = \arg \min_{\xbar \in \A^B_K} \|\x-\xbar\|_2. \nonumber
\end{equation}
We note that $\aalg(\x,K)=\aalg_1(\x,K)$.
Note also that for many structured sparsity models, we will have 
$\A^B_K \subset \A_{BK}$, and so the algorithm $\aalg(\x,BK)$ 
will provide a strictly better approximation than $\aalg_B(\x,K)$.

\subsection{Model-based CoSaMP}

We choose to modify the CoSaMP algorithm \cite{CoSaMP} for two
reasons.  First, it has robust recovery guarantees that are on par
with the best convex optimization-based approaches.  Second, it has
a simple iterative, greedy structure based on a best $BK$-term
approximation (with $B$ a small integer) that is easily modified to
incorporate a best $BK$-term structured sparse approximation
$\aalg_B(K,\x)$. These properties also make the IHT and SP algorithms 
amenable to modification; see Appendix~\ref{app:addalgs} for 
details on IHT. Pseudocode for the modified CoSaMP algorithm 
is given in Algorithm~\ref{alg:Mcosamp}, where $A\pinv$ denotes the 
Moore-Penrose pseudoinverse of $A$.

\begin{algorithm*}[t]
\caption{Model-based CoSaMP}
\label{alg:Mcosamp}
\begin{tabbing}
Inputs: CS matrix $\Phi$, measurements $\y$, structured sparse 
approximation algorithm $\aalg$ \\
Output: $K$-sparse approximation $\xhat$ to true signal $x$ \\
$\xhat_0=0$ , $d = \y$; $i = 0$ \hspace{32.5mm} \{initialize\} \\
{\bf while} \= halting criterion false {\bf do} \hspace{20mm}\= \\
\> 1. $i \leftarrow i+1$ \\
\> 2. $\e \leftarrow \Phi\trans d$ \> \{form signal residual estimate\} \\
\> 3.  $\Omega \leftarrow \mathrm{supp}(\aalg_2(\e,K))$ \> \{prune residual estimate according to structure\} \\
\> 4. $T \leftarrow \Omega \cup \mathrm{supp}(\xhat_{i-1})$ \> \{merge supports\} \\
\> 5. $\b\vert_T \leftarrow \Phi_T\pinv y$, $\b\vert_{T^C}$ \> \{form signal estimate\} \\
\> 6. $\xhat_i \leftarrow \aalg(\b,K)$ \> \{prune signal estimate according to structure\} \\
\> 7. $d \leftarrow \y - \Phi \xhat_i$ \> \{update measurement residual\} \\
{\bf end while} \\
return $\xhat \leftarrow \xhat_i$
\end{tabbing}
\end{algorithm*}

\subsection{Performance of structured sparse signal recovery}
\label{sec:msr}

We now study the performance of model-based CoSaMP signal recovery
on structured sparse and structured compressible signals.
A robustness guarantee for noisy measurements of structured sparse
signals can be obtained using the model-based
RIP~\eqref{eq:modelrip}. 
Our performance guarantee for structured sparse signal recovery will
require that the measurement matrix $\Phi$ be a near-isometry for
all subspaces in $\A^B_K$ for some $B > 1$. This requirement is a
direct generalization of the $2K$-RIP, $3K$-RIP, and higher-order
RIPs from the conventional CS theory.
The following theorem is proven in
Appendix~\ref{app:rcosamp}.

\begin{THEO}
Let $\x \in \A_K$ and let $\y
= \Phi \x + \n$ be a set of noisy CS measurements. If $\Phi$ has an
$\A^4_K$-RIP constant of $\delta_{\A^4_K} \le 0.1$,
then the signal estimate $\xhat_i$ obtained from iteration $i$ of
the model-based CoSaMP algorithm satisfies
\begin{equation}
\|\x - \xhat_i\|_2 \le 2^{-i} \|\x\|_2 + 15\|\n\|_2. \label{eq:rcosamp}
\end{equation}
\label{theo:rcosamp}
\end{THEO}
This guarantee matches that of the  
CoSaMP algorithm~\cite[Theorem 4.1]{CoSaMP}; however, our 
guarantee is only for structured sparse signals rather than for all 
sparse signals.

\subsection{Performance of structured compressible signal recovery}
\label{sec:compresult}

Using the new tools introduced in Section~\ref{sec:models}, we
can provide a robustness guarantee for noisy measurements of
structured compressible signals, using the RAmP as a condition on the measurement matrix $\Phi$.

\begin{THEO}
Let $\x \in \aaprox_s$ be an $s$-structured compressible signal from a
structured sparsity model $\A$ that obeys the NAP, and let 
$\y = \Phi \x + \n$ be a set of noisy CS measurements. 
If $\Phi$ has the $\A^4_K$-RIP with $\delta_{\A^4_K} \le 0.1$
and the $(\epsilon_K,r)$-RAmP with $\epsilon_K \le 0.1$ and $r=s-1$, 
then the signal estimate $\xhat_i$ obtained from iteration $i$ of the
model-based CoSaMP algorithm satisfies
\begin{eqnarray}
\|\x - \xhat_i\|_2 &\le& 2^{-i} \|\x\|_2 + 35\|\n\|_2\nonumber\\
&&+35C_s|\x|_{\aaprox_s}K^{-s}(1+\ln \lceil N/K \rceil).
\label{eq:comptheorem}
\end{eqnarray}
\label{theo:compressible}
\end{THEO}
{\em Proof sketch.} To prove the theorem, we first bound the optimal 
structured sparse recovery error for an $s$-structured compressible signal 
$\x \in \aaprox_s$ when the matrix $\Phi$ has the $(\epsilon_K,r)$-RAmP 
with $r \le s-1$ (see Theorem~\ref{theo:phiresidual}). Then, using 
Theorem~\ref{theo:rcosamp}, we can easily prove the result by following 
the analogous proof in \cite{CoSaMP}. \qed

The standard CoSaMP algorithm also features a similar guarantee for structured compressible signals, with the constant changing from 35 to 20.

\subsection{Robustness to model mismatch}

We now analyze the robustness of model-based CS recovery to {\em
model mismatch}, which occurs when the signal being recovered from
compressive measurements does not conform exactly to the 
structured sparsity model used in the recovery algorithm.

We begin with optimistic results for signals that are ``close'' to
matching the recovery structured sparsity model.  First consider a signal $x$ that is
not $K$-structured sparse as the recovery algorithm assumes but rather
$(K+\kappa)$-structured sparse for some small integer $\kappa$.  This signal
can be decomposed into $x_K$, the signal's $K$-term structured sparse
approximation, and $x-x_K$, the error of this approximation. For
$\kappa\leq K$, we have that $x-x_K\in \R_{2,K}$. If the matrix
$\Phi$ has the $(\epsilon_K,r)$-RAmP, then it follows than
\begin{equation}
\|\Phi(x-x_K)\|_2 \le 2^r\sqrt{1+\epsilon_K}\|x-x_K\|_2. \label{eq:almostsparse}
\end{equation}
Using equations (\ref{eq:rcosamp}) and (\ref{eq:almostsparse}), we
obtain the following guarantee for the $i^{th}$ iteration of
model-based CoSaMP:
\begin{equation}
\|x-\xhat_i\|_2 \le 2^{-i}\|x\|_2 + 16\cdot 2^r\sqrt{1+\epsilon_K}\|x-x_K\|_2+15\|n\|_2. \nonumber
\end{equation}
By noting that $\|x-x_K\|_2$ is small, we obtain a guarantee that is
close to (\ref{eq:rcosamp}).

Second, consider a signal $x$ that is not $s$-structured compressible as
the recovery algorithm assumes but rather $(s-\epsilon)$-structured
compressible. The following bound can be obtained under the
conditions of Theorem~\ref{theo:compressible} by modifying the
argument in Appendix~\ref{app:phiresidual}:
\begin{eqnarray*}
\|\x - \xhat_i\|_2 &\le& 2^{-i} \|\x\|_2 + 35\Bigg(\|\n\|_2\\
&&\left.+C_s|\x|_{\aaprox_s}K^{-s}\left(1+\frac{\lceil \frac{N}{K}\rceil^{\epsilon}-1}{\epsilon}\right)\right).
\end{eqnarray*}
As $\epsilon$ becomes smaller, the factor $\frac{\lceil
N/K\rceil^{\epsilon}-1}{\epsilon}$ approaches $\log\lceil N/K
\rceil$, matching (\ref{eq:comptheorem}).
In summary, as long as the deviations from the structured sparse and
structured compressible classes are small, our model-based recovery
guarantees still apply within a small bounded constant factor.

We end with an intuitive worst-case result for signals that
are arbitrarily far away from structured sparse or structured compressible.
Consider such an arbitrary $x\in \real^N$ and compute its nested
structured sparse approximations $x_{jK} = \aalg(\x,jK)$,
$j=1,\ldots,\lceil N/K \rceil$.  If $x$ is not structured compressible,
then the structured sparse approximation error $\sigma_{jK}(\x)$ is not
guaranteed to decay as $j$ decreases. Additionally, the number of
residual subspaces $\R_{j,K}$ could be as large as $N \choose K$;
that is, the $j^{\rm th}$ difference between subsequent structured sparse
approximations $\x_{T_j} = \x_{jK}-\x_{(j-1)K}$ might lie in any
arbitrary $K$-dimensional subspace. This worst case is equivalent to
setting $r = 0$ and $R_j = {N \choose K}$ in
Theorem~\ref{theo:gaussiancomp}. It is easy to see that the resulting
condition on the number of measurements $M$ matches that of the
standard RIP for CS. Hence, if we inflate the number of measurements to
$M=\bigo{K \log(N/K)}$ (the usual number for conventional CS), the
performance of model-based CoSaMP recovery on an arbitrary signal
$x$ follows the distortion of the best $K$-term {\em structured sparse 
approximation} error of $x$ within a bounded constant factor.

\subsection{Computational complexity of model-based recovery}

The computational complexity of a structured signal recovery
algorithm differs from that of a standard algorithm by two factors.
The first factor is the reduction in the number of measurements $M$
necessary for recovery: since most current recovery algorithms have
a computational complexity that is linear in the number of
measurements, any reduction in $M$ reduces the total complexity. The
second factor is the cost of the structured sparse approximation. The
$K$-term approximation used in most current recovery algorithms can
be implemented with a simple sorting operation ($\bigo{N \log N}$
complexity, in general).  Ideally, the structured sparsity model should 
support a similarly efficient approximation algorithm.

To validate our theory and algorithms and demonstrate their general
applicability and utility, we now present two specific instances of
model-based CS and conduct a range of simulation experiments.

\section{Example: Wavelet Tree Model}
\label{sec:trees}

Wavelet decompositions have found wide application in the analysis,
processing, and compression of smooth and piecewise smooth signals
because these signals are $K$-sparse and compressible, respectively
\cite{Mallatbook}. Moreover, the wavelet coefficients can be
naturally organized into a tree structure, and for many kinds of
natural and manmade signals the largest coefficients cluster along
the branches of this tree.  This motivates a connected tree model
for the wavelet coefficients \cite{treekernel,optimaltree,BDKY}.

While CS recovery for wavelet-sparse signals has been considered
previously~\cite{MarcoSPARS05,LaDoICIP,DuarteICASSP08,DoLaCAMSAP,WaveletBCS}, the
resulting algorithms integrated the tree constraint in an ad-hoc
fashion. Furthermore, the algorithms provide no recovery guarantees
or bounds on the necessary number of compressive measurements.

\subsection{Tree-sparse signals}

We first describe tree sparsity in the context of sparse wavelet
decompositions.  We focus on one-dimensional signals and binary
wavelet trees, but all of our results extend directly to
$d$-dimensional signals and $2^d$-ary wavelet trees.

Consider a signal $\x$ of length $N = 2^I$, for an integer value of
$I$.
The wavelet representation of $x$ is given by
\begin{equation}
\x = v_0\bnu + \sum_{i=0}^{I-1}\sum_{j=0}^{2^i-1}w_{i,j}\bpsi_{i,j}, \nonumber
\end{equation}
where $\bnu$ is the scaling function and $\bpsi_{i,j}$ is the
wavelet function at scale $i$ and offset $j$. The wavelet transform
consists of the scaling coefficient $v_0$ and wavelet coefficients
$w_{i,j}$ at scale $i$, $0 \le i \le I-1$, and position $j$, $0 \le
j \le 2^i-1$. In terms of our earlier matrix notation, $\x$ has the
representation $\x = \Psi \balpha$, where $\Psi$ is a matrix
containing the scaling and wavelet functions as columns, and
$\balpha = [v_0~w_{0,0}~w_{1,0}~w_{1,1}~w_{2,0}\ldots]^T$ is the
vector of scaling and wavelet coefficients.  We are, of course,
interested in sparse and compressible $\alpha$.

\begin{figure}[bt]
\begin{center}
\epsfig{file=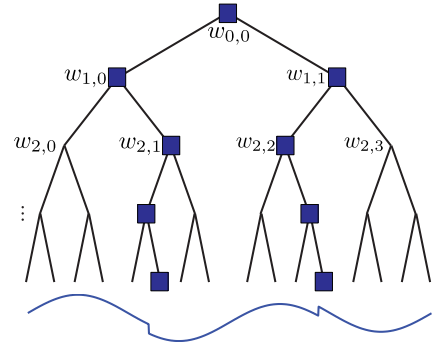,height=45mm}
\end{center}
\caption{\sl \label{fig:bintree} Binary wavelet tree for a
one-dimensional signal. The squares denote the large wavelet
coefficients that arise from the discontinuities in the piecewise
smooth signal drawn below; the support of the large coefficients
forms a rooted, connected tree.}
\end{figure}

The nested supports of the wavelets at different scales create a
parent/child relationship between wavelet coefficients at different
scales. We say that $w_{i-1,\lfloor j/2 \rfloor}$ is the {\em
parent} of $w_{i,j}$ and that $w_{i+1, 2j}$ and $w_{i+1,2j+1}$ are
the {\em children} of $w_{i,j}$. These relationships can be
expressed graphically by the wavelet coefficient tree in
Figure~\ref{fig:bintree}.

Wavelet functions act as local discontinuity detectors, and using
the nested support property of wavelets at different scales, it is
straightforward to see that a signal discontinuity will give rise to
a chain of large wavelet coefficients along a branch of the wavelet
tree from a leaf to the root. Moreover, smooth signal regions will
give rise to regions of small wavelet coefficients. This ``connected
tree'' property has been well-exploited in a number of wavelet-based
processing \cite{HMT,RombergHMT,GSM} and compression \cite{EZW,CDDD}
algorithms.  In this section, we will specialize the theory
developed in Sections \ref{sec:models} and \ref{sec:recovery} to a
connected tree model $\T$.

A set of wavelet coefficients $\Omega$ forms a {\em connected
subtree} if,
whenever a coefficient $w_{i,j} \in \Omega$, then its parent
$w_{i-1,\lceil j/2 \rceil} \in \Omega$ as well. Each such set
$\Omega$ defines a subspace of signals whose support is contained in
$\Omega$; that is, all wavelet coefficients outside $\Omega$ are
zero. In this way, we define the structured sparsity model $\T_K$ as the
union of all $K$-dimensional subspaces corresponding to supports
$\Omega$ that form connected subtrees.
\begin{DEFI}
Define the set of {\em $K$-tree sparse signals} as
\begin{eqnarray}
\T_K &=& \left\{ \x = v_0\bnu +
\sum_{i=0}^{I-1}\sum_{j=1}^{2^i}w_{i,j}\bpsi_{i,j} :
\w\vert_{\Omega^C} = 0,\right.\nonumber \\ 
&& |\Omega| = K, \Omega \textrm{~forms~a~connected~subtree}\Bigg\}. \nonumber
\end{eqnarray}
\end{DEFI}

To quantify the number of subspaces in $\T_K$, it suffices to count
the number of distinct connected subtrees of size $K$ in a binary
tree of size $N$.   We prove the following result in
Appendix~\ref{app:treecount}.
\begin{PROP}
The number of subspaces in $\T_K$ obeys $T_K
\le\frac{4^{K+4}}{Ke^2}$ for $K \ge \log_2 N$ and $T_K \le
\frac{(2e)^K}{K+1}$ for $K < \log_2 N$. \label{prop:treecount}
\end{PROP}
To simplify the presentation in the sequel, we will simply use the weaker bound $T_K \le
\frac{(2e)^K}{K+1}$ for all values of $K$ and $N$.

\subsection{Tree-based approximation}

To implement tree-based signal recovery, we seek an efficient
algorithm $\mathbb{T}(\x,K)$ to solve the optimal approximation
\begin{equation}
\x^\T_K = \arg \min_{\xbar \in \T_K} \|\x-\xbar\|_2.
\label{eq:opt}
\end{equation}
Fortuitously, an efficient solver exists, called the {\em condensing
sort and select algorithm} (CSSA)
\cite{treekernel,optimaltree,BDKY}. Recall that subtree
approximation coincides with standard $K$-term approximation (and
hence can be solved by simply sorting the wavelet coefficients) when
the wavelet coefficients are monotonically nonincreasing along the
tree branches out from the root.
The CSSA solves (\ref{eq:opt}) in the case of general wavelet coefficient values by {\em condensing} the nonmonotonic segments of the tree branches using an iterative sort-and-average routine during a greedy search through the nodes. For each node in the tree, the algorithm calculates the average wavelet coefficient magnitude for each subtree rooted at that node, and records the largest average among all the subtrees as the energy for that node. The CSSA then searches for the unselected node with the largest energy and adds the subtree corresponding to the node's energy to the estimated support as a {\em supernode}: a single node that provides a condensed representation of the corresponding subtree~\cite{BDKY}. Condensing a large coefficient far down the tree accounts for the potentially large cost (in terms of the total budget of tree nodes $K$) of growing the tree to that point.

Since the first step of the CSSA involves sorting all of the wavelet
coefficients, overall it requires $\bigo{N\log N}$ computations.
However, once the CSSA grows the optimal tree of size $K$, it is
trivial to determine the optimal trees of size $<K$ and
computationally efficient to grow the optimal trees of size $>K$
\cite{treekernel}.

The constrained optimization ({\ref{eq:opt}) can be rewritten as an
unconstrained problem by introducing the Lagrange multiplier
$\lambda$ \cite{DonohoCART}:
\begin{equation}
 \min_{\xbar \in \bar{\T}} \|\x-\xbar\|_2^2 + \lambda(\|\bar{\alpha}\|_0-K),
\label{eq:opt2} \nonumber
\end{equation}
where $\overline{\T} = \cup_{n=1}^N \T_n$ and $\bar{\alpha}$ are the
wavelet coefficients of $\xbar$.
Except for the inconsequential $\lambda K$ term, this optimization
coincides with Donoho's {\em complexity penalized sum of squares}
\cite{DonohoCART}, which can be solved in only $\bigo{N}$
computations using coarse-to-fine dynamic programming on the tree.
Its primary shortcoming is the nonobvious relationship between the
tuning parameter $\lambda$ and and the resulting size $K$ of the
optimal connected subtree.

\subsection{Tree-compressible signals}

Specializing Definition \ref{def:model} from Section~\ref{sec:mcomp}
to $\T$, we make the following definition.

\begin{DEFI}
Define the set of {\em $s$-tree compressible signals} as
\begin{eqnarray*}
\taprox_s &=& \{\x \in \real^N: \|\x-\mathbb{T}(\x,K)\|_2 \le G
K^{-s},\\
&&1\le K \le N, G < \infty\}.
\end{eqnarray*}
Furthermore, define $|\x|_{\taprox_s}$ as the smallest value of $G$
for which this condition holds for $\x$ and $s$.
\end{DEFI}

Tree approximation classes contain signals whose wavelet
coefficients have a loose (and possibly interrupted) decay from
coarse to fine scales. These classes have been well-characterized
for wavelet-sparse signals~\cite{CDDD,BDKY,optimaltree} and are
intrinsically linked with the Besov spaces $B^s_q(L_p([0,1]))$.
Besov spaces contain functions of one or more continuous variables
that have (roughly speaking) $s$ derivatives in $L_p([0,1])$; the
parameter $q$ provides finer distinctions of smoothness.  When a
Besov space signal $x_a \in B^s_p(L_p([0,1]))$ with $s> 1/p-1/2$
is sampled uniformly and
converted to a length-$N$ vector $x$, its wavelet coefficients
belong to the tree approximation space $\taprox_s$, with
\begin{equation}
|\x_N|_{\taprox_s} \asymp \|x_a\|_{L_p([0,1])} +
\|x_a\|_{B^s_q(L_p([0,1]))}, \nonumber
\end{equation}
where ``$\asymp$'' denotes an equivalent norm. The same result holds
if $s=1/p-1/2$ and $q \le p$.

\subsection{Stable tree-based recovery from compressive measurements}

For tree-sparse signals, by applying Theorem~\ref{theo:munion} and
Proposition~\ref{prop:treecount}, we find that a subgaussian random
matrix has the $\T_K$-RIP property with constant $\delta_{\T_K}$ and
probability $1-e^{-t}$ if the number of measurements obeys
\begin{equation}
M \ge \frac{2}{c \delta^2_{\T_K}}\left(K \ln \frac{48}{\delta_{\T_K}}+
\ln \frac{512}{Ke^{2}} +t \right).
 \nonumber
\end{equation}
Thus, the number of measurements necessary for stable recovery of
tree-sparse signals is linear in $K$, without the dependence on $N$
present in conventional non-model-based CS recovery.

For tree-compressible signals, we must quantify the number of
subspaces $R_j$ in each residual set $\R_{j,K}$ for the
approximation class. We can then apply the theory of
Section~\ref{sec:compresult} with Proposition~\ref{prop:treecount}
to calculate the smallest allowable $M$ via Theorem
\ref{theo:compressible}.

\begin{PROP}
The number of $K$-dimensional subspaces that comprise $\R_{j,K}$
obeys
\begin{equation}
R_j \le \frac{(2e)^{K(2j+1)}}{(Kj+K+1)(Kj+1)}.
\label{eq:treercount}
\end{equation}
\label{prop:treercount}
\end{PROP}
Using Proposition~\ref{prop:treercount} and
Theorem~\ref{theo:compressible}, we obtain the following condition
for the matrix $\Phi$ to have the RAmP, which is proved in
Appendix~\ref{app:treecomp}.
\begin{PROP}
Let $\Phi$ be an $M\times N$ matrix with i.i.d.\ subgaussian entries.
If
\begin{equation}
M \ge \frac{2\left(10K+2\ln \frac{N}{K(K+1)(2K+1)} +t\right)}{\left(\sqrt{1+\epsilon_K}-1\right)^2}, \nonumber
\label{eq:treecomp}
\end{equation}
then the matrix $\Phi$ has the $(\epsilon_K,s)$-RAmP for the structured
sparsity model $\T$ and all $s > 0.5$ with probability $1-e^{-t}$.
\label{prop:treecomp}
\end{PROP}

Both cases give a simplified bound on the number of measurements
required as $M=\bigo{K}$, which is a substantial improvement over
the $M=\bigo{K\log(N/K)}$ required by conventional CS recovery
methods.   Thus, when $\Phi$ satisfies
Proposition~\ref{prop:treecomp}, we have the guarantee
(\ref{eq:comptheorem}) for sampled Besov space signals from
$B^s_q(L_p([0,1]))$.

\subsection{Experiments}

We now present the results of a number of numerical experiments that
illustrate the effectiveness of a tree-based recovery algorithm. Our
consistent observation is that explicit incorporation of the structured
sparsity model in the recovery process significantly improves the
quality of recovery for a given number of measurements. In addition,
model-based recovery remains stable when the inputs are no longer
tree-sparse, but rather are tree-compressible and/or corrupted with
differing levels of noise. We employ the Daubechies-6 wavelet basis 
for sparsity, and recover the signal using model-based CoSaMP 
(Algorithm~\ref{alg:Mcosamp}) with a CSSA-based structured sparse
approximation step in all experiments.

We first study one-dimensional signals that match the connected
wavelet-tree model described above.  Among such signals is the class
of piecewise smooth functions, which are commonly encountered in
analysis and practice.

Figure~\ref{fig:wvcomparison} illustrates the results of recovering
the tree-compressible \emph{HeaviSine} signal of length $N=1024$
from $M=80$ noise-free random Gaussian measurements using CoSaMP,
$\ell_1$-norm minimization using the \texttt{l1\_eq} solver from the
$\ell_1$-Magic toolbox,\footnote{{\em
http://www.acm.caltech.edu/l1magic}.} and our tree-based recovery
algorithm. It is clear that the number of measurements ($M = 80$) is
far fewer than the minimum number required by CoSaMP and $\ell_1$-norm
minimization to accurately recover the signal. In contrast,
tree-based recovery using $K=26$ is accurate and uses fewer
iterations to converge than conventional CoSaMP. Moreover, the
normalized magnitude of the squared error for tree-based recovery
is equal to 0.037, which is remarkably close to the error between
the noise-free signal and its {\em best} $K$-term tree-approximation
(0.036).

Figure \ref{fig:rcosparse}(a) illustrates the results of a Monte Carlo
simulation study on the impact of the number of measurements $M$ on
the performance of model-based and conventional recovery for a class
of tree-sparse piecewise polynomial signals. Each data point was
obtained by measuring the normalized recovery error of 500 sample
trials. Each sample trial was conducted by generating a new
piecewise polynomial signal of length $N=1024$ with five polynomial
pieces of cubic degree and randomly placed discontinuities, computing
its best $K$-term tree-approximation using the CSSA, and then
measuring the resulting signal using a matrix with i.i.d.\ Gaussian entries.
Model-based recovery attains near-perfect recovery at $M=3K$
measurements, while CoSaMP only matches this performance at $M=5K$.

\begin{figure*}
\begin{center}
\begin{tabular}{cc}
\epsfig{file=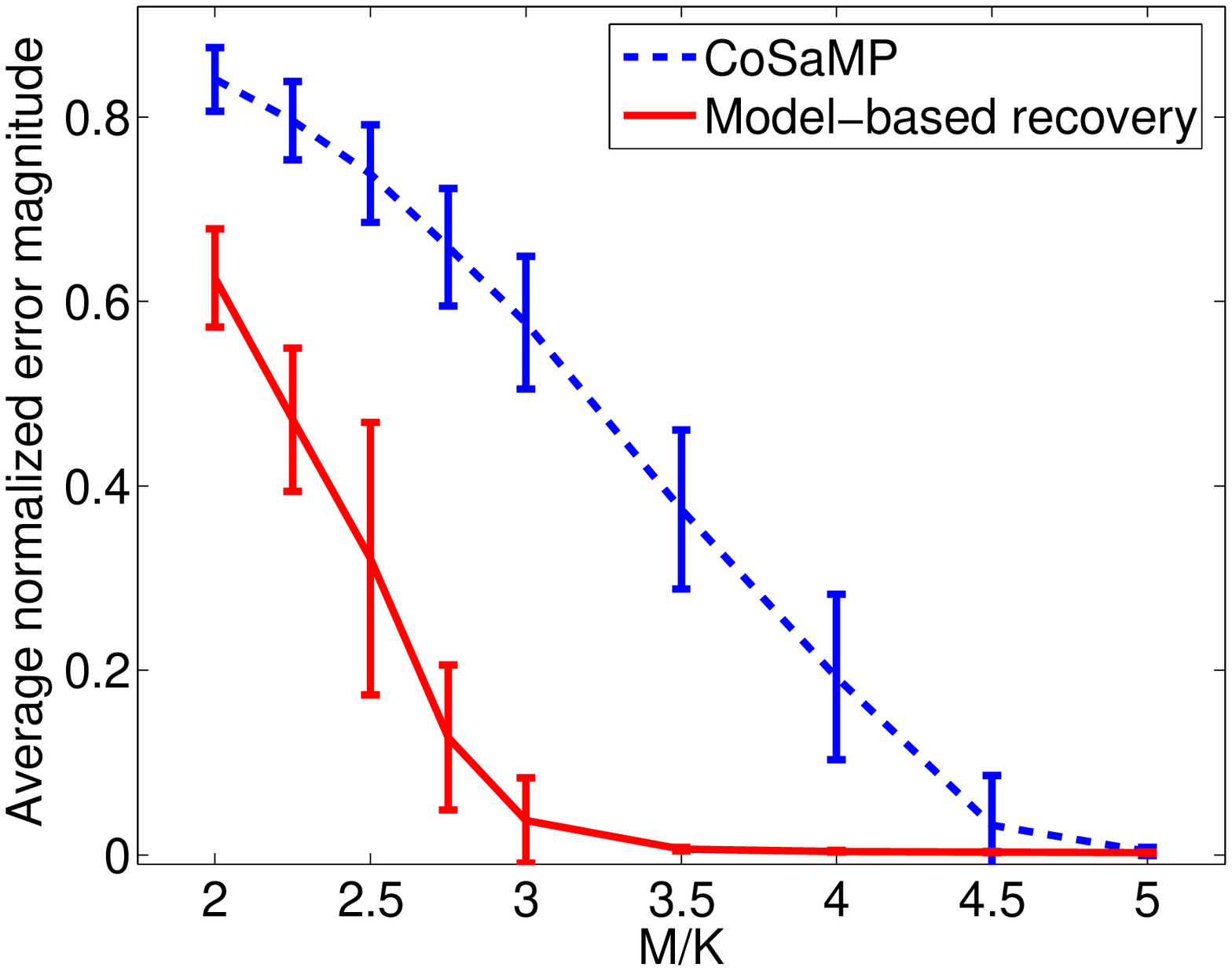,height=50mm} & \epsfig{file=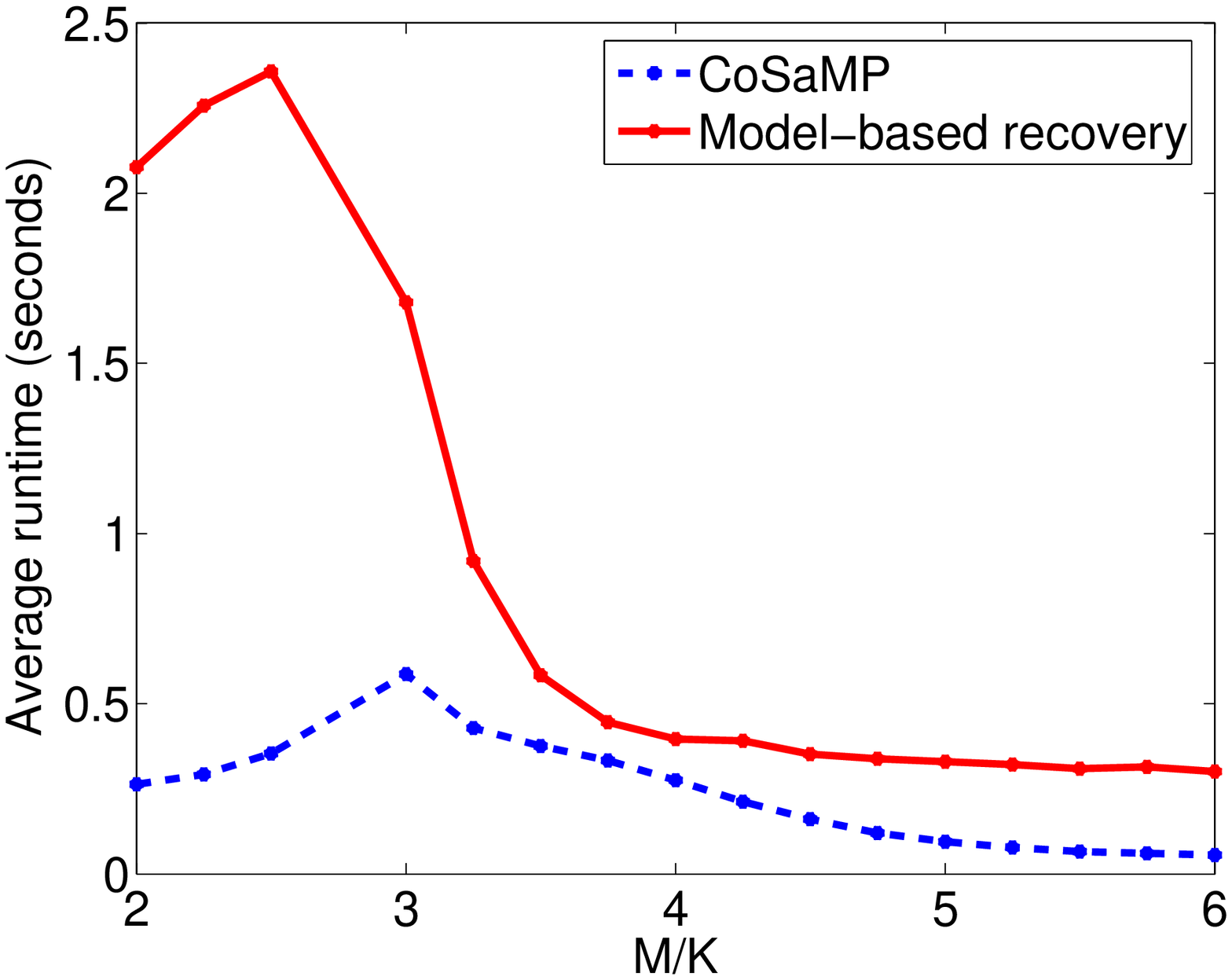,height=51mm} \\
(a) & (b)
\end{tabular}
\end{center}
\caption{\sl \label{fig:rcosparse} Performance of CoSaMP vs.\
wavelet tree-based recovery on a class of tree-sparse signals. 
(a) Average normalized recovery error and (b) average runtime 
for each recovery algorithm as a function of the overmeasuring factor $M/K$.
The number of measurements $M$ for which the wavelet tree-based algorithm
obtains near-perfect recovery is much smaller than that required by CoSaMP. 
The penalty paid for this improvement is a modest increase in the 
runtime.}
\end{figure*}

For the same class of signals, we empirically compared the recovery 
times of our proposed algorithm with those of the standard approach 
(CoSaMP). Experiments were conducted on a Sun workstation with 
a 1.8GHz AMD Opteron dual-core processor and 2GB memory 
running UNIX, using non-optimized Matlab code and a 
function-handle based implementation of the random projection 
operator $\Phi$. As is evident from Figure~\ref{fig:rcosparse}(b), 
wavelet tree-based recovery is in general slower than CoSaMP. This 
is due to the fact that the CSSA step in the iterative procedure is more 
computationally demanding than simple $K-$term approximation. 
Nevertheless, the highest benefits of model-based CS recovery are 
obtained around $M=3K$; in this regime, the runtimes of the two 
approaches are comparable, with tree-based recovery requiring 
fewer iterations and yielding much smaller recovery error than standard 
recovery.

Figure~\ref{fig:mvsn} shows the growth of the overmeasuring factor $M/K$
with the signal length $N$ for conventional CS and model-based recovery.
We generated 50 sample piecewise cubic signals and numerically
computed the minimum number of measurements $M$ required for the
recovery error $\|x-\widehat{x}\|_2 \le 2.5 \sigma_{\T_K}(\x)$, the {\em best}
tree-approximation error, for every sample signal. The figure shows that
while doubling the signal length increases the number of measurements
required by standard recovery  by $K$, the number of measurements
required by model-based recovery is constant for all $N$. These
experimental results verify the theoretical performance described in
Proposition~\ref{prop:treecomp}.
\begin{figure}[bt]
\begin{center}
\epsfig{file=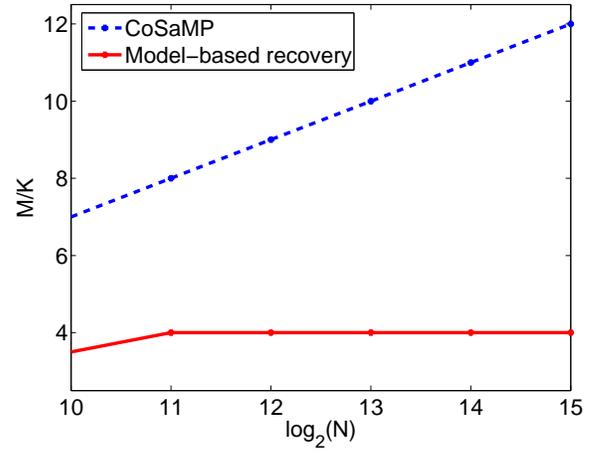,height=60mm}
\end{center}
\caption{\sl \label{fig:mvsn} Required
overmeasuring factor $M/K$ to achieve a target recovery error
$\|\x - \widehat{x}\|_2 \le 2.5 \sigma_{\T_K}(x)$ as a function of
the signal length $N$ for standard and model-based recovery
of piecewise smooth signals. While standard recovery requires
$M$ to increase logarithmically with $N$, the required $M$
is essentially constant for model-based recovery.}
\end{figure}

Further, we demonstrate that model-based recovery performs stably in
the presence of measurement noise. We generated sample
piecewise polynomial signals as above, computed their best $K$-term
tree-approximations, computed $M$ measurements of each
approximation, and finally added Gaussian noise of variance $\sigma$ to
each measurement, so that the expected variance $E[\|n\|_2] = \sigma \sqrt{M}$. 
We emphasize that this noise model implies that the energy of the noise 
added will be larger as $M$ increases. Then, we recovered the signals 
using CoSaMP and model-based recovery and measured the recovery 
error in each case. For comparison purposes, we also tested the recovery 
performance of a $\ell_1$-norm minimization algorithm that accounts for the 
presence of noise, which has been implemented as the \texttt{l1\_qc} solver 
in the $\ell_1$-Magic toolbox.
First, we determined the lowest value of $M$ for which the
respective algorithms provided near-perfect recovery in the absence
of noise in the measurements. This corresponds to $M=3.5K$ for
model-based recovery,
$M=5K$ for CoSaMP, and $M=4.5K$ for $\ell_1$-norm minimization. Next, we
generated 200 sample tree-structured signals, computed $M$ {\em noisy}
measurements, recovered the signal using the given algorithm and
recorded the recovery error. Figure~\ref{fig:rcosnoisy} illustrates
the growth in maximum normalized recovery error (over the 200 sample
trials) as a function of the expected measurement signal-to-noise
ratio for the tree algorithms. We observe similar stability curves
for all three algorithms, while noting that model-based recovery
offers this kind of stability using significantly fewer
measurements.

\begin{figure}[bt]
\begin{center}
\epsfig{file=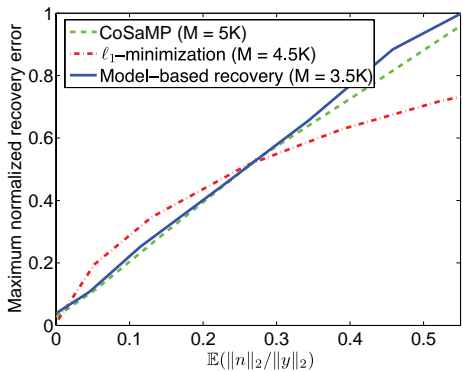,height=50mm}
\end{center}
\caption{\sl \label{fig:rcosnoisy}  Robustness to measurement
noise for standard and wavelet tree-based CS recovery algorithms. We
plot the maximum normalized recovery error over 200 sample trials as
a function of the expected signal-to-noise ratio. The linear growth
demonstrates that model-based recovery possesses the same robustness
to noise as CoSaMP and $\ell_1$-norm minimization.}
\end{figure}

Finally, we turn to two-dimensional images and a wavelet quadtree
model. The connected wavelet-tree model has proven useful for
compressing natural images~\cite{optimaltree}; thus, our algorithm
provides a simple and provably efficient method for recovering a
wide variety of natural images from compressive measurements. An
example of recovery performance is given in
Figure~\ref{fig:peppers}. The test image ({\em Peppers}) is of size
$N=128 \times 128 = 16384$ pixels, and we computed $M=5000$ random
Gaussian measurements.  Model-based recovery again offers higher
performance than standard signal recovery algorithms like CoSaMP,
both in terms of recovery mean-squared error and visual quality.

\begin{figure}[!t]
\centering
\begin{tabular}{ccccc}
{\includegraphics[width=0.25\hsize]{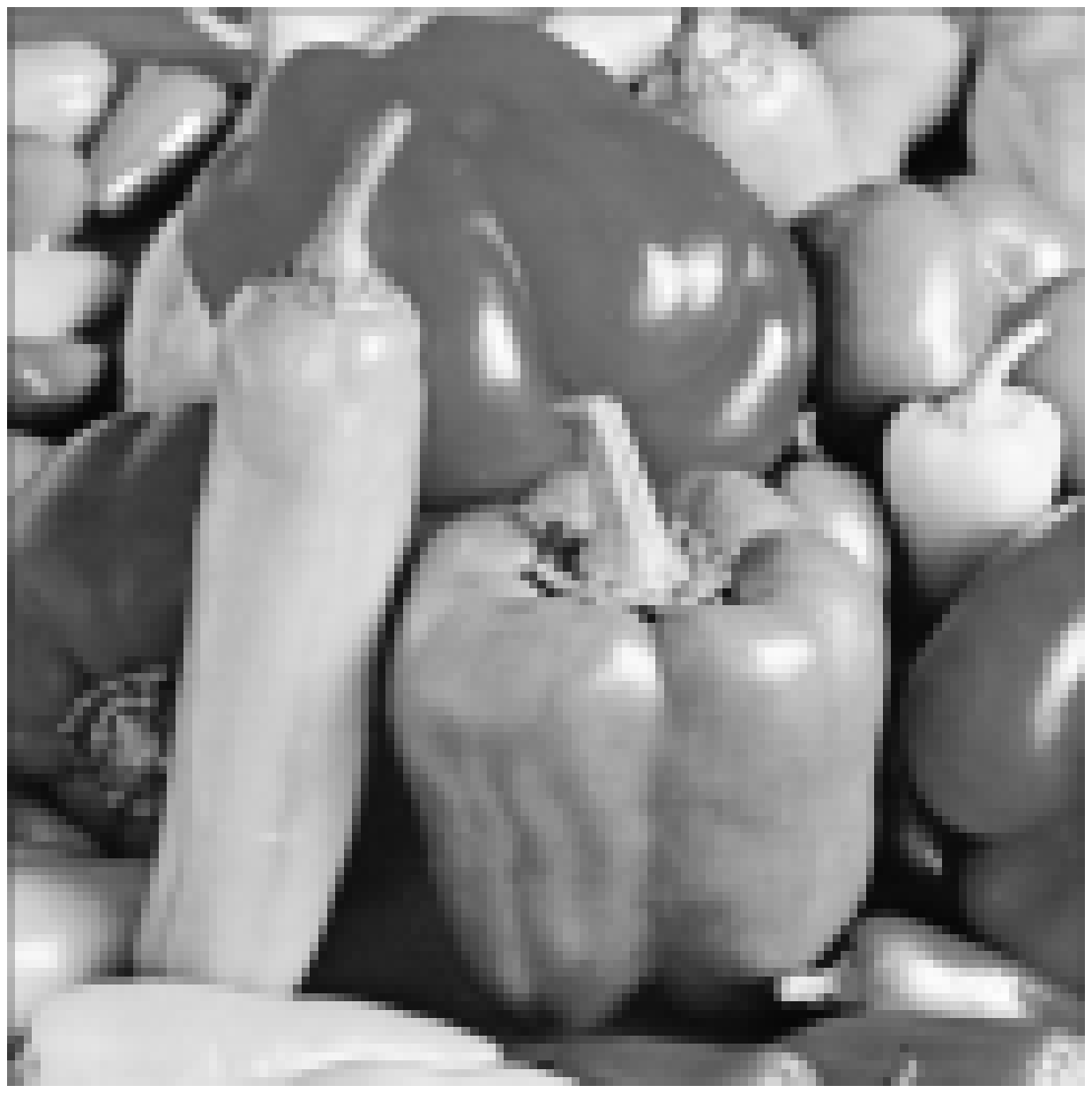}}&
{\includegraphics[width=0.25\hsize]{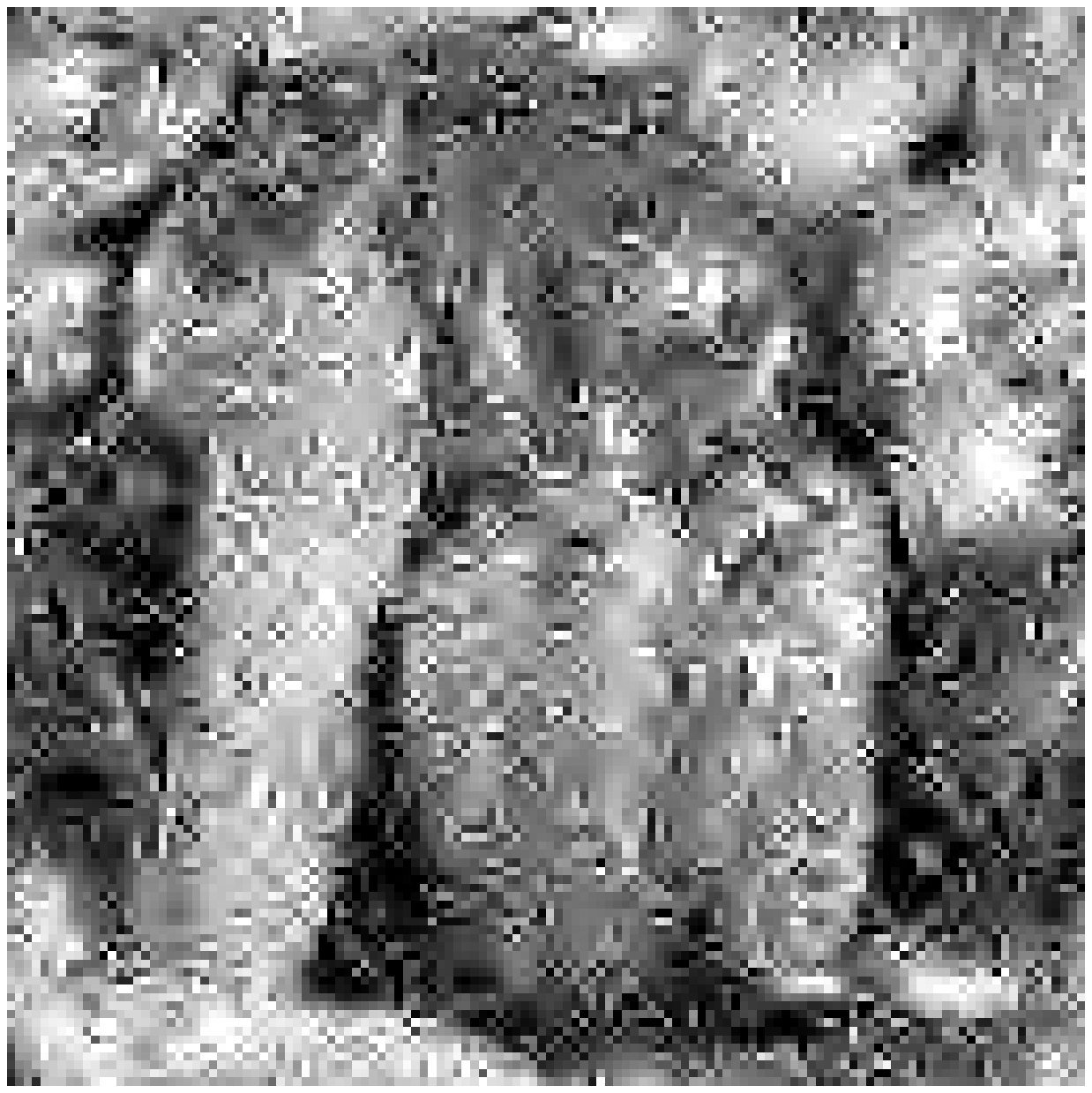}} &
{\includegraphics[width=0.25\hsize]{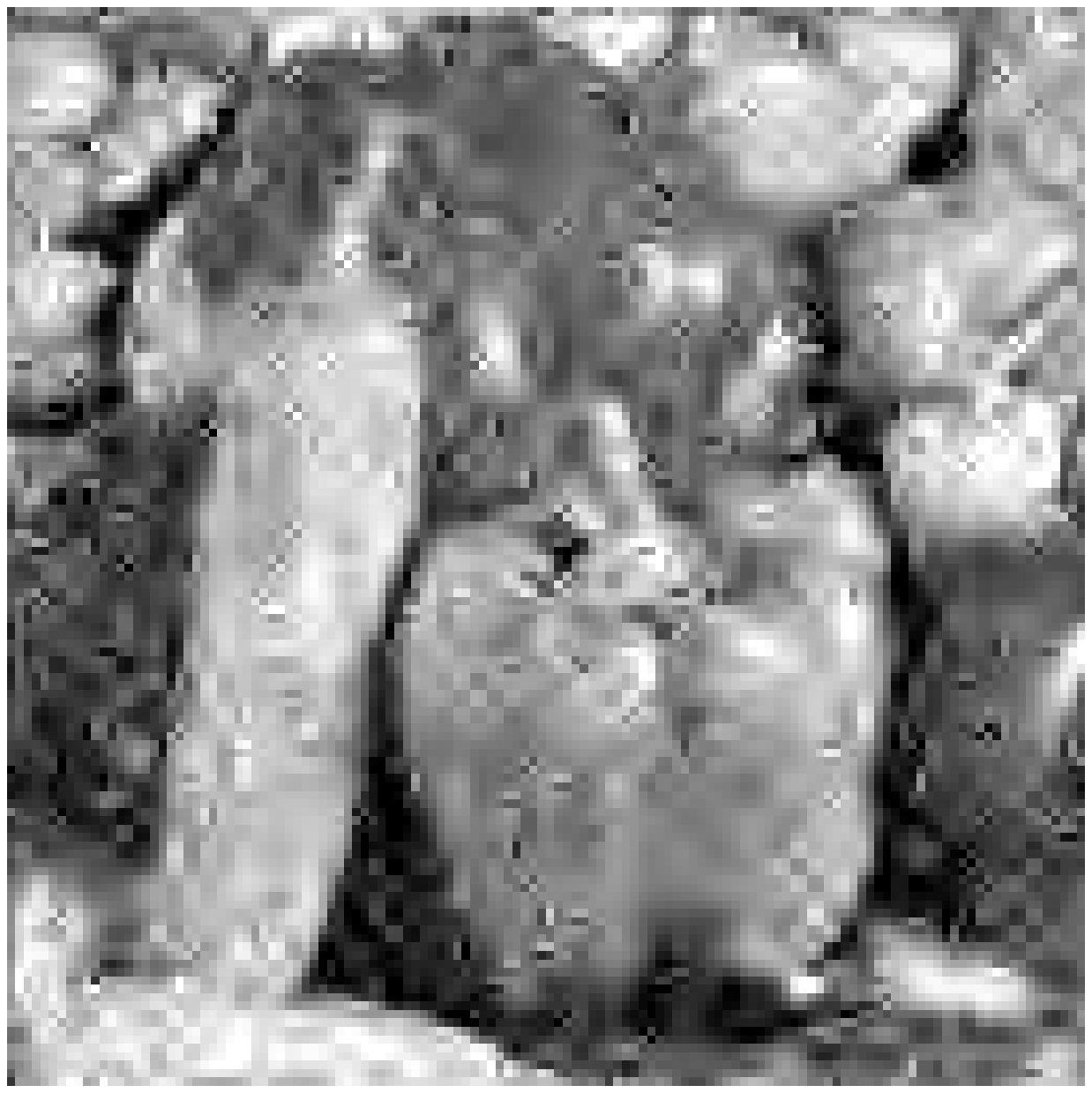}} \\
(a) {\em Peppers} &
(b) CoSaMP & (c) model-based rec. \\
 & (RMSE $=22.8$) & (RMSE $=11.1$)
\end{tabular}
\caption{\sl Example performance of standard and model-based
recovery on images. (a) $N=128 \times 128=16384$-pixel {\em Peppers}
test image.
Image recovery from $M=5000$ compressive measurements using (b)
conventional CoSaMP and (c) our wavelet tree-based algorithm.
\label{fig:peppers}}
\end{figure}

\section{Example: Block-Sparse Signals and Signal Ensembles}
\label{sec:ensembles}

In a {\em block-sparse} signal, the locations of the significant
coefficients cluster in blocks under a specific sorting order.
Block-sparse signals have been previously studied in CS
applications, including DNA microarrays and
magnetoencephalography~\cite{Hassibi,EldarUSS}. An equivalent
problem arises in CS for signal ensembles, such as sensor networks
and MIMO communication~\cite{DCS,DCSNIPS,EldarUSS}. In this case, several
signals share a common coefficient support set.  For example, when a
frequency-sparse acoustic signal is recorded by an array of
microphones, then all of the recorded signals contain the same
Fourier frequencies but with different amplitudes and delays. Such a
signal ensemble can be re-shaped as a single vector by
concatenation, and then the coefficients can be rearranged so that
the concatenated vector exhibits block sparsity.

It has been shown that the block-sparse structure enables signal
recovery from a reduced number of CS measurements, both for the
single signal case~\cite{Hassibi,EldarUSS,SOMP} and the signal
ensemble case~\cite{DCS}, through the use of specially tailored
recovery algorithms. However, the robustness guarantees for the
algorithms \cite{Hassibi,SOMP} either are restricted to exactly
sparse signals and noiseless measurements, do not have explicit
bounds on the number of necessary measurements, or are asymptotic in
nature. An optimization-based algorithm introduced
in~\cite{EldarUSS} provides similar recovery guarantees to those
obtained by the algorithm we present in this chapter; thus, our
method can be interpreted as a greedy-based counterpart to that
provided in~\cite{EldarUSS}.

In this section, we formulate the block sparsity model as a
union of subspaces and pose an approximation algorithm on this union
of subspaces. The approximation algorithm is used to implement
block-based signal recovery. We also define the corresponding class
of block-compressible signals and quantify the number of
measurements necessary for robust recovery.

\subsection{Block-sparse signals}

Consider a class $\S$ of signal vectors $\x \in \real^{JN}$, with
$J$ and $N$ integers. This signal can be reshaped into a $J \times
N$ matrix $\X$, and we use both notations interchangeably in the
sequel. We will restrict entire columns of $\X$ to be part of the
support of the signal as a group. That is, signals $\X$ in a
block-sparse model have entire columns as zeros or nonzeros. The
measure of sparsity for $\X$ is its number of nonzero columns. More
formally, we make the following definition.

\begin{DEFI} \cite{Hassibi,EldarUSS}
Define the set of {\em $K$-block sparse signals} as
\begin{eqnarray}
\S_K &=& \{\X = [\x_1~\ldots~\x_N] \in \real^{J \times N} \textrm{
such that} \nonumber\\
&& \x_n = 0 \textrm{ for } n \notin \Omega, \Omega \subseteq \{1, \ldots,
N\}, |\Omega| = K\}. \nonumber
\end{eqnarray}
\end{DEFI}

It is important to note that a $K$-block sparse signal has sparsity
$KJ$, which is dependent on the size of the block $J$. We can extend
this formulation to ensembles of $J$, length-$N$ signals with common
support. Denote this signal ensemble by
$\{\xtilde_1,\ldots,\xtilde_J\}$, with $\xtilde_j \in \real^N$, $1\le j\le
J$. We formulate a matrix representation $\Xtilde$ of the ensemble
that features the signal $\xtilde_j$ in its $j^{th}$ row: $\Xtilde =
[\xtilde_1~\ldots \xtilde_J]^T$. The matrix $\Xtilde$ features the same
structure as the matrix $X$ obtained from a block-sparse signal;
thus, the matrix $\Xtilde$ can be converted into a block-sparse vector
$\xtilde$ that represents the signal ensemble.

\subsection{Block-based approximation}

To pose the block-based approximation algorithm, we need to define
the mixed norm of a matrix.

\begin{DEFI}
The $(p,q)$ {\em mixed norm} of the matrix $\X = [\x_1~\x_2~\ldots~\x_N]$
is defined as
\begin{equation}\nonumber
\|\X\|_{(p,q)} = \left(\sum_{n=1}^N\|\x_n\|_p^q\right)^{1/q}.
\end{equation}
When $q = 0$, $\|X\|_{(p,0)}$ simply counts the number of nonzero
columns in $\X$.
\end{DEFI}

We immediately find that $\|\X\|_{(p,p)} = \|\x\|_p$, with $\x$ the
vectorization of $\X$. Intuitively, we pose the algorithm
$\mathbb{S}(\X,K)$ to obtain the best block-based approximation of
the signal $\X$ as follows:
\begin{equation}
\X_K^\S = \arg \min_{\Xbar \in \real^{J\times N}} \|\X - \Xbar\|_{(2,2)}
\textrm{ s.t. } \|\Xbar\|_{(2,0)} \le K. \label{eq:blockapprox}
\end{equation}
It is easy to show that to obtain the approximation, it suffices to
perform column-wise hard thresholding: let $\rho$ be the
$K^{\textrm{th}}$ largest $\ell_2$-norm among the columns of $\X$.
Then our approximation algorithm is $\mathbb{S}(\X,K) = \X_K^\S =
[\x_{K,1}^\S~\ldots\x_{K,N}^\S]$, where
\begin{equation}
\x_{K,n}^\S = \left\{ \begin{array}{ll}
\x_n & \|\x_n\|_2 \ge \rho, \\
0 & \|\x_n\|_2 < \rho,
\end{array}
\right. \nonumber
\end{equation}
for each $1 \le n \le N$. Alternatively, a
recursive approximation algorithm can be obtained by sorting the
columns of $\X$ by their $\ell_2$ norms, and then selecting the
columns with largest norms. The complexity of this sorting process
is $\bigo{NJ + N\log N}$.

\subsection{Block-compressible signals}

The approximation class under the block-compressible model
corresponds to signals with blocks whose $\ell_2$ norm has a
power-law decay rate.
\begin{DEFI}
We define the set of $s$-block compressible signals as
\begin{eqnarray}
\saprox_s &=& \{\X =[\x_1~\ldots~\x_N] \in \real^{J\times N}~\textrm{such that} \nonumber\\
&&\|\x_{\mathcal{I}(i)}\|_2 \le Gi^{-s-1/2}, 1\le i \le N, S < \infty\}, \nonumber
\end{eqnarray}
where $\mathcal{I}$ indexes the sorted column norms.
\end{DEFI}
We say that $\X$ is an $s$-block compressible signal if $\X \in
\saprox_s$. For such signals, we have $\|\X-\X_K\|_{(2,2)} =
\sigma_{\S_K}(\x) \le G_1K^{-s}$, and $\|\X-\X_K\|_{(2,1)} \le
G_2K^{1/2-s}$.
Note that the block-compressible model does not impart a
structure to the decay of the signal coefficients, so that the sets
$\R_{j,K}$ are equal for all values of $j$; due to this property,
the $(\delta_{\S_K},s)$-RAmP is implied by the $\S_K$-RIP. Taking
this into account, we can derive the following result
from~\cite{CoSaMP}, which is proven similarly to
Theorem~\ref{theo:rcosamp}.

\begin{THEO}
Let $\x$ be a signal from the structured sparsity model $\S$, and let
$\y = \Phi \x + \n$ be a set of noisy CS measurements. If $\Phi$ has
the $\S_K^4$-RIP with $\delta_{\S_K^4} \le 0.1$,
then the estimate obtained from iteration $i$ of block-based CoSaMP,
using the approximation algorithm~(\ref{eq:blockapprox}), satisfies
\begin{eqnarray*}
\|\x - \xhat_i\|_2 &\le& 2^{-i} \|\x\|_2 +
20\Bigg(\|\X-\X_K^\S\|_{(2,2)}\\
&&\left.+\frac{1}{\sqrt{K}}
\|\X-\X_K^\S\|_{(2,1)} + \|\n\|_2\right).
\end{eqnarray*}
\label{theo:jsm2comp}
\end{THEO}

Thus, the algorithm provides a recovered signal of similar quality
to approximations of $\X$ by a small number of nonzero columns. When
the signal $\x$ is $K$-block sparse, we have that
$||\X-\X_K^\S\|_{(2,2)} = ||\X-\X_K^\S\|_{(2,1)} = 0$, obtaining the
same result as Theorem~\ref{theo:rcosamp}, save for a constant
factor.

\subsection{Stable block-based recovery from compressive measurements}

Since Theorem~\ref{theo:jsm2comp} poses the same requirement on the
measurement matrix $\Phi$ for sparse and compressible signals, the
same number of measurements $M$ is required to provide performance
guarantees for block-sparse and block-compressible signals. The
class $\S_K$ contains $S = {N \choose K}$ subspaces of dimension
$JK$. Thus, a subgaussian random matrix has the $\S_K$-RIP property
with constant $\delta_{\S_K}$ and probability $1-e^{-t}$ if the number of measurements obeys
\begin{equation}
M \ge \frac{2}{c \delta^2_{\S_K}}\left(K\left(\ln\frac{2N}{K} +
J \ln \frac{12}{\delta_{\S_K}}\right)+t \right). \label{eq:blockM}
\end{equation}
To compare with the standard CS measurement bound, the number of
measurements required for robust recovery scales as
$M=\bigo{JK+K\log (N/K)}$, which is a substantial improvement over
the $M=\bigo{JK\log(N/K)}$ that would be required by conventional
CS recovery methods. When the size of the block  $J$ is larger than
$\log(N/K)$, then this term becomes $\bigo{KJ}$; that is, it is linear on
the total sparsity of the block-sparse signal.

We note in passing that the bound on the number of measurements
(\ref{eq:blockM}) assumes a dense subgaussian measurement matrix,
while the measurement matrices used in~\cite{DCS} have a
block-diagonal structure. To obtain measurements from an $M\times JN$
dense matrix in a distributed setting, it suffices to
partition the matrix into $J$ pieces of size $M\times N$ and
calculate the CS measurements at each sensor with the corresponding
matrix; these individual measurements are then summed to obtain the
complete measurement vector. For large $J$, (\ref{eq:blockM}) implies
that the total number of measurements required for recovery of the signal
ensemble is lower than the bound for the case where each signal recovery is
performed independently for each signal ($M = \bigo{JK \log (N/K)}$).

\subsection{Experiments}

We conducted several numerical experiments comparing model-based
recovery to CoSaMP in the context of block-sparse signals. We employ
the model-based  CoSaMP recovery of Algorithm~\ref{alg:Mcosamp} with
the block-based approximation algorithm~(\ref{eq:blockapprox}) in
all cases.  For brevity, we exclude a thorough comparison of our
model-based algorithm with $\ell_1$-norm minimization and defer it
to future work. In practice, we observed that our algorithm performs
several times faster than convex optimization-based procedures.

Figure~\ref{fig:blocksparse_recon} illustrates an $N=4096$ signal
that exhibits block sparsity, and its recovered version from $M=960$
measurements using CoSaMP and model-based recovery. The block
size $J=64$ and there were $K=6$ active blocks in the signal. We
observe the clear advantage of using the block-sparsity model in
signal recovery.

\begin{figure*}[!t]
\centering
\begin{tabular}{ccc}
{\includegraphics[width=0.2\hsize]{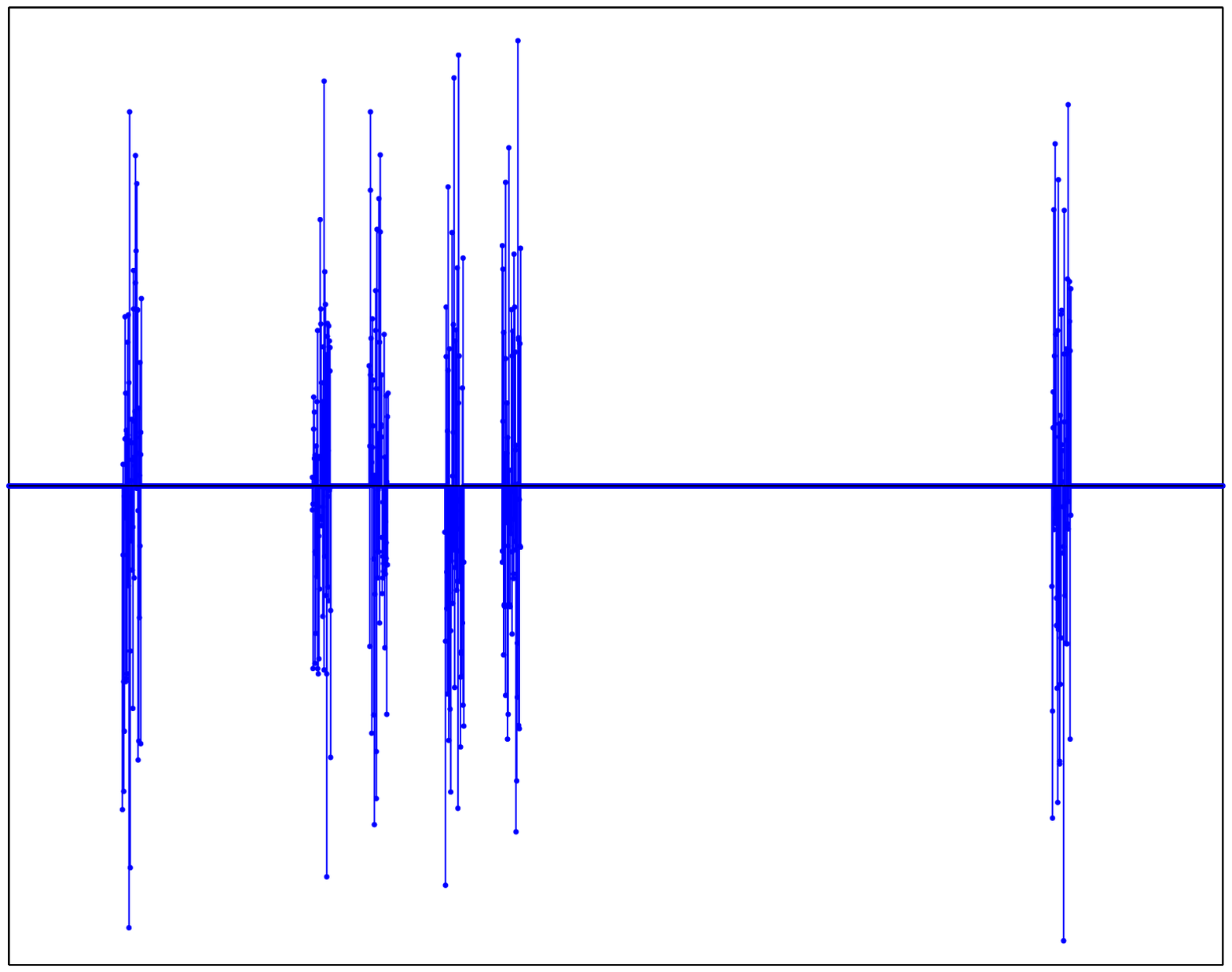}}&
{\includegraphics[width=0.2\hsize]{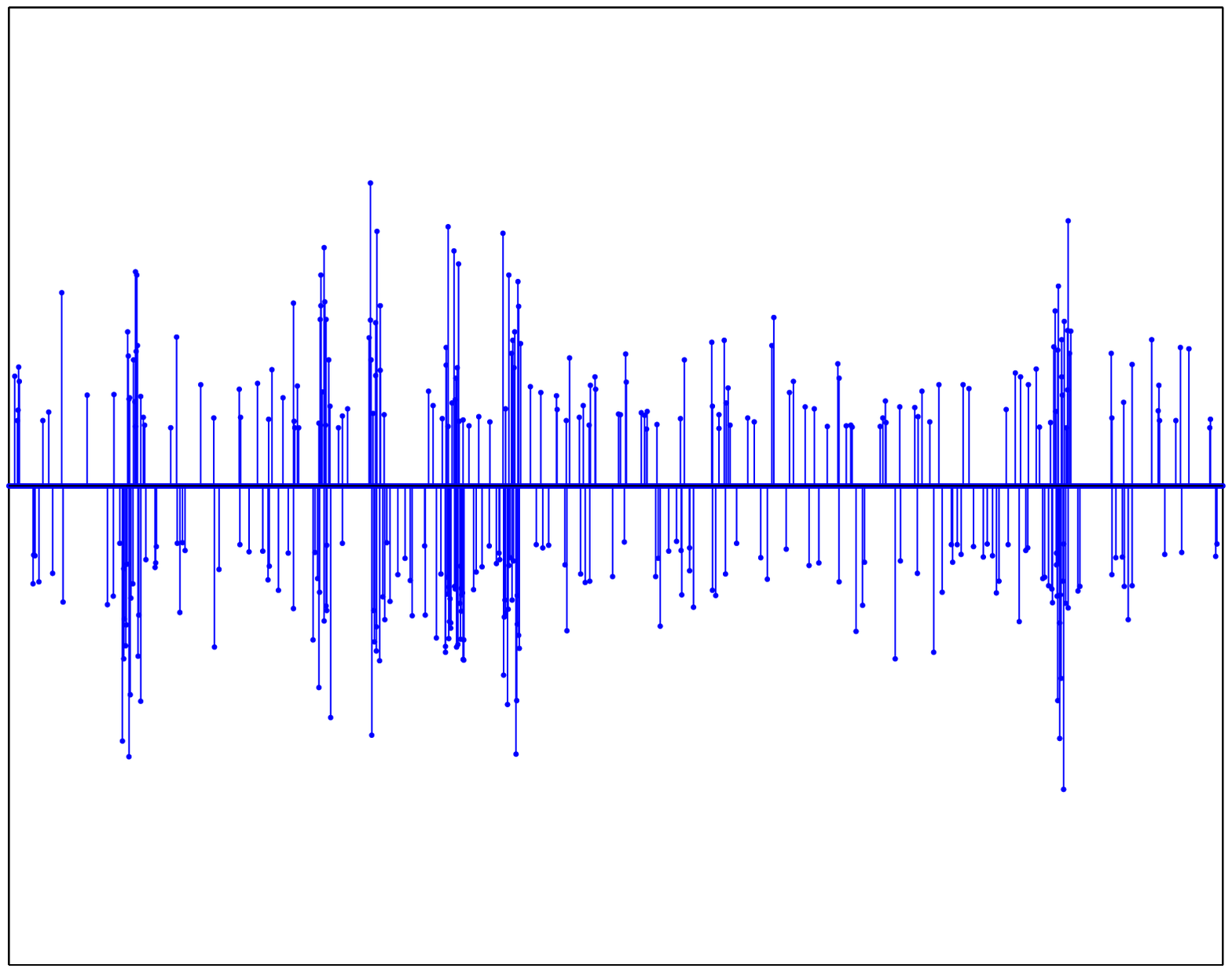}} &
{\includegraphics[width=0.2\hsize]{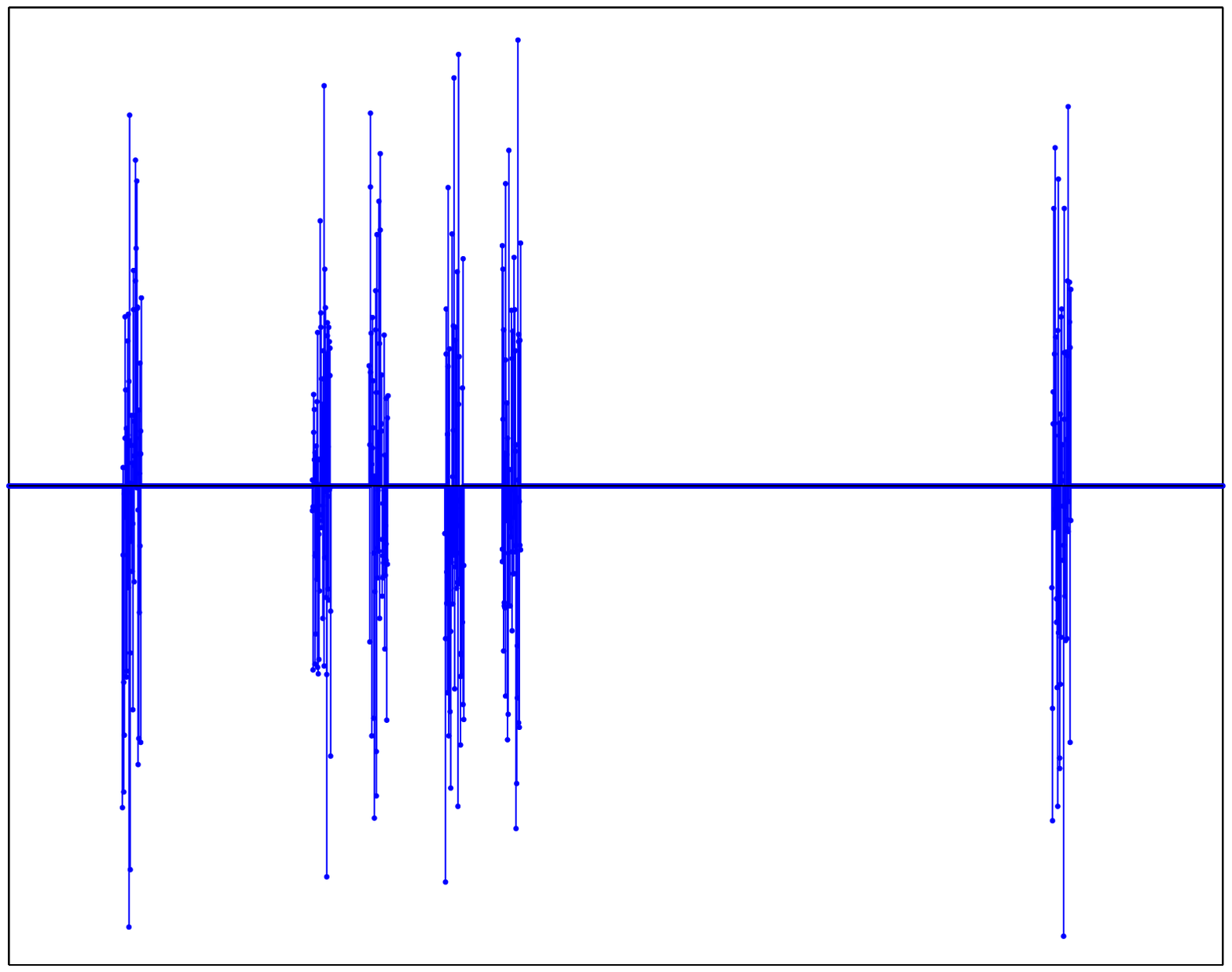}} \\
(a) original block-sparse signal & (b) CoSaMP & (c) model-based recovery \\
 &  (RMSE $= 0.723$) & (RMSE $= 0.015$)
\end{tabular}
\caption{\small\sl Example performance of structured signal
recovery for a block-sparse signal. (a) Example block-sparse
signal of length $N=4096$ with $K=6$ nonzero blocks of size $J=64$.
Recovered signal from $M=960$ measurements using (b) conventional
CoSaMP recovery and (c) block-based recovery. Standard recovery not 
only recovers spurious nonzeros, but also attenuates the magnitude of 
the actual nonzero entries.
\label{fig:blocksparse_recon}}
\end{figure*}

We now consider block-compressible signals. An example recovery is
illustrated in Figure~\ref{fig:blockcmprs_recon}. In this case, the
$\ell_2$-norms of the blocks decay according to a power law, as
described above. Again, the number of measurements is far below the
minimum number required to guarantee stable recovery through
conventional CS recovery. However, enforcing the structured sparsity
model in the approximation process results in a solution that is very
close to the best 5-block approximation of the signal.

\begin{figure*}[!t]
\centering
\begin{tabular}{cccc}
{\includegraphics[width=0.2\hsize]{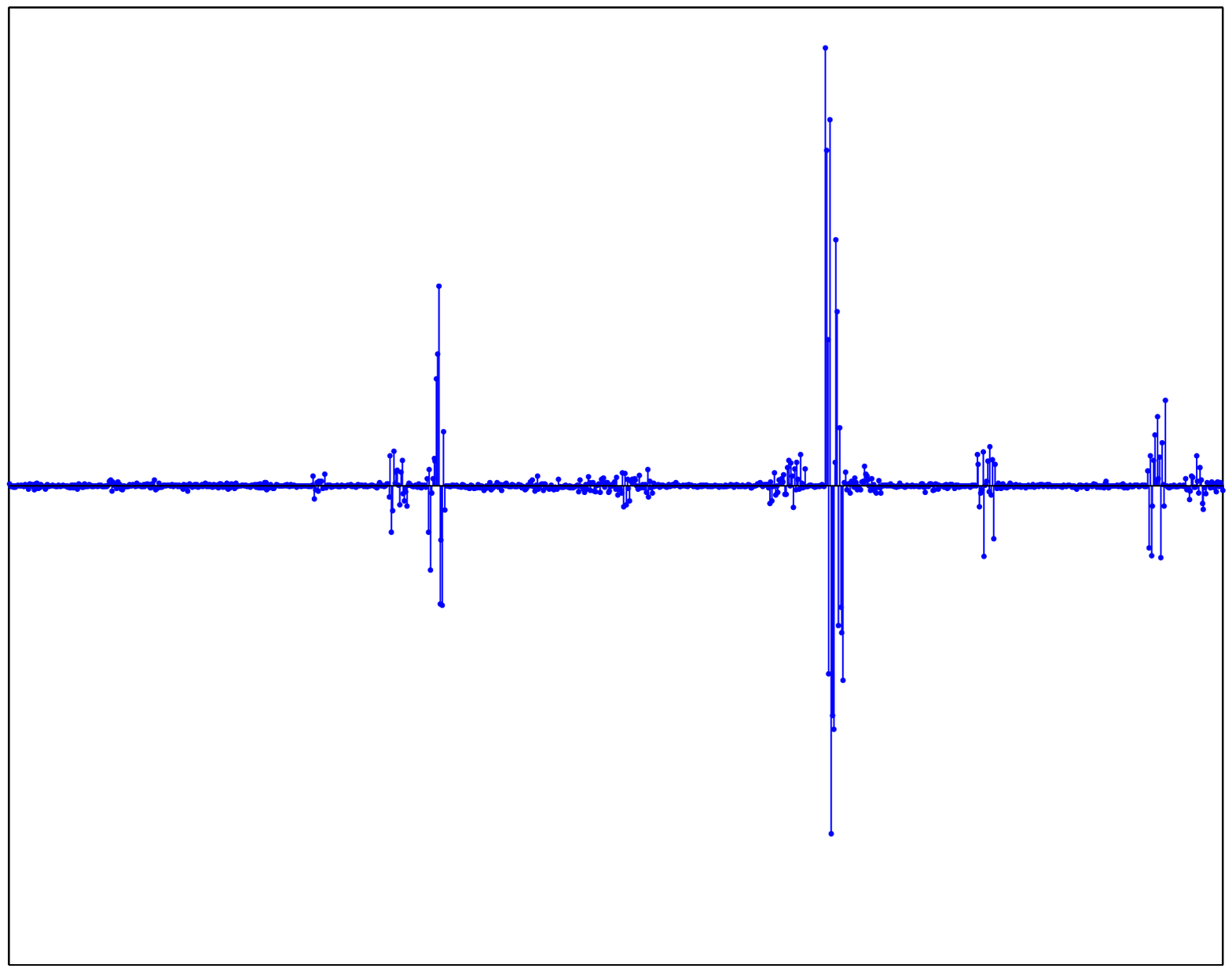}}&
{\includegraphics[width=0.2\hsize]{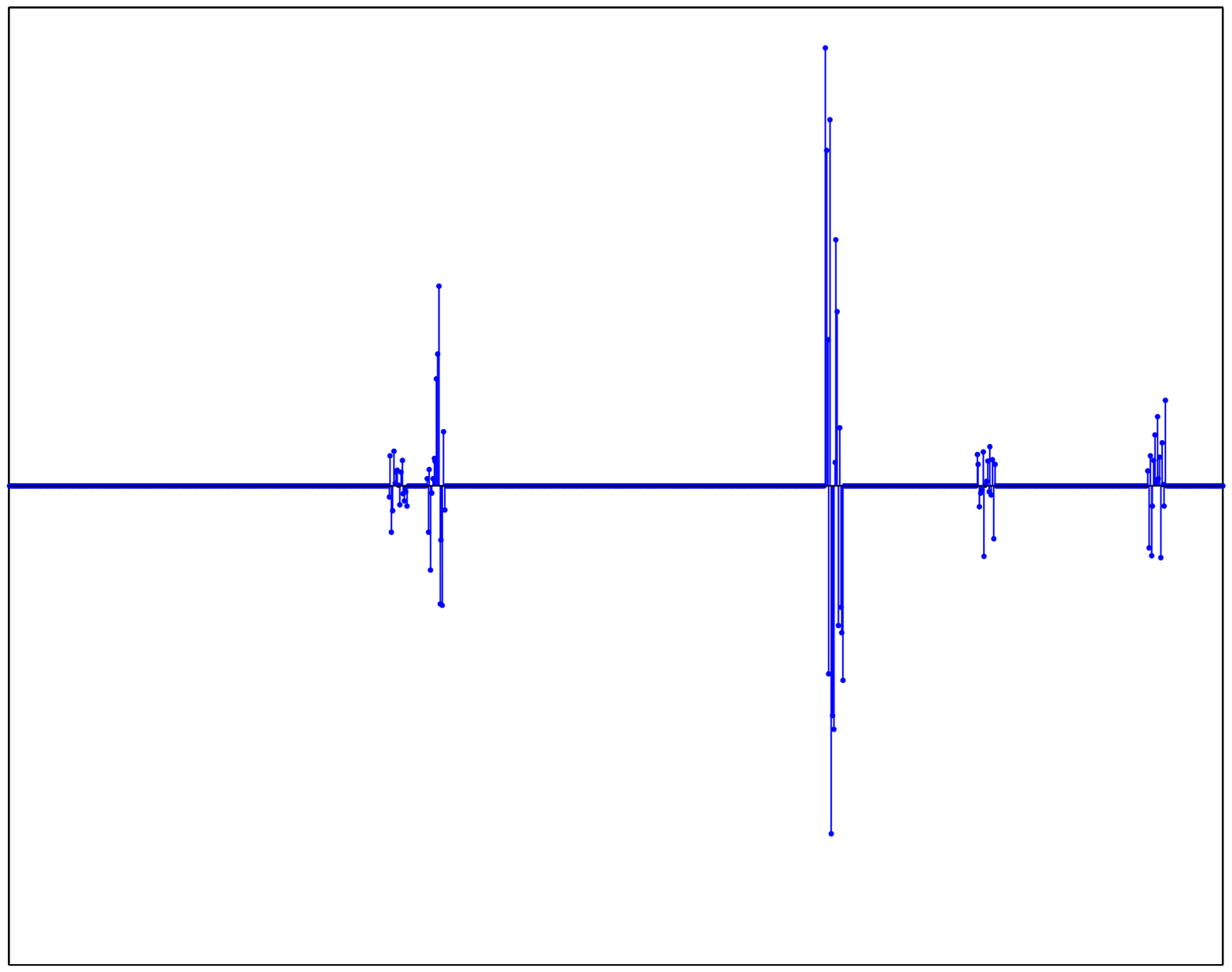}}&
{\includegraphics[width=0.2\hsize]{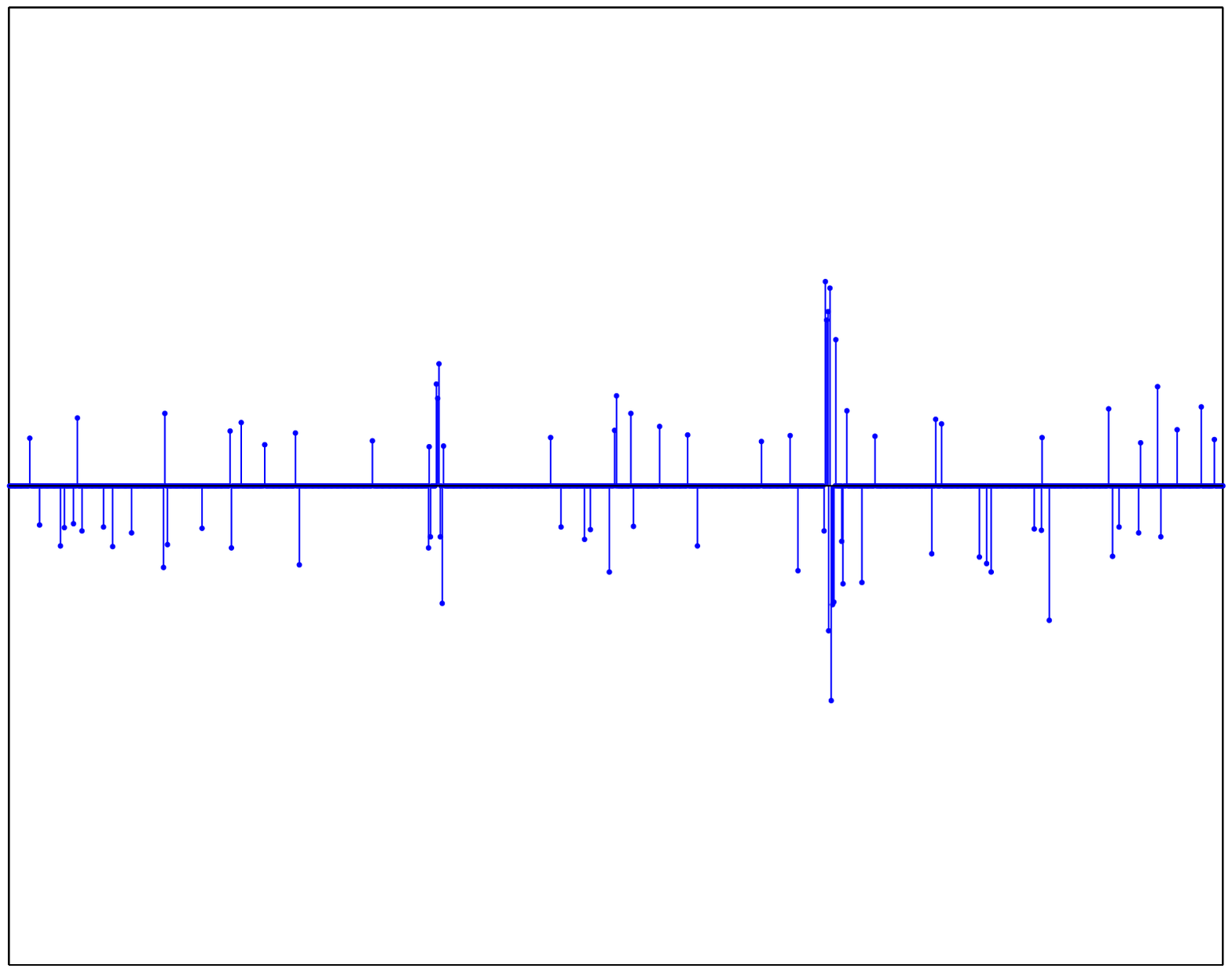}} &
{\includegraphics[width=0.2\hsize]{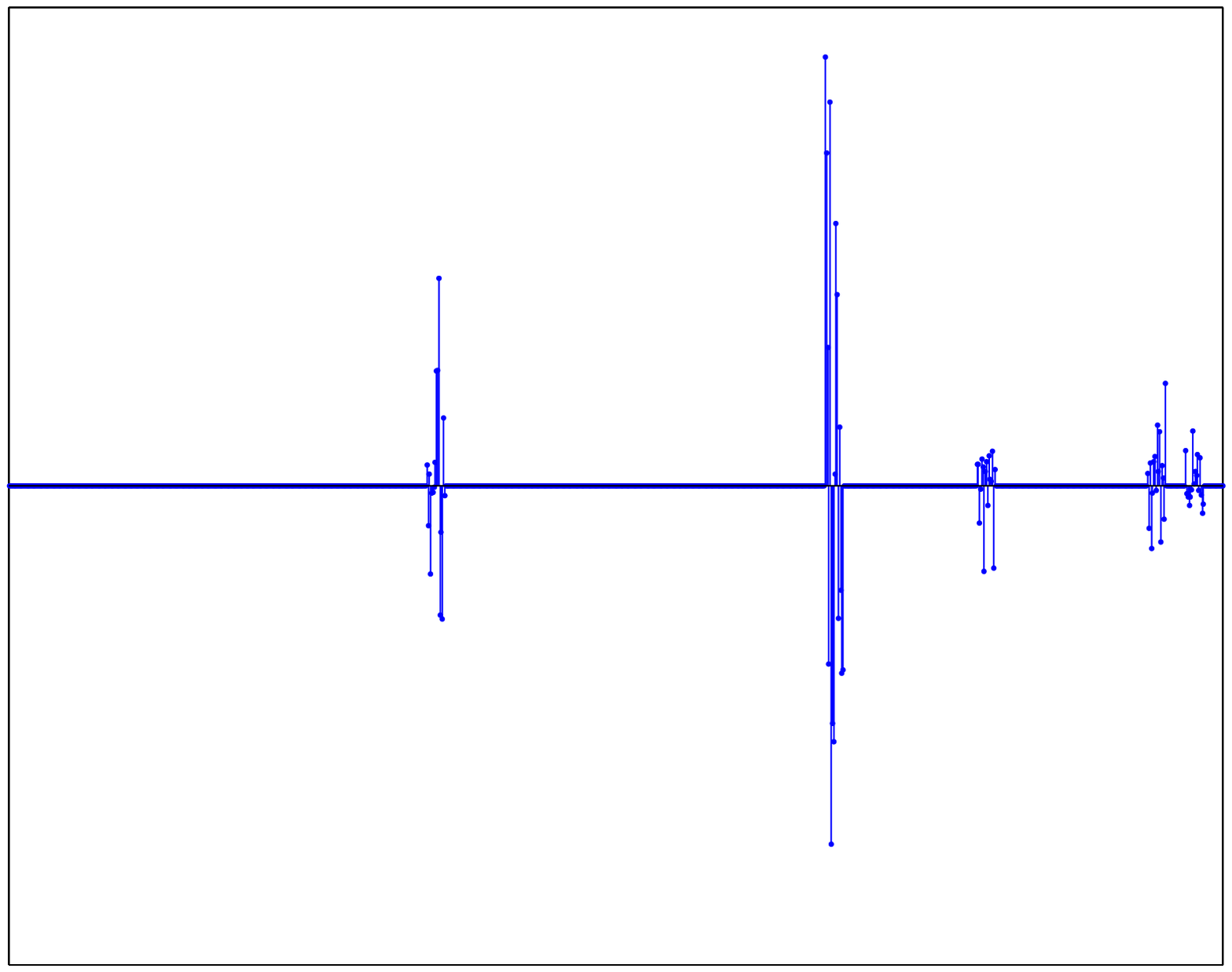}}\\
(a) signal &  (b) best 5-block approximation & (c) CoSaMP &  (d) model-based recovery  \\
 & (RMSE $=0.116$) &
(RMSE $=0.711$) & (RMSE $=0.195$)
\end{tabular}
\caption{\small\sl Example performance of structured signal
recovery for block-compressible signals. (a) Example
block-compressible signal, length $N=1024$. (b) Best block-based
approximation with $K = 5$ blocks. Recovered signal from $M=200$
measurements using both (c) conventional CoSaMP recovery and (d)
block-based recovery. Standard recovery not only recovers spurious 
significant entries, but also attenuates the magnitude of the actual 
significant entries
\label{fig:blockcmprs_recon}}
\end{figure*}

Figure~\ref{fig:blocksp_mc}(a) indicates the decay in recovery error as
a function of the numbers of measurements for CoSaMP and model-based
recovery. We generated sample block-sparse signals as follows: we
randomly selected a set of $K$ blocks, each of size $J$, and endow them with
coefficients that follow an i.i.d.\ Gaussian distribution. Each sample point
in the curves is generated by performing 200 trials of the
corresponding algorithm. As in the connected wavelet-tree case, we
observe clear gains using model-based recovery, particularly for
low-measurement regimes; CoSaMP matches model-based recovery only
for $M \geq 5K.$

\begin{figure*}[bt]
\begin{center}
\begin{tabular}{cc}
\epsfig{file=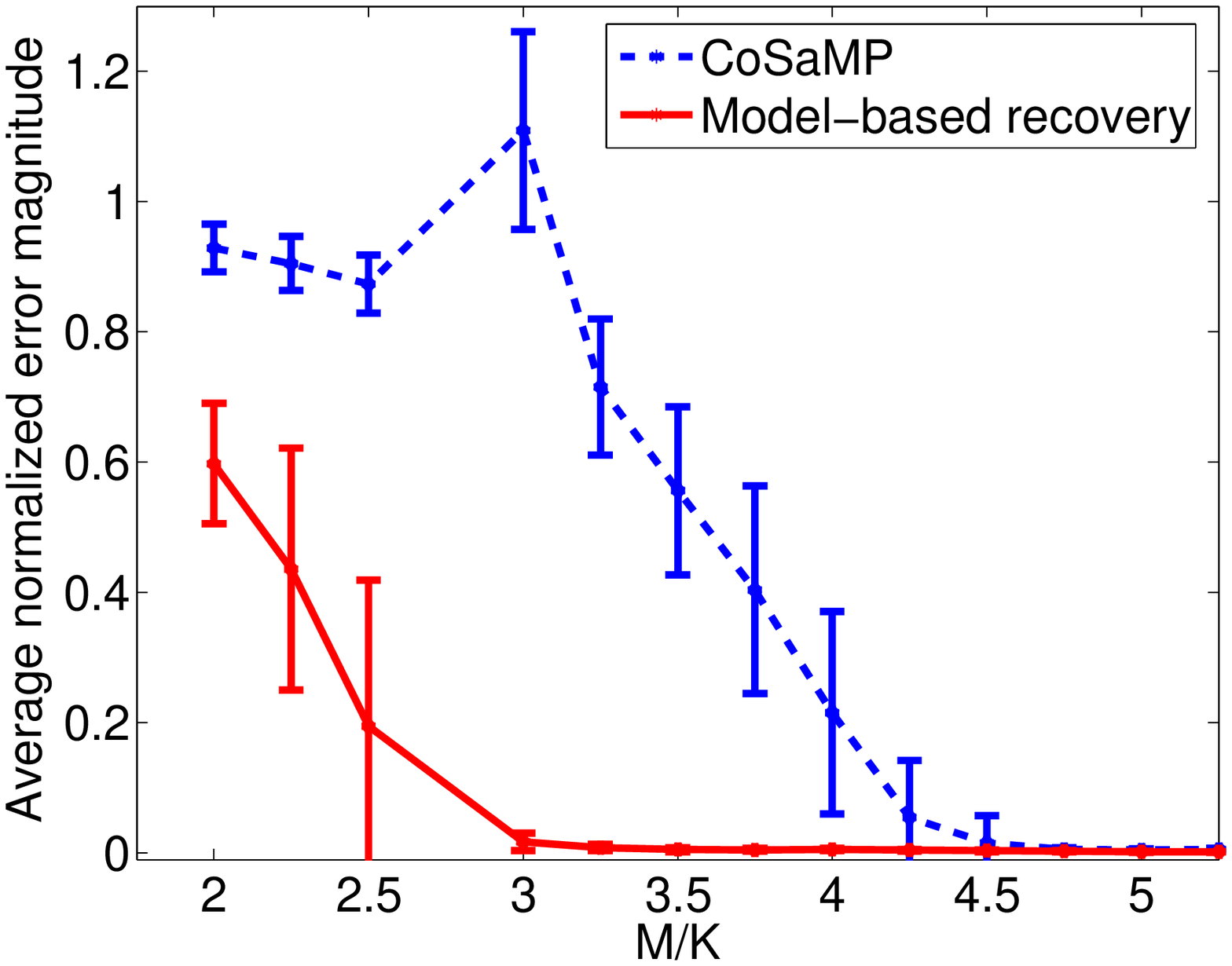,height=50mm} &
\epsfig{file=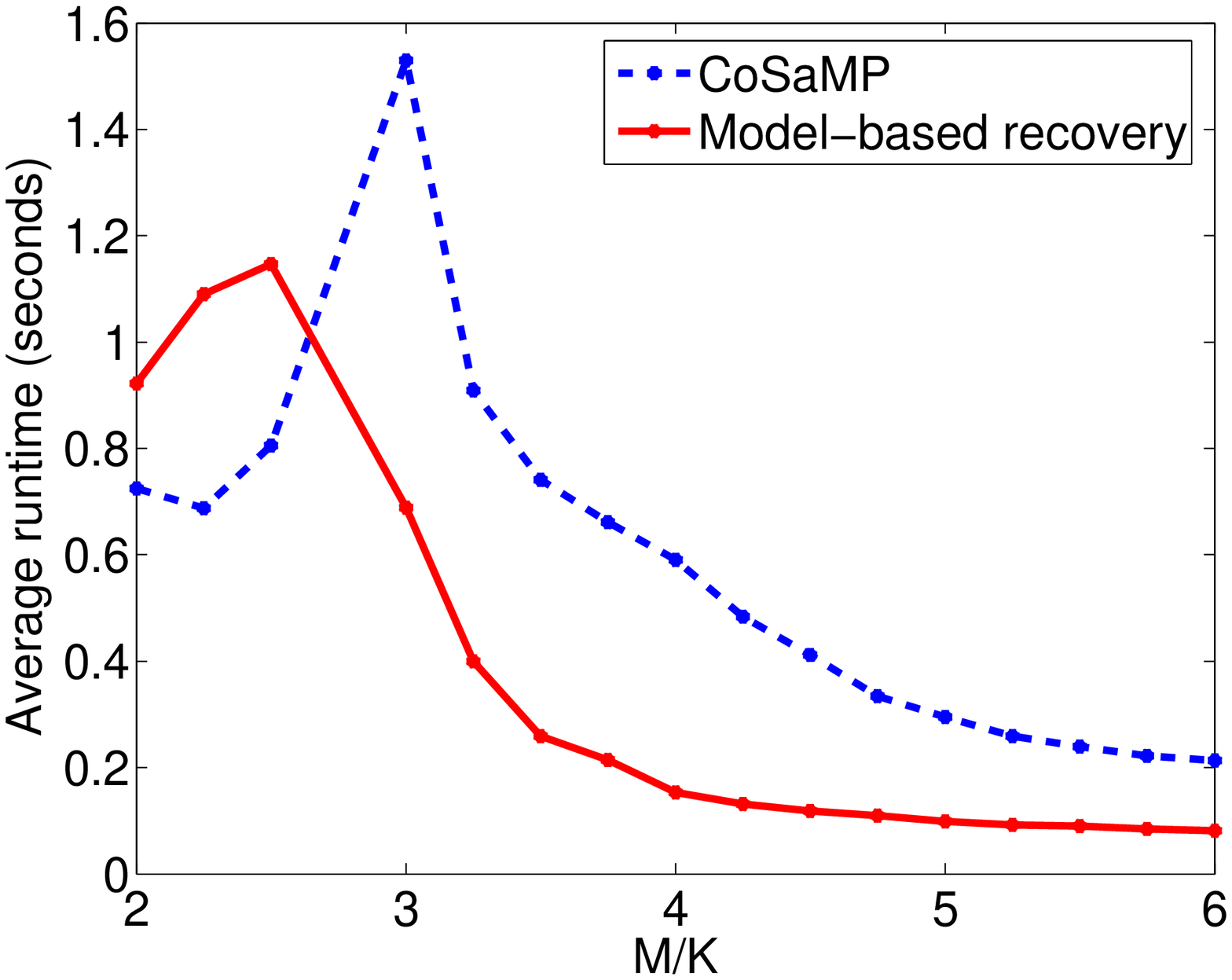,height=51mm} \\
(a) & (b)
\end{tabular}
\end{center}
\caption{\small\sl \label{fig:blocksp_mc} Performance of CoSaMP vs.\
block-based recovery on a class of block-sparse signals.
(a) Average normalized recovery error and (b) average runtime for
each recovery algorithm as a function of the overmeasuring factor
$M/K$. CoSaMP does not match the performance of the block-based
algorithm until $M = 5K$. Furthermore, the block-based algorithm
has faster convergence time than CoSaMP.}
\end{figure*}

Figure~\ref{fig:blocksp_mc}(b) compares the recovery times of the
two approaches. For this particular model, we observe that our
proposed approach is in general much faster than CoSaMP. This
is because of two reasons: a) the block-based approximation step
involves sorting fewer coefficients, and thus is faster than
$K-$term approximation; b) block-based recovery requires fewer
iterations to converge to the true solution.

\section{Conclusions}
\label{sec:concl}

In this paper, we have aimed to demonstrate that there are
significant performance gains to be made by exploiting more
realistic and richer signal models beyond the simplistic sparse and
compressible models that dominate the CS literature. Building on the
unions of subspaces results of \cite{samplingunion} and the proof
machinery of \cite{CoSaMP}, we have taken some first steps
towards what promises to be a general theory for model-based CS by
introducing the notion of a structured compressible signal and the
associated restricted amplification property (RAmP) condition it
imposes on the measurement matrix $\Phi$.
Our analysis poses the nested approximation property (NAP) as a 
sufficient condition that is satisfied by many structured sparsity models.

For the volumes of natural and manmade signals and images that are
wavelet-sparse or compressible, our tree-based CoSaMP and IHT
algorithms offer performance that significantly exceeds today's
state-of-the-art while requiring only $M=\bigo{K}$ rather than
$M=\bigo{K\log(N/K)}$ random measurements. For block-sparse signals
and signal ensembles with common sparse support, our block-based 
CoSaMP and IHT algorithms offer not only excellent performance but 
also require just $M = \bigo{JK}$ measurements, where $JK$ is the 
signal sparsity. Furthermore, block-based recovery can recover signal 
ensembles using fewer measurements than the number required when 
each signal is recovered independently; we have shown such advantages 
using real-world data from environmental sensor networks~\cite{ModelDCS}. 
Additional structured sparsity models have been developed using our 
general framework in~\cite{ModelCSSPARS} and~\cite{ModelCSSAMPTA}; 
we have also released a Matlab toolbox containing the corresponding 
model-based CS recovery algorithms, available at 
\texttt{http://dsp.rice.edu/software}.

There are many avenues for future work on model-based CS. We have
only considered the recovery of signals from models that can be
geometrically described as a union of subspaces; possible extensions
include other, more complex geometries such as
high-dimensional polytopes and nonlinear manifolds. We also expect
that the core of our proposed algorithms --- a structured sparse
approximation step --- can be integrated into other iterative
algorithms, such as relaxed $\ell_1$-norm minimization methods.
Furthermore, our framework will benefit from the formulation of new
structured sparsity models that are endowed with efficient structured
sparse approximation algorithms.

\appendices

\section{Proof of Theorem~\ref{theo:gaussiancomp}}
\label{app:gaussiancomp}

To prove this theorem, we will study the distribution of the maximum singular value of a submatrix $\Phi_T$ of a matrix with i.i.d.\ Gaussian entries $\Phi$ corresponding to the columns indexed by $T$. From this we obtain the probability that RAmP does not hold for a fixed support $T$. We will then evaluate the same probability for all supports $T$ of  elements of $\R_{j,K}$, where the desired bound on the amplification is dependent on the value of $j$. This gives us the probability that the RAmP does not hold for a given residual subspace set $\R_{j,K}$. We fix the probability of failure on each of these sets; we then obtain probability that the matrix $\Phi$ does not have the RAmP using a union bound. We end by obtaining conditions on the number of rows $M$ of $\Phi$ to obtain a desired probability of failure.

We begin from the following concentration of measure for the largest singular value of a $M \times K$ submatrix $\Phi_T$, $|T| = K$, of an $M\times N$ matrix $\Phi$ with i.i.d.\ subgaussian entries that are properly normalized~\cite{CandesDLP,ledoux,subgaussian}:
$$P\left(\sigma_{\max}(\Phi_T) > 1+\sqrt{\frac{K}{M}}+\tau+\beta\right) \le e^{-M\tau^2/2}.$$
For large enough $M$, $\beta \ll 1$; thus we ignore this small constant in the sequel.
By letting $\tau = j^r\sqrt{1+\epsilon_K}-1-\sqrt{\frac{K}{M}}$ (with the appropriate value of $j$ for $T$), we obtain
$$P\left(\sigma_{\max}(\Phi_T) > j^r\sqrt{1+\epsilon_K}\right) \le e^{-\frac{M}{2}\left(j^r\sqrt{1+\epsilon_K}-1-\sqrt{\frac{K}{M}}\right)^2}.$$
We use a union bound over all possible $R_j$ supports for $\u \in \R_{j,K}$  to obtain the probability that $\Phi$ amplifies the norm of some $\u$ by more than $j^r\sqrt{1+\epsilon_K}$:
\begin{eqnarray*}
P\left(\|\Phi\u\|_2 > \left(j^r\sqrt{1+\epsilon_K}\right)\|\u\|_2~\mathrm{for~some}~\u \in \R_{j,K}\right) \\ \le R_je^{-\frac{1}{2}\left(\sqrt{M}(j^r\sqrt{1+\epsilon_K}-1)-\sqrt{K}\right)^2}.
\end{eqnarray*}
Bound the right hand side by a constant $\mu$; this requires
\begin{equation}
R_j \le e^{\frac{1}{2}\left(\sqrt{M}(j^r\sqrt{1+\epsilon_K}-1)-\sqrt{K}\right)^2}\mu
\label{eq:rmu}
\end{equation}
for each $j$.
We use another union bound among the residual subspaces $\R_{j.K}$ to measure the probability that the RAmP does not hold:
\begin{eqnarray*}
P\left(\frac{\|\Phi\u\|_2}{\|\u\|_2} > j^r\sqrt{1+\epsilon_K},~\u \in \R_{j,K},~1 \le j \le \lceil N/K \rceil\right) \\
\le \left\lceil \frac{N}{K} \right\rceil \mu.
\end{eqnarray*}
To bound this probability by $e^{-t}$, we need $\mu = \frac{K}{N}e^{-t}$; plugging this into (\ref{eq:rmu}), we obtain
$$R_j \le e^{\frac{1}{2}\left(\sqrt{M}(j^r\sqrt{1+\epsilon_K}-1)-\sqrt{K}\right)^2}\frac{K}{N}e^{-t}
$$
for each $j$. Simplifying, we obtain that for $\Phi$ to posess the RAmP with probability $1-e^{-t}$, the following must hold for all $j$:
\begin{equation}
M \ge \left(\frac{\sqrt{2\left(\ln \frac{R_jN}{K}+t\right)}+\sqrt{K}}{j^r\sqrt{1+\epsilon_K}-1}\right)^2.
\label{eq:rmu2}
\end{equation}
Since $(\sqrt{a}+\sqrt{b})^2 \le 2a+2b$ for $a,b > 0$, then the hypothesis (\ref{eq:gaussiancomp}) implies (\ref{eq:rmu2}), proving the theorem. \qed

\section{Proof of Theorem~\ref{theo:phiresidual}}
\label{app:phiresidual}

In this proof, we denote $\aalg(\x,K) = \x_K $ for brevity.
To bound $\|\Phi(\x-\x_K)\|_2, $ we write $\x$ as
$$\x = \x_K+ \sum_{j=2}^{\lceil N/K \rceil} \x_{T_j},$$
where
$$\x_{T_j} = \x_{jK}-\x_{(j-1)K}, j = 2,\ldots,\lceil N/K \rceil$$
is the difference between the best $jK$ structured sparse 
approximation and the best $(j-1)K$ structured sparse 
approximation. Additionally, each piece $\x_{T_j} \in \R_{j,K}$. 
Therefore, since $\Phi$ satisfies the $(\epsilon_K,s-1)$-RAmP, we obtain
\begin{eqnarray*}
\|\Phi(\x -\x_K)\|_2 &=& \left\|\Phi\left(\sum_{j=2}^{\lceil N/K \rceil} \x_{T_j}\right)\right\|_2 \le \sum_{j=2}^{\lceil N/K \rceil}\|\Phi\x_{T_j}\|_2 \\
&\le& \sum_{j=2}^{\lceil N/K \rceil}\sqrt{1+\epsilon_K}j^{s-1}\|\x_{T_j}\|_2.
\label{eq:sumpieces}
\end{eqnarray*}
Since $\x \in \aaprox_s$, the norm of each piece can be bounded as
\begin{eqnarray*}
\|\x_{T_j}\|_2 &=& \|\x_{jK}-\x_{(j-1)K}\|_2 \\
&\le& \|\x-\x_{(j-1)K}\|_2 + \|\x-\x_{jK}\|_2 \\
&\le& |\x|_{\aaprox_s} K^{-s}\left((j-1)^{-s}+j^{-s}\right).
\end{eqnarray*}
Applying this bound in (\ref{eq:sumpieces}), we obtain
\begin{eqnarray}
\|\Phi(\x -\x_K)\|_2  &\le& \sqrt{1+\epsilon_K}\sum_{j=2}^{\lceil N/K \rceil}j^{s-1}\|\x_{T_j}\|_2, \nonumber \\
&\le& \frac{\sqrt{1+\epsilon_K}}{K^s}|\x|_{\aaprox_s} \sum_{j=2}^{\lceil N/K \rceil}\frac{j^{s-1}}{(j-1)^s}+\frac{j^{s-1}}{j^s}, \nonumber \\
&\le& \frac{\sqrt{1+\epsilon_K}}{K^s}|\x|_{\aaprox_s} \sum_{j=2}^{\lceil N/K \rceil}\frac{1}{j(1-1/j)^s}+\frac{1}{j}, \nonumber \\
&\le& \frac{\sqrt{1+\epsilon_K}}{K^s}|\x|_{\aaprox_s} \sum_{j=2}^{\lceil N/K \rceil}\frac{2^s}{j}+\frac{1}{j}, \nonumber \\
&\le& (2^s+1)\frac{\sqrt{1+\epsilon_K}}{K^s}|\x|_{\aaprox_s} \sum_{j=2}^{\lceil N/K \rceil}j^{-1}. \nonumber
\nonumber
\end{eqnarray}
It is easy to show, using Euler-Maclaurin summations, that $\sum_{j=2}^{\lceil N/K \rceil} j^{-1} \le  \ln \lceil N/K \rceil$; we then obtain
$$
\|\Phi(\x -\x_K)\|_2 \le (2^s+1)\sqrt{1+\epsilon_K}K^{-s}\ln \left\lceil \frac{N}{K} \right\rceil|\x|_{\aaprox_s}, $$
which proves the theorem.
\qed

\section{Model-based Iterative Hard Thresholding}
\label{app:addalgs}

Our proposed model-based iterative hard thresholding (IHT) is given
in Algorithm~\ref{alg:MIHT}.
For this algorithm, Theorems~\ref{theo:rcosamp},
\ref{theo:compressible}, and \ref{theo:jsm2comp} can be proven with
only a few modifications: $\Phi$ must have the $\A_K^3$-RIP with
$\delta_{\A_K^3} \le 0.1$, and the constant factor in the bound
changes from 15 to 4 in Theorem~\ref{theo:rcosamp}, from 35 to 10 in
Theorem~\ref{theo:compressible}, and from 20 to 5 in
Theorem~\ref{theo:jsm2comp}.

\begin{algorithm*}[t]
\caption{Model-based Iterative Hard Thresholding}
\label{alg:MIHT}
\begin{tabbing}
Inputs: CS matrix $\Phi$, measurements $\y$, structured sparse 
approximation algorithm $\aalg_K$ \\
Output: $K$-sparse approximation $\xhat$ to true signal $x$ \\
$\xhat_0=0$ , $d = \y$; $i = 0$ \hspace{32.5mm} \{initialize\} \\
{\bf while} \= halting criterion false {\bf do} \hspace{20mm}\= \\
\> 1. $i \leftarrow i+1$ \\
\> 2. $\b \leftarrow \xhat_{i-1} + \Phi\trans d$ \> \{form signal estimate\} \\
\> 3.  $\xhat_i \leftarrow \aalg(\b,K)$ \> \{prune residual estimate according to structure\} \\
\> 4. $d \leftarrow \y - \Phi \xhat_i$ \> \{update measurement residual\} \\
{\bf end while} \\
return $\xhat \leftarrow \xhat_i$
\end{tabbing}
\end{algorithm*}

To illustrate the performance of the algorithm, we repeat the {\em
HeaviSine} experiment from Figure~\ref{fig:wvcomparison}. Recall
that $N = 1024$, and $M = 80$ for this example. The advantages of
using our tree-structured sparse approximation step (instead of mere hard
thresholding) are evident from Figure~\ref{fig:iht_recon}. In
practice, we have observed that our model-based algorithm converges
in fewer steps than IHT and yields much more accurate results in
terms of recovery error.

\begin{figure}[!t]
\centering
\begin{tabular}{ccc}
{\includegraphics[width=0.25\hsize]{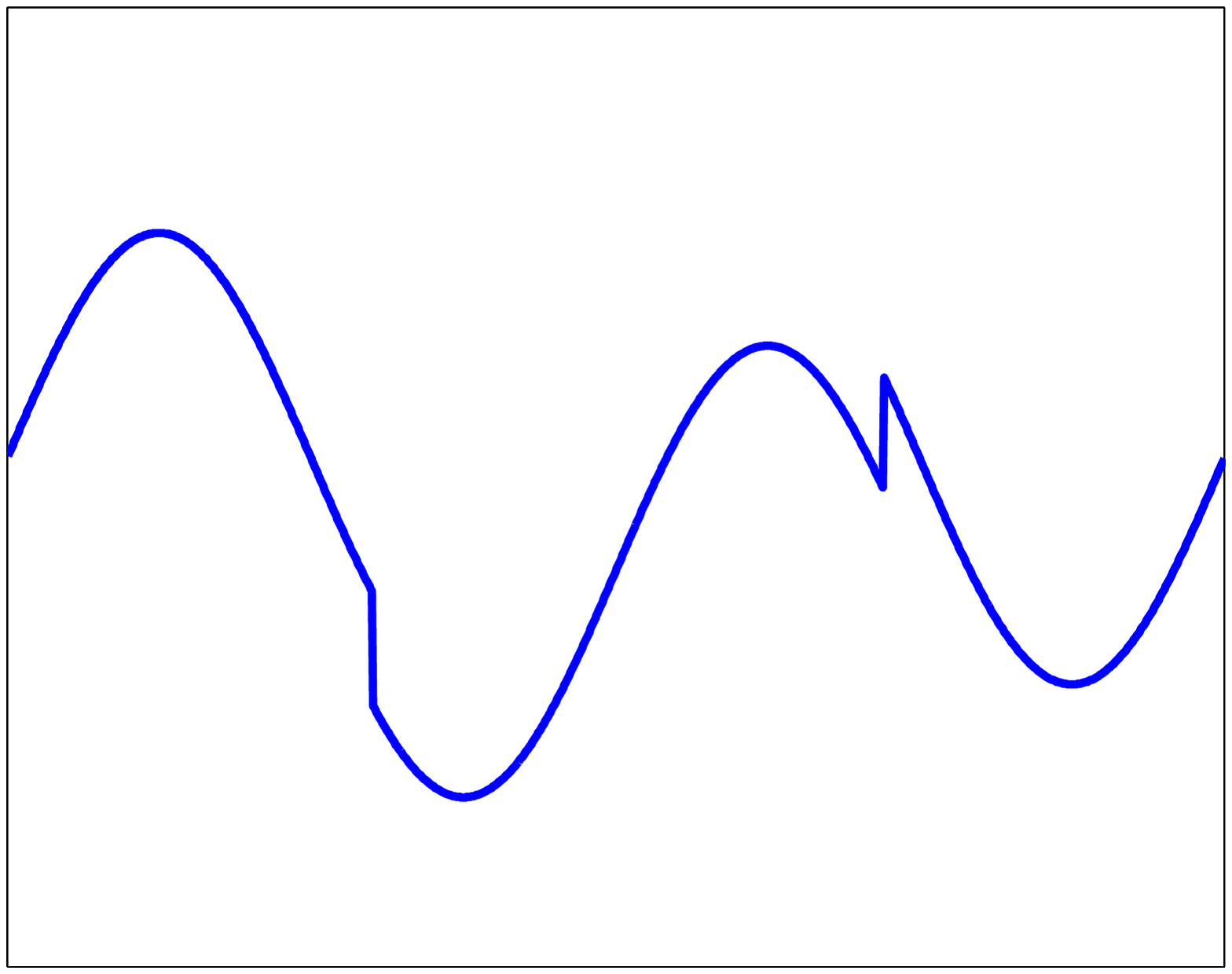}}&
{\includegraphics[width=0.25\hsize]{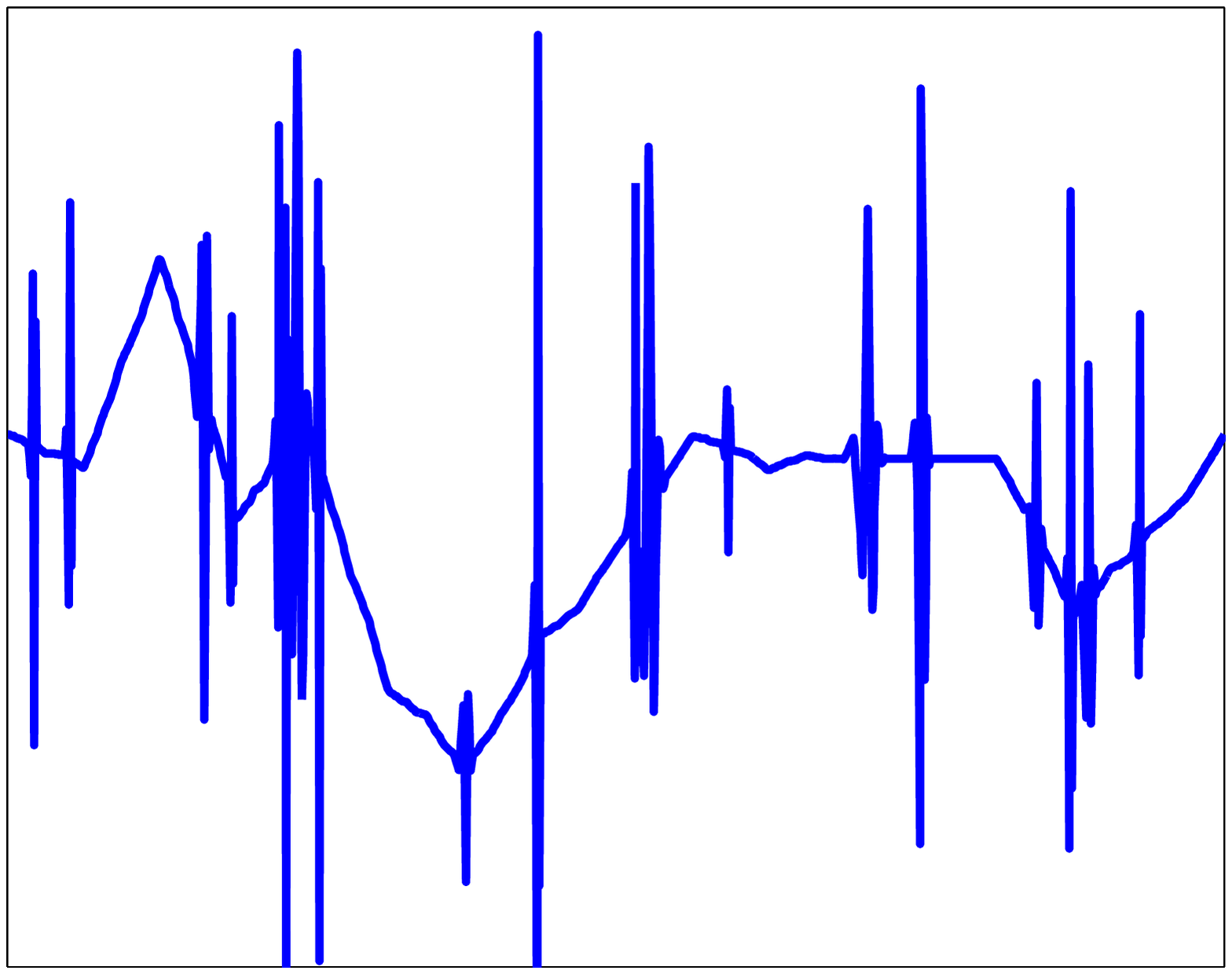}} &
{\includegraphics[width=0.25\hsize]{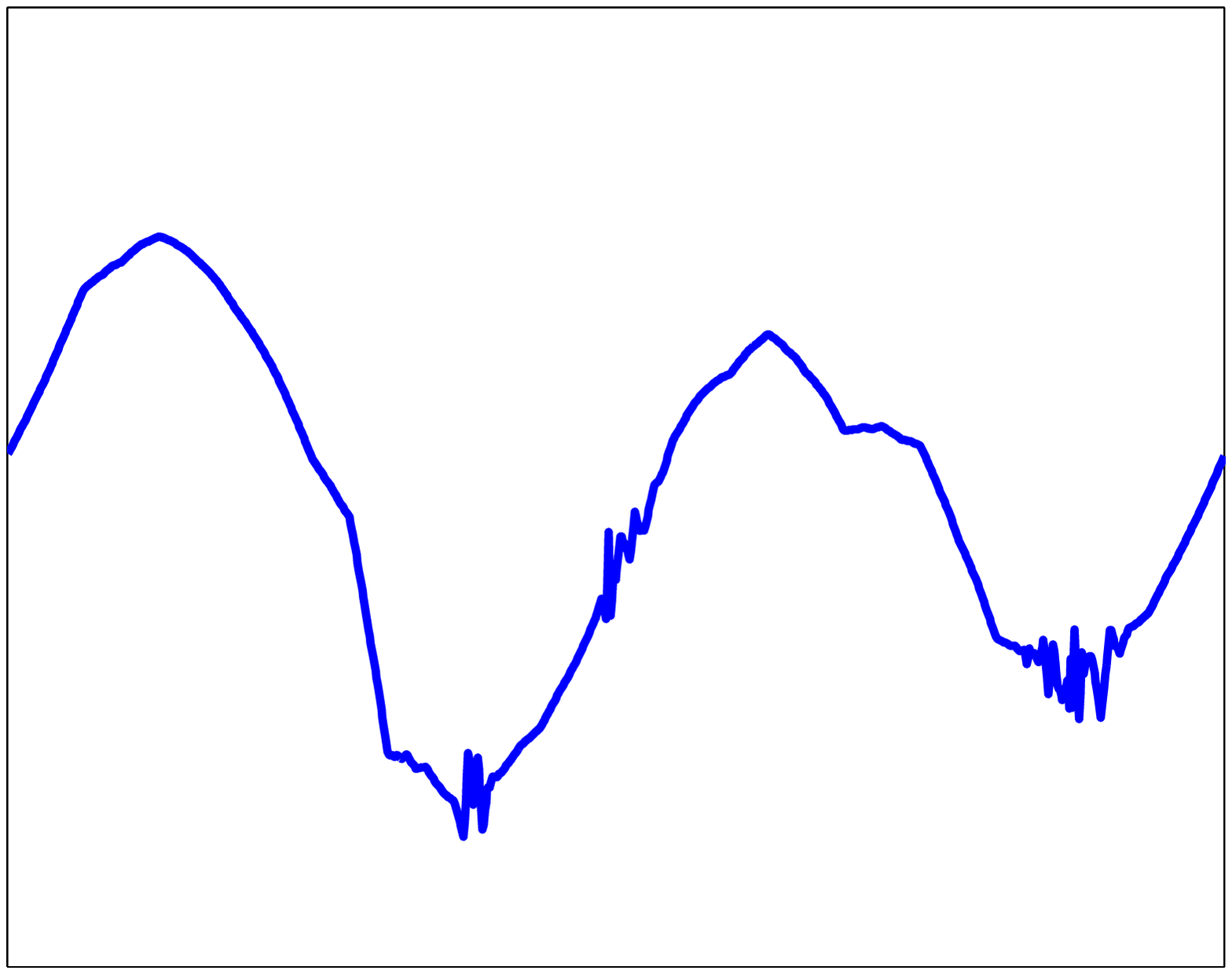}} \\
(a) original & (b) IHT & (c) model-based IHT \\
 & (RMSE $= 0.627$) & (RMSE $= 0.080$)
\end{tabular}
\caption{\small\sl Example performance of model-based IHT. (a)
Piecewise smooth {\em HeaviSine} test signal, length $N=1024$.
Signal recovered from $M=80$ measurements using both (b) standard
and (c) model-based IHT recovery.  Root mean-squared error (RMSE)
values are normalized with respect to the $\ell_2$ norm of the
signal. \label{fig:iht_recon}}
\end{figure}

\section{Proof of Theorem~\ref{theo:rcosamp}}
\label{app:rcosamp}

The proof of this theorem is identical to that of the CoSaMP
algorithm in~\cite[Section 4.6]{CoSaMP}, and requires a set of six
lemmas. The sequence of Lemmas~\ref{lemm:prop1}--\ref{lemm:prune} below
are modifications of the lemmas in~\cite{CoSaMP} that are restricted
to the structured sparsity model. Lemma~\ref{lemm:merge} does not need any
changes from~\cite{CoSaMP}, so we state it without proof.
The proof of Lemmas~\ref{lemm:id}--\ref{lemm:prune} use the properties in Lemmas~\ref{lemm:prop1} and \ref{lemm:prop2}, which are simple to prove.
\begin{LEMM}
Suppose $\Phi$ has $\A$-RIP with constant $\delta_\A$. Let $\Omega$
be a support corresponding to a subspace in $\A$. Then we have the
following handy bounds.
\begin{eqnarray}
\|\Phi_\Omega\trans \u\|_2 &\le& \sqrt{1+\delta_\A}\|\u\|_2, \nonumber \\
\|\Phi_\Omega\pinv \u\|_2 &\le& \frac{1}{\sqrt{1-\delta_\A}}\|\u\|_2, \nonumber \\
\|\Phi_\Omega\trans \Phi_\Omega \u\|_2 &\le& (1 + \delta_\A)\|\u\|_2, \nonumber \\
\|\Phi_\Omega\trans \Phi_\Omega \u\|_2 &\ge& (1 - \delta_\A)\|\u\|_2, \nonumber \\
\|(\Phi_\Omega\trans \Phi_\Omega)^{-1} \u\|_2 &\le& \frac{1}{1 - \delta_\A}\|\u\|_2, \nonumber \\
\|(\Phi_\Omega\trans \Phi_\Omega)^{-1} \u\|_2 &\ge& \frac{1}{1 + \delta_\A}\|\u\|_2.
\nonumber
\end{eqnarray}
\label{lemm:prop1}
\end{LEMM}
\begin{LEMM}
Suppose $\Phi$ has $\A^2_K$-RIP with constant $\delta_{\A^2_K}$. Let $\Omega$ be a support corresponding to a subspace in $\A_K$, and let $\x \in \A_K$. Then
$\|\Phi_\Omega\trans \Phi \x\vert_{\Omega^C}\|_2 \le \delta_{\A^2_K}\|\x\vert_{\Omega^C}\|_2.$
\label{lemm:prop2}
\end{LEMM}

We begin the proof of Theorem~\ref{theo:rcosamp} by fixing an iteration $i \ge 1$ of model-based CoSaMP. We write $\xhat = \xhat_{i-1}$ for the
signal estimate at the beginning of the $i^{th}$ iteration. Define
the signal residual $\s = \x - \xhat$, which implies that $\s \in
\A_K^2$. We note that we can write $\r = \y - \Phi \xhat = \Phi (\x
- \xhat) + \n = \Phi\s + \n$.

\begin{LEMM}
{\em (Identification)} The set $\Omega =
\mathrm{supp}(\aalg_2(\e,K))$, where $\e = \Phi\trans \r$,
identifies a subspace in $\A_K^2$, and obeys
\begin{equation}
\|\s\vert_{\Omega^C}\|_2 \le 0.2223 \|\s\|_2 + 2.34 \|\n\|_2. \nonumber
\end{equation}
\label{lemm:id}
\end{LEMM}

{\em Proof of Lemma~\ref{lemm:id}:} Define the set $\Pi = \mathrm{supp}(\s)$. Let $\e_\Omega = \aalg_2(\e,K)$ be the structured sparse approximation to $\e$ with support $\Omega$, and similarly let $\e_\Pi$ be the approximation to $\e$ with support $\Pi$. Each approximation is equal to $\e$ for the coefficients in the support, and zero elsewhere. Since $\Omega$ is the support of the best approximation in $\A_K^2$, we must have:
\begin{eqnarray}
\|\e-\e_\Omega\|_2^2 &\le& \|\e-\e_\Pi\|_2^2, \nonumber \\
\sum_{n=1}^N (\e[n]-\e_\Omega[n])^2 &\le& \sum_{n=1}^N (\e[n]-\e_\Pi[n])^2, \nonumber \\
\sum_{n\notin \Omega} \e[n]^2 &\le& \sum_{n \notin \Pi} \e[n]^2, \nonumber \\
\sum_{n=1}^N \e[n]^2 - \sum_{n\notin \Omega} \e[n]^2 &\ge& \sum_{n=1}^N \e[n]^2 - \sum_{n \notin \Pi} \e[n]^2, \nonumber \\
\sum_{n\in \Omega} \e[n]^2 &\ge& \sum_{n \in \Pi} \e[n]^2, \nonumber \\
\sum_{n\in \Omega\setminus\Pi} \e[n]^2 &\ge& \sum_{n \in \Pi\setminus\Omega} \e[n]^2, \nonumber \\
\|\e\vert_{\Omega\setminus\Pi}\|_2^2 &\ge& \|\e\vert_{\Pi\setminus\Omega}\|_2^2,
\nonumber
\end{eqnarray}
where $\Omega \setminus \Pi$ denotes the set difference of $\Omega$ and $\Pi$. These signals are in $\A_K^4$ (since they arise as the difference of two elements from $\A_K^2$); therefore, we can apply the $\A^4_K$-RIP constants and Lemmas~\ref{lemm:prop1} and~\ref{lemm:prop2} to provide the following bounds on both sides (see~\cite{CoSaMP} for details):
\begin{eqnarray}
\|\e\vert_{\Omega\setminus\Pi}\|_2 &\le& \delta_{\A_K^4} \|s\|_2+\sqrt{1+\delta_{\A_K^2}} \|\n\|_2, \label{eq:ideq1} \\
\|\e\vert_{\Pi\setminus\Omega}\|_2 &\ge& (1-\delta_{\A_K^2})\|\s\vert_{\Omega^C}\|_2-\delta_{\A_K^2}\|s\|_2\nonumber\\
&&-\sqrt{1+\delta_{\A_K^2}} \|\n\|_2. \label{eq:ideq2}
\end{eqnarray}
Combining (\ref{eq:ideq1}) and (\ref{eq:ideq2}), we obtain
$$ \|\s\vert_{\Omega^C}\|_2 \le \frac{(\delta_{\A_K^2} + \delta_{\A_K^4}) \|s\|_2 + 2 \sqrt{1+\delta_{\A_K^2}}\|\n\|_2}{1-\delta_{\A_K^2}}.$$
The argument is completed by noting that $\delta_{\A_K^2} \le \delta_{\A_K^4} \le 0.1$. \qed
\begin{LEMM}
{\em (Support Merger)} Let $\Omega$ be a set of at
most $2K$ indices. Then the set $\Lambda = \Omega \cup
\mathrm{supp}(\xhat)$ contains at most $3K$ indices, and
$\|\x\vert_{\Lambda^C}\|_2 \le \|\s\vert_{\Omega^C}\|_2$.
\label{lemm:merge}
\end{LEMM}
\begin{LEMM}
{\em (Estimation)} Let $\Lambda$ be a support corresponding to a
subspace in $\A_K^3$, and define the least squares signal estimate
$\b$ by $\b\vert_T = \Phi_T\pinv y$, $\b\vert_{T^C} = 0$. Then
\begin{equation}
\|\x-\b\|_2 \le 1.112 \|\x\vert_{\Lambda^C}\|_2 + 1.06\|\n\|_2. \nonumber
\end{equation}
\label{lemm:est}
\end{LEMM}
{\em Proof of Lemma~\ref{lemm:est}:} It can be shown~\cite{CoSaMP} that
$$\|\x-\b\|_2 \le \|\x\vert_{\Lambda^C}\|_2 + \|(\Phi_\Lambda\trans \Phi_\Lambda)^{-1}\Phi_\Lambda\trans \Phi\x\vert_{\Pi^C}\|_2+\|\Phi_\Pi\pinv\n\|_2.$$
Since $\Lambda$ is a support corresponding to a subspace in $\A_K^3$ and $\x \in \A_K$, we use Lemmas~\ref{lemm:prop1} and~\ref{lemm:prop2} to obtain
\begin{eqnarray}
\|\x-\b\|_2 &\le& \|\x\vert_{\Lambda^C}\|_2 + \frac{\|\Phi_\Lambda\trans \Phi\x\vert_{\Pi^C}\|_2}{1-\delta_{\A_K^3}}+ \frac{\|\n\|_2}{\sqrt{1-\delta_{\A_K^3}}} \nonumber, \\
& \le & \left(1+\frac{\delta_{\A_K^4}}{1-\delta_{\A_K^3}}\right)\|\x\vert_{\Pi^C}\|_2+ \frac{\|\n\|_2}{\sqrt{1-\delta_{\A_K^3}}}. \nonumber
\end{eqnarray}
Finally, note that $\delta_{\A_K^3} \le \delta_{\A_K^4} \le 0.1$. \qed
\begin{LEMM}
{\em (Pruning)} The pruned approximation $\xhat_i =
\aalg(\b,K)$ is such that
\begin{equation}
\|\x-\xhat_i\|_2 \le 2 \|\x - \b\|_2. \nonumber
\end{equation}
\label{lemm:prune}
\end{LEMM}
{\em Proof of Lemma~\ref{lemm:prune}: } Since $\xhat_i$ is the best approximation in $\A_K$ to $\b$, and $\x \in \A_K$, we obtain
$$\|\x-\xhat_i\|_2 \le \|\x-\b\|_2 + \|\b-\xhat_i\|_2 \le 2 \|\x - \b\|_2.$$ \qed

We use these lemmas in reverse sequence for the inequalities below:
\begin{eqnarray}
\|\x-\xhat_i\|_2 &\le&  2 \|\x - \b\|_2 \nonumber, \\
& \le & 2(1.112 \|\x\vert_{\Lambda^C}\|_2 + 1.06\|\n\|_2), \nonumber \\
& \le & 2.224 \|\s\vert_{\Omega^C}\|_2 + 2.12 \|\n\|_2, \nonumber \\
&\le & 2.224(0.2223 \|\s\|_2 + 2.34 \|\n\|_2) + 2.12 \|\n\|_2, \nonumber \\
&\le & 0.5 \|\s\|_2+7.5\|\n\|_2, \nonumber \\
&\le & 0.5 \|\x - \xhat_{i-1}\|_2+7.5\|\n\|_2. \nonumber
\end{eqnarray}
From the recursion on $\xhat_{i}$, we obtain $\|\x-\xhat_i\|_2 \le
2^{-i}\|\x\|_2+15\|\n\|_2$. This completes the proof of
Theorem~\ref{theo:rcosamp}.\qed

\section{Proof of Proposition~\ref{prop:treecount}}
\label{app:treecount}

When $K < \log_2 N$, the number of subtrees of size $K$ of a binary tree of size $N$ is the Catalan number~\cite{RandomTrees}
$$T_{K,N} = \frac{1}{K+1}{2K \choose K} \le \frac{(2e)^K}{K+1},$$
using Stirling's approximation. When $K > \log_2 N$, we partition this count of subtrees into the numbers of subtrees $t_{K,h}$ of size $K$ and height $h$, to obtain
$$T_{K,N} = \sum_{h=\lfloor\log_2K\rfloor+1}^{\log_2 N} t_{K,h}$$
We obtain the following asymptotic identity from~\cite[page 51]{RandomTrees}:
\begin{eqnarray}
t_{K,h} &=& \frac{4^{K+1.5}}{h^4}\sum_{m\ge 1}\left[\frac{2K}{h^2}(2\pi m)^4 - 3(2\pi m)^2\right]e^{-\frac{K(2\pi m)^2}{h^2}} \nonumber \\
&&+4^K\bigo{e^{-\ln^2h}} + 4^K\bigo{\frac{\ln^8h}{h^5}}+4^K\bigo{\frac{\ln^8h}{h^4}}, \nonumber \\
&\le& \frac{4^{K+2}}{h^4}\sum_{m\ge 1}\left[\frac{2K}{h^2}(2\pi m)^4 - 3(2\pi m)^2\right]e^{-\frac{K(2\pi m)^2}{h^2}}. \label{eq:tkh}
\end{eqnarray}

We now simplify the formula slightly: we seek a bound for the sum term (which we denote by $\beta_h$ for brevity):
\begin{eqnarray}
\beta_h &=& \sum_{m\ge 1}\left[\frac{2K}{h^2}(2\pi m)^4 - 3(2\pi m)^2\right]e^{-\frac{K(2\pi m)^2}{h^2}}\nonumber\\
&\le& \sum_{m\ge 1}\frac{2K}{h^2}(2\pi m)^4 e^{-\frac{K(2\pi m)^2}{h^2}}. \label{eq:hsum}
\end{eqnarray}
Let $m_{\max} = \frac{h}{\pi\sqrt{2K}}$, the value of $m$ for which the term inside the sum (\ref{eq:hsum}) is maximum; this is not necessarily an integer. Then,
\begin{eqnarray}
\beta_h & \le& \sum_{m=1}^{\left\lfloor m_{\max} \right\rfloor-1}\frac{2K}{h^2}(2\pi m)^4 e^{-\frac{K(2\pi m)^2}{h^2}} \nonumber \\
&&+\sum_{m=\left\lfloor m_{\max} \right\rfloor}^{\left\lceil m_{\max} \right\rceil}\frac{2K}{h^2}(2\pi m)^4 e^{-\frac{K(2\pi m)^2}{h^2}} \nonumber \\
&&+\sum_{m\ge \left\lceil m_{\max} \right\rceil+1}\frac{2K}{h^2}(2\pi m)^4 e^{-\frac{K(2\pi m)^2}{h^2}}, \nonumber \\
& \le & \int_1^{\left\lfloor m_{\max} \right\rfloor} \frac{2K}{h^2}(2\pi x)^4 e^{-\frac{K(2\pi x)^2}{h^2}}dx\nonumber\\ 
&&+\sum_{m=\left\lfloor m_{\max} \right\rfloor}^{\left\lceil m_{\max} \right\rceil}\frac{2K}{h^2}(2\pi m)^4 e^{-\frac{K(2\pi m)^2}{h^2}} \nonumber \\
&&+ \int_{\left\lceil m_{\max} \right\rceil}^\infty \frac{2K}{h^2}(2\pi x)^4 e^{-\frac{K(2\pi x)^2}{h^2}}dx, \nonumber
\end{eqnarray}
where the second inequality comes from the fact that the series in the sum is strictly increasing for $m\le\left\lfloor m_{\max} \right\rfloor$ and strictly decreasing for $m > \left\lceil m_{\max} \right\rceil$.
One of the terms in the sum can be added to one of the integrals. If we have that
\begin{equation}
(2\pi \left\lfloor m_{\max} \right\rfloor)^4 e^{-\frac{K(2\pi \left\lfloor m_{\max} \right\rfloor)^2}{h^2}} < (2\pi \left\lceil m_{\max} \right\rceil)^4 e^{-\frac{K(2\pi \left\lceil m_{\max} \right\rceil)^2}{h^2}},
\label{eq:twoterms}
\end{equation}
then we can obtain
\begin{eqnarray}
\beta_h & \le & \int_1^{\left\lceil m_{\max} \right\rceil} \frac{2K}{h^2}(2\pi x)^4 e^{-\frac{K(2\pi x)^2}{h^2}}dx\nonumber\\ 
&&+\frac{2K}{h^2}(2\pi \left\lceil m_{\max} \right\rceil)^4 e^{-\frac{K(2\pi \left\lceil m_{\max} \right\rceil)^2}{h^2}} \nonumber \\
&&+ \int_{\left\lceil m_{\max} \right\rceil}^\infty \frac{2K}{h^2}(2\pi x)^4 e^{-\frac{K(2\pi x)^2}{h^2}}dx. \nonumber
\end{eqnarray}
When the opposite of (\ref{eq:twoterms}) is true, we have that
\begin{eqnarray}
\beta_h & \le & \int_1^{\left\lfloor m_{\max} \right\rfloor} \frac{2K}{h^2}(2\pi x)^4 e^{-\frac{K(2\pi x)^2}{h^2}}dx\nonumber\\
&&+\frac{2K}{h^2}(2\pi \left\lfloor m_{\max} \right\rfloor)^4 e^{-\frac{K(2\pi \left\lfloor m_{\max} \right\rfloor)^2}{h^2}} \nonumber \\
&&+ \int_{\left\lfloor m_{\max} \right\rfloor}^\infty \frac{2K}{h^2}(2\pi x)^4 e^{-\frac{K(2\pi x)^2}{h^2}}dx. \nonumber
\end{eqnarray}
Since the term in the sum reaches its maximum for $m_{\max}$, we will have in all three cases that
\begin{equation}
\beta_h \le  \int_1^{\infty} \frac{2K}{h^2}(2\pi x)^4 e^{-\frac{K(2\pi x)^2}{h^2}}dx +\frac{8h^2}{Ke^2}. \nonumber
\end{equation}
We perform a change of variables $u = 2\pi x$ and define $\sigma = h/\sqrt{2K}$ to obtain
\begin{eqnarray*}
\beta_h &\le& \frac{1}{2\pi}\int_0^\infty \frac{1}{\sigma^2}u^4 e^{-u^2/2\sigma^2}dx +\frac{8h^2}{Ke^2}\\
&\le& \frac{1}{2\sigma\sqrt{2\pi}}\int_{-\infty}^\infty \frac{1}{\sqrt{2\pi}\sigma}u^4 e^{-u^2/2\sigma^2}dx+\frac{8h^2}{Ke^2}.
\end{eqnarray*}
Using the formula for the fourth central moment of a Gaussian distribution:
\begin{equation}
\int_{-\infty}^\infty \frac{1}{\sqrt{2\pi}\sigma}u^4 e^{-u^2/2\sigma^2}dx=3\sigma^4, \nonumber
\end{equation}
we obtain
\begin{equation}
\beta_h \le \frac{3\sigma^3}{2\sqrt{2\pi}} +\frac{8h^2}{Ke^2}
= \frac{3h^3}{8\sqrt{\pi K^3}}+\frac{8h^2}{Ke^2}. \nonumber
\end{equation}
Thus, (\ref{eq:tkh}) simplifies to
\begin{equation}
t_{K,h} \le \frac{4^K}{K}\left(\frac{6}{h\sqrt{\pi K}}+\frac{128}{h^2e^2}\right). \nonumber
\end{equation}
Correspondingly, $T_{K,N}$ becomes
\begin{eqnarray*}
T_{K,N} &\le& \sum_{h=\lfloor\log_2K\rfloor+1}^{\log_2 N} \frac{4^K}{K}\left(\frac{6}{h\sqrt{\pi K}}+\frac{128}{h^2e^2}\right), \nonumber \\
&\le& \frac{4^K}{K}\left(\frac{6}{\sqrt{\pi K}}\sum_{h=\lfloor\log_2K\rfloor+1}^{\log_2 N}\frac{1}{h}\right.\\
&&\left.+\frac{128}{e^2}\sum_{h=\lfloor\log_2K\rfloor+1}^{\log_2 N}\frac{1}{h^2}\right). 
\end{eqnarray*}
It is easy to show, using Euler-Maclaurin summations, that
$$\sum_{j=a}^{b} j^{-1} \le  \ln\frac{b}{a-1} \textrm{ and } \sum_{j=a}^{b} j^{-2} \le  \frac{1}{a-1};$$
we then obtain
\begin{eqnarray*}
T_{K,N} &\le& \frac{4^K}{K}\left(\frac{6}{\sqrt{\pi K}}\ln\frac{\log_2 N}{\lfloor\log_2K\rfloor}+\frac{128}{e^2\lfloor\log_2K\rfloor}\right)\\ 
&\le& \frac{4^{K+4}}{Ke^2\lfloor\log_2K\rfloor} \le \frac{4^{K+4}}{Ke^2}. \nonumber
\end{eqnarray*}
This proves the proposition. \qed

\section{Proof of Proposition~\ref{prop:treecomp}}
\label{app:treecomp}

We wish to find the value of the bound (\ref{eq:gaussiancomp}) for the subspace count given in (\ref{eq:treercount}). We obtain $M \ge \max_{1\le j \le \lceil N/K \rceil} M_j$, where
\begin{eqnarray*}
M_j &=& \frac{1}{\left(j^r\sqrt{1+\epsilon_K}-1\right)^2}\\
&&\left(2K+4\ln \frac{(2e)^{K(2j+1)}N}{K(Kj+1)(Kj+K+1)} +2t\right).
\end{eqnarray*}
We separate the terms that are linear on $K$ and $j$, and obtain
\begin{eqnarray*}
M_j &=& \frac{1}{\left(j^r\sqrt{1+\epsilon_K}-1\right)^2}\Bigg(K(3+4\ln 2)+8Kj(1+\ln 2)\\
&&\left.+4\ln\frac{N}{K(Kj+1)(Kj+K+1)} +2t\right), \nonumber \\
&=& \frac{1}{\left(j^{s-0.5}\sqrt{1+\epsilon_K}-j^{-0.5}\right)^2}\\
&&\left(8K(1+\ln 2)+\frac{K(3+4\ln 2)}{j}\right.\\
&&\left.+\frac{4}{j}\ln \frac{N}{K(Kj+1)(Kj+K+1)} +\frac{2t}{j}\right).
\end{eqnarray*}
The sequence $\{M_j\}_{j=1}^{\left \lceil \frac{N}{K} \right \rceil}$ is a decreasing sequence, since the denominators are decreasing sequences whenever $s > 0.5$.
We then have
\begin{eqnarray*}
M &\ge& \frac{1}{\left(\sqrt{1+\epsilon_K}-1\right)^2}\Bigg(K(11+12\ln 2)\\
&&\left.+4\ln\frac{N}{K(K+1)(2K+1)} +2t\right).
\end{eqnarray*}
This completes the proof of Proposition~\ref{prop:treecomp}.
\qed

\section*{Acknowledgements}
We thank Petros Boufounos, Mark Davenport, Yonina Eldar, 
Moshe Mishali, and Robert Nowak for helpful discussions.


\begin{thebibliography}{10}

\bibitem{Mallatbook}
S.~Mallat,
\newblock {\em A Wavelet Tour of Signal Processing},
\newblock Academic Press, San Diego, 1999.

\bibitem{DonohoCS}
D.~L. Donoho,
\newblock ``Compressed sensing,''
\newblock {\em IEEE Trans. Info. Theory}, vol. 52, no. 4, pp. 1289--1306, Sept.
  2006.

\bibitem{CandesCS}
E.~J. Cand\`{e}s,
\newblock ``Compressive sampling,''
\newblock in {\em Proc. International Congress of Mathematicians}, Madrid,
  Spain, 2006, vol.~3, pp. 1433--1452.

\bibitem{richbCS}
R.~G. Baraniuk,
\newblock ``Compressive sensing,''
\newblock {\em {IEEE} Signal Processing Mag.}, vol. 24, no. 4, pp. 118--120,
  124, July 2007.

\bibitem{csmrf}
V.~Cevher, M.~F. Duarte, C.~Hegde, and R.~G. Baraniuk,
\newblock ``Sparse signal recovery using {M}arkov {R}andom {F}ields,''
\newblock in {\em Proc. Workshop on Neural Info. Proc. Sys. (NIPS)}, Vancouver,
  Canada, Dec. 2008.

\bibitem{samplingunion}
T.~Blumensath and M.~E. Davies,
\newblock ``Sampling theorems for signals from the union of finite-dimensional
  linear subspaces,''
\newblock {\em IEEE Trans. Info. Theory}, vol. 55, no. 4, pp. 1872--1882, Apr.
  2009.

\bibitem{dosamplingunion}
Y.~M. Lu and M.~N. Do,
\newblock ``Sampling signals from a union of subspaces,''
\newblock {\em {IEEE} Signal Processing Mag.}, vol. 25, no. 2, pp. 41--47, Mar.
  2008.

\bibitem{Hassibi}
M.~Stojnic, F.~Parvaresh, and B.~Hassibi,
\newblock ``On the reconstruction of block-sparse signals with an optimal
  number of measurements,''
\newblock {\em {IEEE} Trans. Signal Processing}, vol. 57, no. 8, pp.
  3075--3085, Aug. 2009.

\bibitem{EldarUSS}
Y.~Eldar and M.~Mishali,
\newblock ``Robust recovery of signals from a structured union of subspaces,''
\newblock {\em IEEE Trans. Info. Theory}, vol. 55, no. 11, pp. 5302--5316, Nov.
  2009.

\bibitem{DCS}
D.~Baron, M.~F. Duarte, S.~Sarvotham, M.~B Wakin, and R.~G. Baraniuk,
\newblock ``Distributed compressive sensing,''
\newblock 2005,
\newblock Preprint.

\bibitem{CoSaMP}
D.~Needell and J.~Tropp,
\newblock ``{CoSaMP}: {I}terative signal recovery from incomplete and
  inaccurate samples,''
\newblock {\em Applied and Computational Harmonic Analysis}, vol. 26, no. 3,
  pp. 301--321, May 2009.

\bibitem{NowakEM}
M.A.T. Figueiredo and R.D. Nowak,
\newblock ``{An EM algorithm for wavelet-based image restoration},''
\newblock {\em IEEE Trans. Image Processing}, vol. 12, no. 8, pp. 906--916,
  2003.

\bibitem{DaubechiesThresholding}
I.~Daubechies, M.~Defrise, and C.~{De Mol},
\newblock ``An iterative thresholding algorithm for linear inverse problems
  with a sparsity constraint,''
\newblock {\em Comm. Pure Appl. Math.}, vol. 57, pp. 1413--1457, 2004.

\bibitem{CandesPSR}
Emmanuel~J. Cand\`{e}s and Justin~K. Romberg,
\newblock ``Signal recovery from random projections,''
\newblock in {\em Proc. Computational Imaging {III} at {SPIE} Electronic
  Imaging}, San Jose, CA, Jan. 2005, vol. 5674, pp. 76--86.

\bibitem{IHT2}
T.~Blumensath and M.~E. Davies,
\newblock ``Iterative threhsolding for sparse approximations,''
\newblock {\em Journal of Fourier Analysis and Applications}, vol. 14, no. 5,
  pp. 629--654, Dec. 2008.

\bibitem{IHT}
T.~Blumensath and M.~E. Davies,
\newblock ``Iterative hard thresholding for compressed sensing,''
\newblock {\em Applied and Computational Harmonic Analysis}, vol. 27, no. 3,
  pp. 265--274, Nov. 2009.

\bibitem{HMT}
M.~S. Crouse, R.~D. Nowak, and R.~G. Baraniuk,
\newblock ``Wavelet-based statistical signal processing using hidden {M}arkov
  models,''
\newblock {\em IEEE Trans. Signal Processing}, vol. 46, no. 4, pp. 886--902,
  Apr. 1998.

\bibitem{RichBUMD05}
R.~G. Baraniuk,
\newblock ``Fast reconstruction from incoherent projections,'' Workshop on
  Sparse Representations in Redundant Systems, May 2005.

\bibitem{MarcoSPARS05}
M.~F. Duarte, M.~B. Wakin, and R.~G. Baraniuk,
\newblock ``Fast reconstruction of piecewise smooth signals from random
  projections,''
\newblock in {\em Proc. SPARS05}, Rennes, France, Nov. 2005.

\bibitem{LaDoICIP}
C.~La and M.~N. Do,
\newblock ``Tree-based orthogonal matching pursuit algorithm for signal
  reconstruction,''
\newblock in {\em IEEE International Conference on Image Processing (ICIP)},
  Atlanta, GA, Oct. 2006, pp. 1277--1280.

\bibitem{DuarteICASSP08}
M.~F. Duarte, M.~B. Wakin, and R.~G. Baraniuk,
\newblock ``Wavelet-domain compressive signal reconstruction using a hidden
  {Markov} tree model,''
\newblock in {\em IEEE Int. Conf. on Acoustics, Speech and Signal Processing
  (ICASSP)}, Las Vegas, NV, April 2008, pp. 5137--5140.

\bibitem{DoLaCAMSAP}
M.~N. Do and C.~N.~H. La,
\newblock ``Tree-based majorize-minimize algorithm for compressed sensing with
  sparse-tree prior,''
\newblock in {\em International Workshop on Computational Advances in
  Multi-Sensor Adaptive Processing}, Saint Thomas, US Virgin Islands, Dec.
  2007, pp. 129--132.

\bibitem{WaveletBCS}
L.~He and L.~Carin,
\newblock ``Exploiting structure in wavelet-based {B}ayesian compressive
  sensing,''
\newblock {\em {IEEE} Trans. Signal Processing}, vol. 57, no. 9, pp.
  3488--3497, Sep. 2009.

\bibitem{BreslerSIAM}
K.~Lee and Y.~Bresler,
\newblock ``Selecting good {F}ourier measurements for compressed sensing,''
  SIAM Conference on Imaging Science, July 2008.

\bibitem{subgaussian}
S.~Mendelson, A.~Pajor, and N.~Tomczak-Jaegermann,
\newblock ``Uniform uncertainty principle for {B}ernoulli and subgaussian
  ensembles,''
\newblock {\em Constructive Approximation}, vol. 28, no. 3, pp. 277--289, Dec.
  2008.

\bibitem{natarajan}
B.~K. Natarajan,
\newblock ``Sparse approximate solutions to linear systems,''
\newblock {\em SIAM Journal on Computation}, vol. 24, no. 2, pp. 227--234, Apr.
  1995.

\bibitem{BPDN}
S.~S. Chen, D.~L. Donoho, and M.~A. Saunders,
\newblock ``{Atomic Decomposition by Basis Pursuit},''
\newblock {\em SIAM Journal on Scientific Computing}, vol. 20, pp. 33, 1998.

\bibitem{Nowak}
J.~Haupt and R.~Nowak,
\newblock ``Signal reconstruction from noisy random projections,''
\newblock {\em IEEE Trans. Info. Theory}, vol. 52, no. 9, pp. 4036--4048, Sept.
  2006.

\bibitem{CandesDS}
E.~J. Cand\`{e}s and T.~Tao,
\newblock ``The {D}antzig selector: {S}tatistical estimation when $p$ is much
  larger than $n$,''
\newblock {\em Annals of Statistics}, vol. 35, no. 6, pp. 2313--2351, Dec.
  2007.

\bibitem{OMP}
J.~Tropp and A.~C. Gilbert,
\newblock ``Signal recovery from partial information via orthogonal matching
  pursuit,''
\newblock {\em IEEE Trans. Info. Theory}, vol. 53, no. 12, pp. 4655--4666, Dec.
  2007.

\bibitem{STOMP}
D.~L. Donoho, I.~Drori, Y.~Tsaig, and J.~L. Starck,
\newblock ``Sparse solution of underdetermined linear equations by stagewise
  orthogonal matching pursuit,''
\newblock 2006,
\newblock Preprint.

\bibitem{SP}
W.~Dai and O.~Milenkovic,
\newblock ``Subspace pursuit for compressive sensing: {C}losing the gap between
  performance and complexity,''
\newblock {\em IEEE Trans. Info. Theory}, vol. 55, no. 5, pp. 2230--2249, May
  2009.

\bibitem{CandesRIP}
E.~J. Cand\`{e}s,
\newblock ``The restricted isometry property and its implications for
  compressed sensing,''
\newblock {\em Compte Rendus de l'Academie des Sciences, Series I}, vol. 346,
  no. 9--10, pp. 589--592, May 2008.

\bibitem{treekernel}
R.~G. Baraniuk and D.~L. Jones,
\newblock ``A signal-dependent time-frequency representation: {F}ast algorithm
  for optimal kernel design,''
\newblock {\em {IEEE} Trans. Signal Processing}, vol. 42, no. 1, pp. 134--146,
  Jan. 1994.

\bibitem{optimaltree}
R.~G. Baraniuk,
\newblock ``Optimal tree approximation with wavelets,''
\newblock in {\em Wavelet Applications in Signal and Image Processing VII},
  Denver, CO, July 1999, vol. 3813 of {\em Proc. SPIE}, pp. 196--207.

\bibitem{BDKY}
R.~G. Baraniuk, R.~A. DeVore, G.~Kyriazis, and X.~M. Yu,
\newblock ``Near best tree approximation,''
\newblock {\em Advances in Computational Mathematics}, vol. 16, no. 4, pp.
  357--373, May 2002.

\bibitem{RombergHMT}
J.~K. Romberg, H.~Choi, and R.~G. Baraniuk,
\newblock ``Bayesian tree-structured image modeling using wavelet-domain hidden
  {M}arkov models,''
\newblock {\em IEEE Trans. Image Processing}, vol. 10, no. 7, pp. 1056--1068,
  July 2001.

\bibitem{GSM}
J.~Portilla, V.~Strela, M.~J. Wainwright, and E.~P. Simoncelli,
\newblock ``Image denoising using a scale mixture of {G}aussians in the wavelet
  domain,''
\newblock {\em {IEEE} Trans. Image Processing}, vol. 12, no. 11, pp.
  1338--1351, Nov. 2003.

\bibitem{EZW}
J.~Shapiro,
\newblock ``Embedded image coding using zerotrees of wavelet coefficients,''
\newblock {\em IEEE Trans. Signal Processing}, vol. 41, no. 12, pp. 3445--3462,
  Dec. 1993.

\bibitem{CDDD}
A.~Cohen, W.~Dahmen, I.~Daubechies, and R.~A. DeVore,
\newblock ``Tree approximation and optimal encoding,''
\newblock {\em Applied and Computational Harmonic Analysis}, vol. 11, no. 2,
  pp. 192--226, Sept. 2001.

\bibitem{DonohoCART}
D.~Donoho,
\newblock ``{CART} and best ortho-basis: {A} connection,''
\newblock {\em Annals of Statistics}, vol. 25, no. 5, pp. 1870--1911, Oct.
  1997.

\bibitem{DCSNIPS}
M.~B. Wakin, S.~Sarvotham, M.~F. Duarte, D.~Baron, and R.~G. Baraniuk,
\newblock ``Recovery of jointly sparse signals from few random projections,''
\newblock in {\em Proc. Workshop on Neural Info. Proc. Sys. (NIPS)}, Vancouver,
  Nov. 2005.

\bibitem{SOMP}
J.~Tropp, A.~C. Gilbert, and M.~J. Strauss,
\newblock ``Algorithms for simultaneous sparse approximation. {P}art {I}:
  Greedy pursuit,''
\newblock {\em Signal Processing}, vol. 86, no. 3, pp. 572--588, Apr. 2006.

\bibitem{ModelDCS}
M.~F. Duarte, V.~Cevher, and R.~G. Baraniuk,
\newblock ``Model-based compressive Sensing for Signal Ensembles,''
\newblock in {\em Allerton Conference on Communication, Control, and Computing}, Monticello, IL, Oct. 2009.

\bibitem{ModelCSSPARS}
C.~Hegde, M.~F. Duarte, and V.~Cevher,
\newblock ``Compressive sensing recovery of spike trains using a structured
  sparsity model,''
\newblock in {\em Workshop on Signal Processing with Adaptive Sparse Structured
  Representations (SPARS)}, Saint Malo, France, Apr. 2009.

\bibitem{ModelCSSAMPTA}
V.~Cevher, P.~Indyk, C.~Hegde, and R.~G. Baraniuk,
\newblock ``Recovery of clustered sparse signals from compressive
  measurements,''
\newblock in {\em Int. Conf. on Sampling Theory and Applications (SAMPTA)},
  Marseille, France, May 2009.

\bibitem{CandesDLP}
E.~J. Cand\`{e}s and T.~Tao,
\newblock ``Decoding by linear programming,''
\newblock {\em IEEE Trans. Info. Theory}, vol. 51, pp. 4203--4215, Dec. 2005.

\bibitem{ledoux}
M.~Ledoux,
\newblock {\em The Concentration of Measure Phenomenon},
\newblock American Mathematical Society, 2001.

\bibitem{RandomTrees}
G.~G. Brown and B.~O. Shubert,
\newblock ``On random binary trees,''
\newblock {\em Mathematics of Operations Research}, vol. 9, no. 1, pp. 43--65,
  Feb. 1984.

\end{thebibliography}
\end{document}